\documentclass[twocolumn,showpacs,preprintnumbers,amsmath,amssymb,prb,nofootinbib]{revtex4}

\usepackage{graphicx}
\usepackage{dcolumn}
\usepackage[tight]{subfigure}
\usepackage{amsmath}
\usepackage{amsfonts}
\usepackage{bbm} 
\usepackage{verbatim}
\usepackage{float}
\usepackage[usenames,dvipsnames]{color}
\usepackage{bm} 
\usepackage{cases}  
\usepackage{mathtools} 
\usepackage{dsfont}
\usepackage{cancel} 

\usepackage{mathrsfs} 
\usepackage{enumitem} 

\usepackage{etoolbox} 

\usepackage{hyperref} 
\usepackage[normalem]{ulem} 
\usepackage{xspace}

\newrobustcmd{\ARNAIVE}{Splettstoesser06,Splettstoesser08a,Winkler09,Reckermann10a,Calvo12a,Avron12,Haupt13,Riwar13,Winkler13,Rojek14}
\newrobustcmd{\INTERACTIONINDUCED}{Sinitsyn07EPL,Sinitsyn07PRB,Sinitsyn09,Reckermann10a,Ren10,Yuge13,Riwar13,Yoshii13,Nakajima15}
\newrobustcmd{\FCS}{Sinitsyn07EPL,Ren10,Yuge12,Liu13,Yuge13,Nakajima15} 

\newrobustcmd{\tot}[1]{{#1}^{\text{tot}}}
\newrobustcmd{\D}[1]{#1}
\newrobustcmd{\env}[1]{{#1}^{\text{res}}}

\newrobustcmd{\unit}{\mathbbm{1}}   
\newrobustcmd{\bra}[1]{\langle #1|}   
\newrobustcmd{\ket}[1]{|#1 \rangle}   
\newrobustcmd{\braket}[1]{\langle #1 \rangle}   
\newrobustcmd{\eval}[2]{\left . #1 \right | _{#2}}

\newrobustcmd{\delR}{\vec{\nabla_R}}

\newrobustcmd{\Ldag}[1]{{{#1}^{\pmb{\dagger}}}}
\newrobustcmd{\Lket}[1]{ \pmb{|} {#1} \pmb{)} }
\newrobustcmd{\Lbra}[1]{ \pmb{(} {#1} \pmb{|} }
\newrobustcmd{\Lbraket}[1]{{\pmb{(}  {#1}  \pmb{)}} }
\newrobustcmd{\expec}[1]{{\langle {#1}  \rangle }}
\newrobustcmd{\pureD}[1] {\ket{#1}\bra{#1}}

\newrobustcmd\scalemath[2]{\scalebox{#1}{\mbox{\ensuremath{\displaystyle #2}}}}

\newrobustcmd{\Tr}[1]{\underset{#1}{\mathsf{Tr}}}   
\newrobustcmd{\tr}{\mathsf{Tr}}   
\newrobustcmd{\nohat}[1]{#1}   
\newrobustcmd{\sdagger}{{\dagger}}   

\newrobustcmd{\vecg}[1]{{\bm #1}}   
\renewrobustcmd{\vec}[1]{{\mathbf{#1}}}   

\newrobustcmd{\timeint}[4]{ \underset{#1 \geq #2 \geq #3 \geq #4}{\int d #2 d #3 } }
\newrobustcmd{\timeintfour}[6]{ \underset{#1 \geq #2 \geq #3 \geq #4 \geq #5 \geq #6}{\int d #2 d #3 d #4 d #5} }

\newrobustcmd{\hl}[1]{{\textcolor{magenta}{#1}}}
\renewrobustcmd{\hl}[1]{{\textcolor{black}{#1}}}

\newrobustcmd{\fcs}{FCS\xspace}
\newrobustcmd{\ase}{ASE\xspace}
\newrobustcmd{\ar}{AR\xspace}

\renewrobustcmd{\tot}{\hl{\text{tot}}} 
\renewrobustcmd{\S}{\hl{\text{S}}}  
\renewrobustcmd{\i}{\hl{\text{i}}} 
\renewrobustcmd{\a}{\text{r}} 

\newrobustcmd{\T}{\hl{\mathcal{T}}} 
\newrobustcmd{\M}{\hl{\text{M}}}  
\newrobustcmd{\R}{\hl{\text{R}}}  
\newrobustcmd{\g}{\hl{\text{g}}}  
\newrobustcmd{\n}{\hl{\text{n}}}  
\newrobustcmd{\p}{\hl{l}}  
\newrobustcmd{\C}{\hl{c}} 
\newrobustcmd{\mom}{\hl{\mathcal{M}}}
\newrobustcmd{\cum}{\hl{\mathcal{C}}}
\newrobustcmd{\curve}{\hl{\mathscr{C}}}

\newrobustcmd{\Ref}[1]{Ref.~\onlinecite{#1}}   
\newrobustcmd{\Refs}[1]{Refs.~\onlinecite{#1}}   
\newrobustcmd{\Tab}[1]{Table~\ref{#1}}   
\newrobustcmd{\tab}[1]{\ref{#1}}   
\newrobustcmd{\Fig}[1]{Fig.~\ref{#1}}   
\newrobustcmd{\Eq}[1]{Eq.~(\ref{#1})}   
\newrobustcmd{\Eqb}[1]{Equation~(\ref{#1})}   
\newrobustcmd{\Eqs}[1]{Eqs.~(\ref{#1})}   
\newrobustcmd{\eq}[1]{(\ref{#1})}   
\newrobustcmd{\App}[1]{App.~\ref{#1}}
\newrobustcmd{\app}[1]{\ref{#1}}   
\newrobustcmd{\Sec}[1]{Sec.~\ref{#1}}   
\renewrobustcmd{\sec}[1]{\ref{#1}}   

\definecolor{myred}{rgb}{0.9,0,0}
\definecolor{myblue}{rgb}{0,0,0.5}
\definecolor{mygreen}{rgb}{0,0.6,0}

\robustify{\sum}
\robustify{\int}
\robustify{\nonumber}

\begin{document}

\title{Meter calibration and the geometric pumping process in open quantum systems}
\author{T. Pluecker$^{(1,2)}$}
\author{M. R. Wegewijs$^{(1,2,3)}$}
\author{J. Splettstoesser$^{(4)}$}

\affiliation{
(1) Institute for Theory of Statistical Physics,
RWTH Aachen, 52056 Aachen,  Germany
\\
(2) JARA-FIT, 52056 Aachen, Germany
\\
(3) Peter Gr{\"u}nberg Institut,
Forschungszentrum J{\"u}lich, 52425 J{\"u}lich,  Germany
\\
(4) Department of Microtechnology and Nanoscience (MC2),
Chalmers University of Technology, SE-41298 G{\"o}teborg, Sweden
}
\date{\today}
\pacs{
  73.63.Kv,
   05.60.Gg,
		72.10.Bg 	
		03.65.Vf
 }
\begin{abstract}
We consider the process of pumping charge through an open quantum system,
motivated by the example of a quantum dot with strong repulsive or attractive electron-electron interaction. Using the geometric formulation of adiabatic \emph{nonunitary} evolution put forward by Sarandy and Lidar, we derive an encompassing approach to ideal charge measurements of time-dependently driven transport, that stays near the familiar approach to closed systems. Following Schaller, Kie{\ss}lich and Brandes we explicitly account for a meter that registers the transported charge outside the system. The gauge freedom underlying geometric pumping effects in \emph{all moments} of the transported charge emerges naturally as the \emph{calibration} of this \emph{meter}. Remarkably, we find that geometric and physical considerations cannot be applied independently as done in closed systems: \emph{physical} recalibrations do not form a group due to constraints of positivity (Bochner's theorem). This complication goes unnoticed when considering only the average charge but it is relevant for understanding the origin of geometric effects in the higher moments of the charge-transport statistics. As an application we derive two prominent existing approaches to pumping, based on full-counting-statistics (FCS) and adiabatic-response (AR), respectively, from our approach in a transparent way. This allows us to reconcile all their apparent incompatibilities, in particular the puzzle how for the average charge the \emph{nonadiabatic stationary-state} AR result can exactly agree with the \emph{adiabatic nonstationary} FCS result. We relate this and other difficulties to a single characteristic of geometric approaches to \emph{open} systems: the system-environment boundary can always be chosen to either include or exclude the ideal charge meter. This leads to a physically motivated relation between the mixed-state Berry phase and the entirely different geometric phase of Landsberg.\end{abstract}

\maketitle

\section{Introduction\label{sec:intro}}
\subsection{Geometric phases in open systems}

In condensed matter physics geometric and topological quantities are generally appreciated for their robustness against perturbations.
In physics, topological quantities often arise from geometric ones and the former have been used to classify phases of isolated solids at zero temperature\cite{Altland97,Read00,Snyder08,Nayak08,Kitaev09Conference,Ryu10,Hasan10,Qi11,Budich15a,Chiu15,Kennedy16}.
Recently, this topological classification has also been addressed for finite temperature and nonequilibrium\cite{Diehl11a,Bardyn13,Iemini16}
which is more challenging since it requires a description in terms of mixed-state density operators\cite{Huang14,Viyuela15,Budich15a,Budich15b} rather than pure-state vectors.
There is, however, a further complicating factor for such \emph{open} systems:
due to the diversity of kinetic equations that describe their nonunitary dynamics
there is a much wider variety of geometric formulations\cite{Uhlmann86,Sjoeqvist00,Sarandy05,Sarandy06} than for closed systems which are all described by the Schr\"odinger equation.
Despite this strong impetus to establish relations between geometric  approaches in open systems,
little attention has been given to such comparisons so far~\cite{Nakajima15,Pluecker17a}.

An important issue that is currently being addressed in several of the above works~\cite{Diehl11a,Bardyn13,Iemini16,Huang14,Viyuela15,Budich15a,Budich15b,Bardyn17a}
is to find observables that are sensitive to given geometric and topological quantities that one defines for an open system.
Some works \cite{Splettstoesser06,Calvo12a,Avron11,Sinitsyn07EPL,Nakajima15}
follow an opposite line of questioning
by first focusing on observable transport effects
and then expressing these in geometric terms.
Identifying geometric quantities this way,
even in relatively simple open systems, seems an essential first step towards finding effects indicative of topological order.
In closed systems\cite{Thouless83,Cohen03}
\emph{pumping}, i.e., transport due to slow periodic driving,
was found to probe geometry and topology
and one might expect the same for open systems.
However, as we recently highlighted~\cite{Pluecker17a}, the geometric phases that one encounters for pumping through open quantum systems
do \emph{not} relate to the reduced quantum \emph{state} of the system (obtained by tracing out the environment):
instead, \emph{external observables} --or related generating functions-- accumulate these geometric phases.
Another issue concerns
two prominent open-system approaches that deal with pumping in this way,
the adiabatic-response approach (\ar)
and the full counting statistics (\fcs):
Despite their completely opposite features
they were shown to give identical results for the average pumped charge~\cite{Nakajima15,Pluecker17a}.
Reconciling these features requires a deeper understanding of the distinction between geometric phases in open and closed quantum systems
and motivates the present work.

\begin{figure}[H] 
	\centering{
		\includegraphics[width=0.85\linewidth]{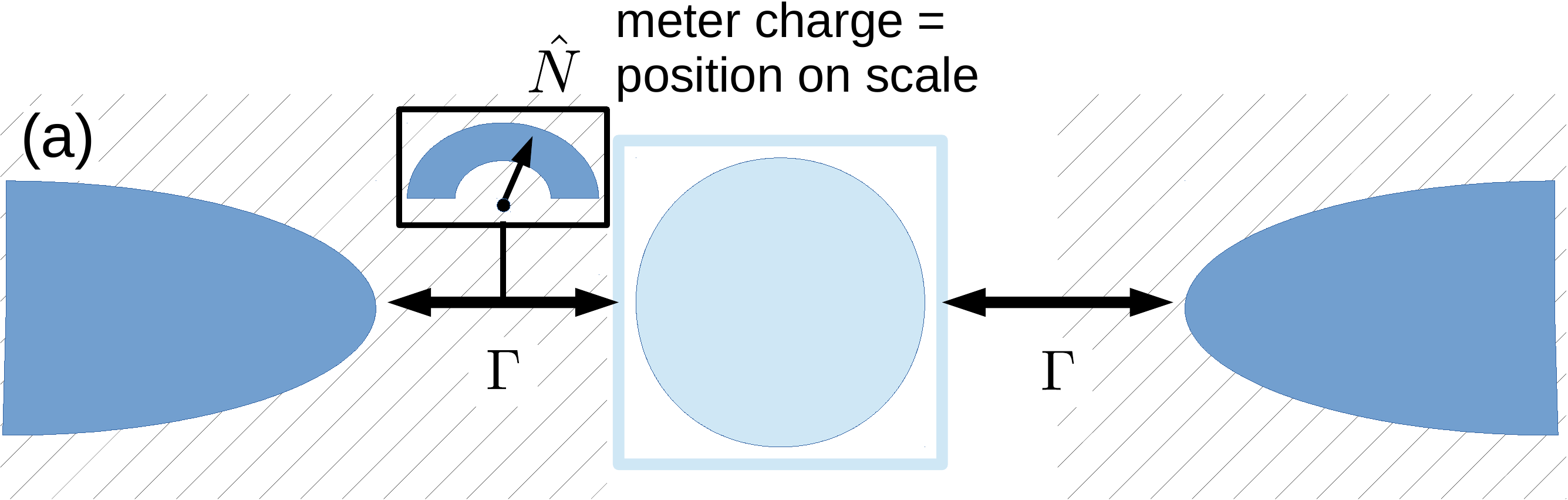}
	}
	\\
	\centering{
		\includegraphics[width=0.85\linewidth]{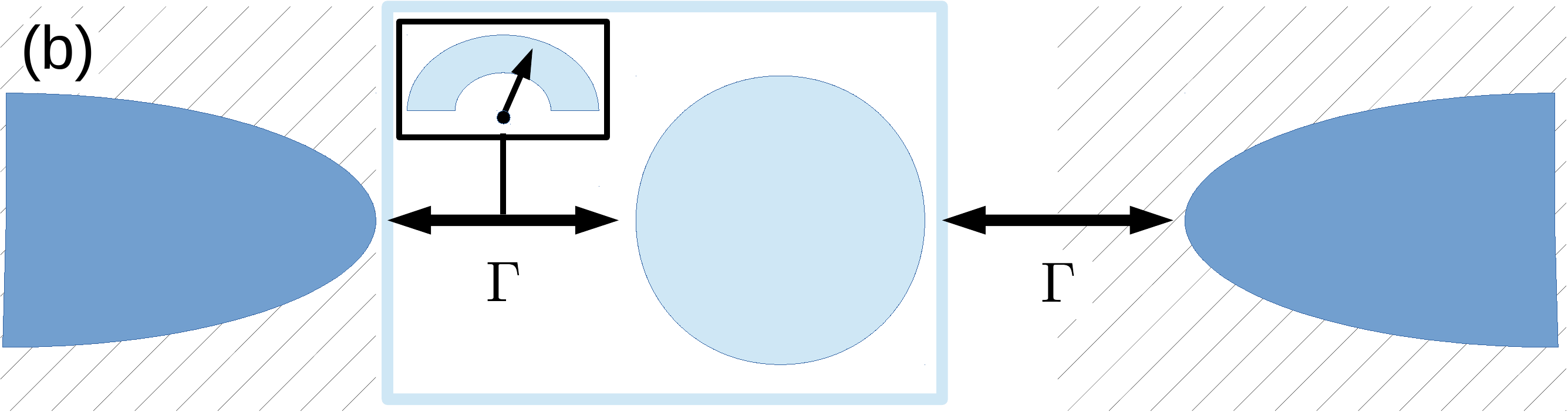}
	}
	\\
	\centering{
		\includegraphics[width=0.85\linewidth]{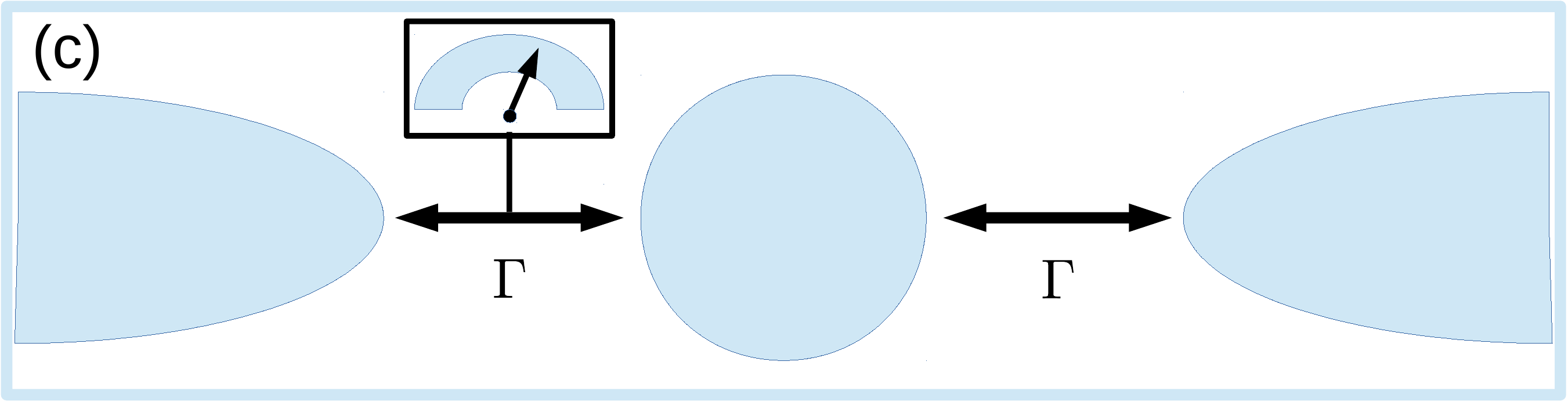}
	}
	\caption{
		Equivalent ways of accounting for charge pumping through an open system (circle)
			with coupling $\Gamma$.
		(a) \emph{Meter outside}: the meter is considered as part of the environment,
		together with the electrodes, all of which are traced out (hatched).
		(b) \emph{Meter inside}: the meter is considered as part of a \emph{composite open system}:
		only the electrodes are integrated out (hatched).
		(c) Total system with nothing integrated out.
	}
	\label{fig:model}  
\end{figure}

The key questions can be outlined further,
guided by the role of a meter that registers the charge $\hat{N}$
without backaction\footnote
	{Integrating out the meter is equivalent to having no meter at all.}:
either the meter lies outside (a) or inside (b) the open-system boundary
as sketched in \Fig{fig:model}.

(a) \emph{Meter outside}.
Steady-state pumping through an open quantum system has been studied in the way sketched in \Fig{fig:model}(a) in several works~\cite{\ARNAIVE}.
In \Ref{Pluecker17a} it was shown, that in this approach geometric pumping effects do not result from a geometric phase of the \emph{state} (reduced density operator).
Indeed, when computing this phase in the steady state limit, it turns out to be zero.
Instead, the pumped \textit{charge} per period
was shown to equal a single Landsberg-type geometric phase~\cite{Landsberg92,Landsberg93,Andersson03thesis,Andersson05,Sinitsyn09} that arises naturally when computing the adiabatic response~\cite{Splettstoesser06} (\ar) of the current that is linear in the velocity of the driving parameters.

\begin{figure}[H]
	\centering\includegraphics[width=0.9\linewidth]{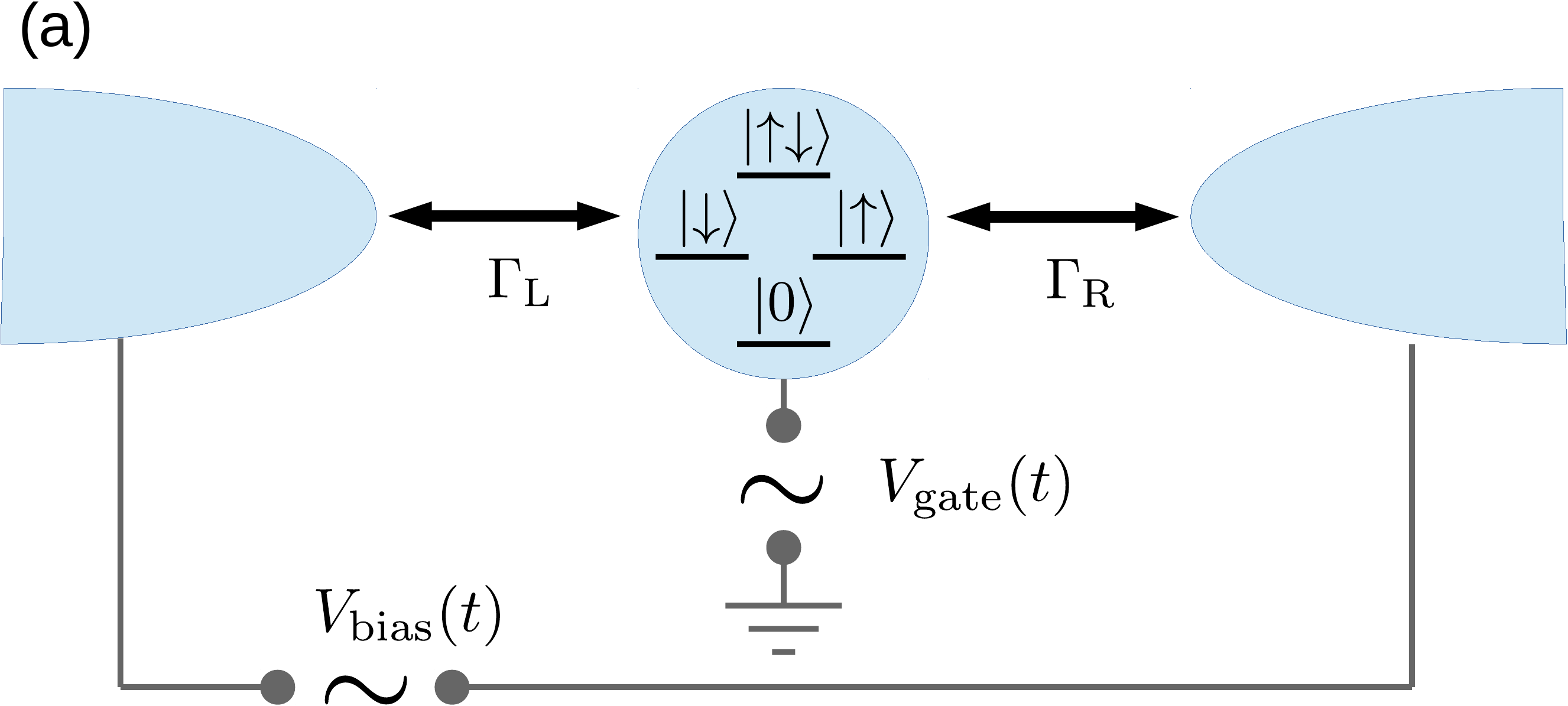}
	\vspace{0.5cm}\\
	\centering\includegraphics[width=0.9\linewidth]{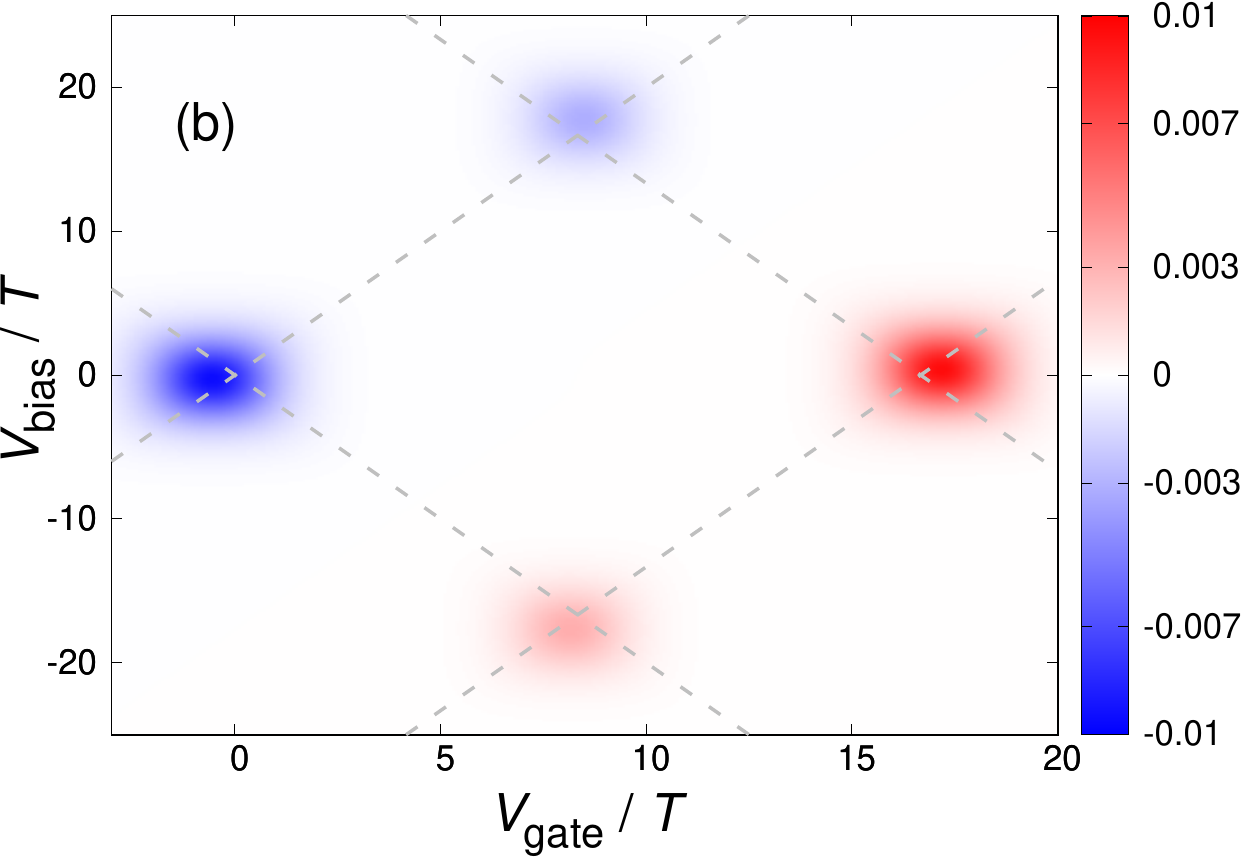}
	\centering\includegraphics[width=0.9\linewidth]{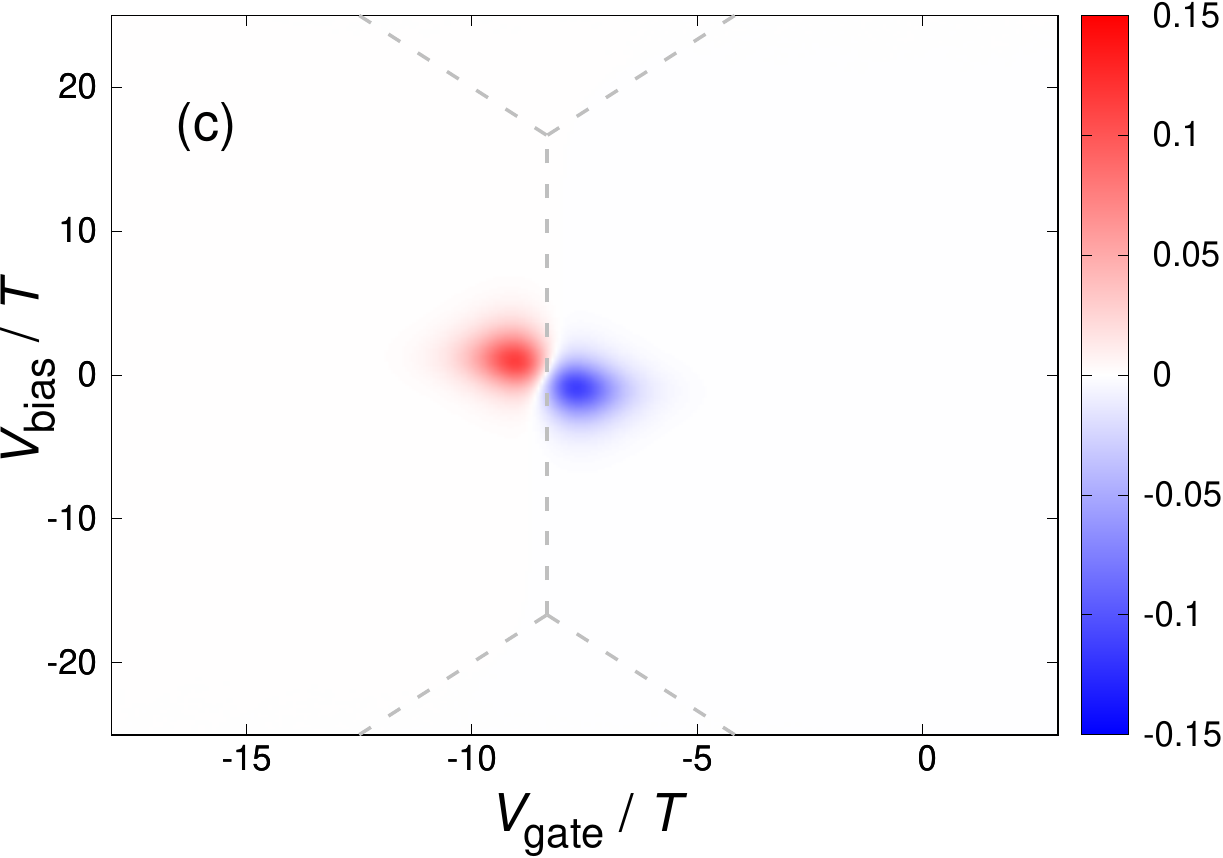}
	\caption{
		(a) 
		Weakly coupled, strongly interacting quantum dot
		hosting up to two electrons
		as introduced in \Sec{sec:model}.
		The time-dependently driven parameters $\vec{R}=(V_\text{gate}/T, V_\text{bias}/T)$ are the gate- and bias-voltage.
		The charge pumped through such an open system in steady state
		can be equivalently computed using either of two very different approaches:
		the geometric part ($\g$) of the transported charge equals
		$\mom^{(1)}_\g = \oint_{\curve} d\vec{R} A
		\text{ [\ar approach, \Eq{eq:DeltaN-ar}] }
		=\cum^{(1)}_\g =
		\oint_{\curve} d\vec{R} \partial_{i\phi} A^\phi |_{\phi=0}
		\text{ [\fcs approach, \Eq{eq:DeltaN-fcs}]}$,
		as discussed in more detail in the paper.
		The two plots illustrate the result (obtained in either way) for two regimes:
		(b)~Repulsive interaction $U=|U| > 0$.
		Plotted is the pumping curvature, the charge pumped through the quantum dot per unit area covered by an infinitesimal driving cycle $\curve$,
		as function of the center $\vec{R}$ of the cycle.
		The pumping signal is nonzero only at points where parametric resonances of the system cross (dashed Coulomb-diamond edges)
		and provides \emph{direct} information~\cite{Reckermann10a,Calvo12a,Pluecker17a} on spin and coupling asymmetry
		that is not provided by the stationary transport.
		(c)~Attractive interaction $U=-|U|<0$, experimentally realized only recently~\cite{Cheng15,Cheng16}.
		At zero bias $V_\text{bias}=0$ a single two-lobed pumping response emerges at the electron-pair resonance $V_\text{gate}=-U/2$ (vertical dashed line).
		Parameters for (b) and (c):
		$\Gamma_\text{L} / \Gamma_\text{R} = 0.5$.
		$|U| = 16.67 \, T \gg \Gamma_\text{L} + \Gamma_\text{R}$, and temperature $T$.
	}
	\label{fig:pump}  
\end{figure}

For concreteness, we show two examples for such charge pumping in \Fig{fig:pump}, which can both be accessed with this formalism.
Interestingly, in these examples pumping is generated by strong interaction effects~\cite{\INTERACTIONINDUCED} which in the weak coupling limit can readily be treated using the open-system (density-operator) approach.
The geometric origin of the pumping was shown to lie in the nonuniqueness of assigning an operator to an observable measured \emph{outside}\footnote
	{Applied to \emph{system} observables this also covers~\cite{Pluecker17a} the Kato-projection approach of Avron, Fraas, and Graf~\cite{Avron11,Avron12}.}
the open system, i.e., in the hatched area in \Fig{fig:model}(a):
One is free to add to the charge operator an offset $g$ that depends on the driving parameters, collected in a vector $\vec{R}$,
\begin{align}
\hat{N}_g(\vec{R})= \hat{N} + g(\vec{R}) \mathds{1}
\label{eq:GaugeFreedom}
,
\end{align}
without altering the pumped charge~\cite{Sinitsyn09,Pluecker17a}.
Physically, this is a literal gauge, i.e., a recalibration of the scale used for reading out a charge meter.
These physical gauge transformations fully explain the emergence of a geometric first moment
$\mom^{(1)} := \braket{\hat{N}}(\T) - \braket{\hat{N}}(0)$
upon driving the parameters during one period $\T$.
However, driven transport also involves geometric effects for higher moments $\mom^{(k)}$ such as the pumping noise~\cite{Riwar13} related to $\braket{\hat{N}^2}$. 
The calibration freedom \eq{eq:GaugeFreedom} does not explain their occurrence
and so far it has remained unclear which further \emph{physical} gauge freedom is responsible for this.

One should note that the computation of the average transported charge requires 
a memory-kernel for the \emph{observable}
(in addition to a state-evolution kernel)
because the reservoirs and meter have been integrated out.
It was shown that
through this latter observable kernel the gauge freedom \Eq{eq:GaugeFreedom}, responsible for the geometric pumping  phase, enters the problem.
Also, the nonzero value of this phase
could be directly tied to the retarded response (``lag'') of the current through the system to the driving parameters as measured by an outside observable:
the pumping current is therefore \emph{nonadiabatic}.

As highlighted in \Ref{Pluecker17a} the \ar approach has the advantage of leading to technically simple calculations, producing such examples as \Fig{fig:pump},
while at the same time tying geometric quantities to concrete transport physics.
For example, the geometric connection \emph{equals} the physical pumping current.
However, even without going into detail, one notices that the features of this approach, listed in Table \tab{tab:compare}, go against virtually all familiar ideas about the appearance of geometric phases in closed quantum systems (adiabaticity, quantum-state phase accumulation, Berry-Simon type connection, etc.).
This paper addresses the important question how and to what extent this familiar picture can be restored by starting from a different, yet equivalent, formulation.
In particular, how does the phase-accumulation of the charge \emph{observable} emerge from the phase of some quantum \emph{state}?

(b) \emph{Meter inside}.
Another prominent approach to pumping proceeds as sketched in \Fig{fig:model}(b):
the meter is part of the open system (not integrated out).
The \fcs approach keeps track of the meter by a formal counting-field parameter $\phi$, a statistical tool\cite{Esposito09rev}.
The charge-transfer statistics is computed from
a generating function $Z^\phi$
via a generating-operator $\rho^\phi$
which describes the first cumulant
$\cum^{(1)} := \braket{\hat{N}}(\T) - \braket{\hat{N}}(0)$
but also all higher cumulants $\cum^{(k)}$.
Sinitsyn and coworkers applied this approach to the problem of pumping~\cite{\FCS}
and by an adiabatic decoupling obtained a result for the cumulant generating function $z^\phi = \ln Z^\phi$
which is nonstationary for fixed parameters.
For each continuous value of $\phi$ it has a pumping contribution that is a geometric phase of the Berry-Simon type\cite{Berry84,Simon83}.
The geometric part of the first cumulant $\cum^{(1)}$ is the measurable pumped charge and it is plotted in \Fig{fig:pump} for our example.

The features of the \fcs approach listed in Table \tab{tab:compare} are similar to the closed-system geometric approach.
However, this similarity is formal: it is not clear to which physical \emph{state} the adiabaticity and geometric phase refer.
This is a serious concern --raised already in~\Ref{Nakajima15}-- since by its adiabatic decoupling approximation the \fcs approach to pumping (correctly) produces the same result as the \ar approach, in which the current is undeniably nonadiabatic\cite{Pluecker17a}, see \Fig{fig:pump}.
In fact, all the features of the \fcs and \ar approach listed in \Tab{tab:compare} seem to be in conflict.
Nevertheless, for the first moment / cumulant they produce identical results within their limits of applicability as shown in detailed analyses~\cite{Nakajima15,Pluecker17a}.
These latter works, however, do not give a simple explanation or, even better, a single systematic approach that rationalizes all their differences in a simple and transparent way.

A further question that received little attention so far relates to the full set of moments / cumulants, i.e., the \emph{entire} transport process.
It is an advantage of the \fcs approach that it gives a formal expression for \emph{all} cumulants $\cum^{(k)}$ of the transported charge which
shows that they all have a geometric contribution. However, their properties have not been discussed much.
In particular, as we will see here,
the geometric gauge transformations may transform proper generating functions to functions that violate Bochner's positivity criterion~\cite{BochnerLecture},
implying their inverse Fourier transforms take negative values
and do \emph{not} correspond to charge-transfer probability distributions anymore, cf. \Ref{Levitov96}.
The nontrivial positivity restriction is typical of the open-system (density-operator) setting
and its interplay with geometric considerations requires clarification if gauge transformations affecting the higher moments are to be understood physically.

\subsection{This paper}

Motivated by the above, we relate in this paper the charge pumping process to geometric phases of the adiabatically evolving \emph{mixed} quantum \emph{state}.
We achieve this by including an ideal \emph{charge meter} in the total system sketched in \Fig{fig:model}(c)
following Schaller et al.\cite{Schaller09}
and
by applying the adiabatic (mixed-)state evolution (\ase) approach,
developed by Sarandy and Lidar~\cite{Sarandy05,Sarandy06}.
Geometric pumping through an open quantum system can then be understood in a formulation that stays near the established intuition of the familiar closed-system formalism.

We hereby go beyond\footnote
	{Although in~\Ref{Pluecker17a} the \ase approach~\cite{Sarandy05,Sarandy06} also played a role,
	it was not applied to the composite system-plus-meter state
	which is the crucial advance that we report here.}
recent discussions of the technical equivalence of pumping approaches\cite{Nakajima15,Pluecker17a}.
From our new vantage point
both the \ar and \fcs approach to pumping can be derived and understood transparently:
all their opposing features, listed in \Tab{tab:compare},
can be tied 
\par
\begingroup \squeezetable
\begin{table*}
	\caption{\label{tab:compare}
		Comparison of approaches to pumping
		guiding the discussion in Sections \sec{sec:ase-review}-\sec{sec:ar}.
		Notation:
		$\hat{n}$ is the \emph{reservoir} charge operator
		(used traditionally to account for the transported charge),
		$\hat{N}$ is the \emph{meter} charge operator
		(needle position in \Fig{fig:model}, used here to register the transported charge).
		The meter phase $\phi$ (FCS counting field) is the momentum conjugate to the charge $\hat{N}$
		(meter needle position)
		as discussed in \Sec{sec:model}.
		In this table the subscript ``g'' indicates that only the geometric or pumping part of a quantity is meant,
		see indicated sections for further notation and details.
		\\
	}
	\begin{ruledtabular}
		\begin{tabular}{ll|l||l}
			\textbf{Feature}
			& \textbf{Adiabatic state-evolution (\ase)}					
			& \textbf{Full counting statistics (\fcs)}
			&  \textbf{Adiabatic response (\ar)}
			\\
			& \Sec{sec:meter} $\leftarrow$ \Sec{sec:model} $+$ \Sec{sec:ase-review}
			& \Sec{sec:fcs}
			 (where $\hat{n}\to \hat{N}$)
			& \Sec{sec:ar}
			(where $\hat{n}\to \hat{N}$)
			\\
			&&
			\\
			\textbf{Object}
			& State $\rho$ of system + meter
			& Counting operator $\rho^\phi$ $+$ counting-field $\phi$
			& State system $\rho^\S$ + reservoir charge $\braket{\hat{n}}$
			\\
			\textbf{Observables}
			& Expectation values: $\braket{\hat{N}^k}(t)$
			& Cumulants $\cum^1=\braket{\hat{n}}(\T)-\braket{\hat{n}}(0)$, $\ldots$
			& Moment $\mom^1=\int^{\T}_0 dt \braket{\hat{I}_n}(t)$
			\\
			\textbf{Driving}
			& Adiabatic \phantom{$N^{k}_{l}$}
			& Adiabatic
			& Nonadiabatic
			\\
			\textbf{Evolution}
			& Nonstationary (parametrically) \phantom{$N^{k}_{l}$}
			& Nonstationary (parametrically)
			& Stationary (parametrically)
			\\
			\textbf{Connection}
			& Berry-Simon: $A^\phi = \Lbraket{\bar{w}^\phi|\delR w^\phi}$
			& Berry-Simon: $A^\phi = \Lbraket{\bar{w}^\phi|\delR w^\phi}$
			& Landsberg: $A=\Lbra{\unit}W_{I_N} \frac{1}{W} \Lket{\delR \rho}$
			\\
			\textbf{Geometric phase}
			& Continuum of phases $\oint d\vec{R} A^\phi$
			& Continuum of phases $\oint d\vec{R} A^\phi$ 
			& Single phase $\oint d\vec{R} A$
			\\
			\textbf{Gauge freedom}
			& Meter calibration: $\Lket{\phi} \to G^\phi \Lket{\phi}$
			& System normalization: $\Lket{w^\phi_0} \to G^\phi \Lket{w^\phi_0} $
			& Reservoir charge shift: $\hat{n} \to \hat{n}+g \unit$
			\\
			\textbf{Parallel transport}
			& Moment currents: $\tfrac{d}{dt} \braket{\hat{N}^k }_\g = 0$ for all $k$
			& Cumulant currents $\tfrac{d}{dt} \cum^{(k)}_\g = 0$ for all $k$
			& Current $\tfrac{d}{dt}\braket{\hat{n}}_\g = 0$
			\\
			& $\iff$
			$A^\phi \dot{\vec{R}} + \tfrac{d}{dt} g^\phi=0$ for all $\phi$
			& $\iff$
			$A^\phi \dot{\vec{R}} + \tfrac{d}{dt} g^\phi=0$ for all $\phi$
			& $\iff$
			$A\dot{\vec{R}} + \tfrac{d}{dt} g=0$
		\end{tabular}
	\end{ruledtabular}
\end{table*}\endgroup
\noindent
to a single characteristic of open-system descriptions,
namely, that
the boundary between system and environment can be redrawn to include [\Fig{fig:model}(a)] or exclude [\Fig{fig:model}(b)] the meter,
thereby completely altering the physical and geometrical description at every step,
but without changing the final technical outcome.
This comprehensive comparison has never been done and
is important for the various reasons that we outlined above.

Most prominently, we directly tie the physical origin of geometric effects on the entire pumping process to the simple fact that a physical meter can be recalibrated.
This generalizes an earlier result~\cite{Pluecker17a} restricted only to the pumped charge (first moment)
to the \emph{entire} transport process (all moments).
We show that this is the origin of geometric pumping effects both in the \fcs and \ar approach, despite all their differences.
In particular, in the \fcs this point becomes clear when treating the counting field as the physical momentum conjugate to the meter charge.
We furthermore find the announced nontrivial conflict between the physical positivity requirements and geometric considerations that one does not encounter in closed quantum systems:
physical meter calibrations do not form a transformation \emph{subgroup}
of the full group of allowed geometric gauges.

On a more general level our findings highlight
that taking measurements explicitly into account is important
for the geometric and --\emph{a fortiori}-- the topological characterization of open systems.
Also, by explicitly accounting for the meter, our approach provides a suitable starting point for consideration of more complicated situations than addressed in this paper
where nonideal aspects of the charge measurement process need to be included, such as backaction and spontaneous charge-symmetry breaking in superconducting systems.

\subsection{Outline.}
The paper is organized as follows.
In \Sec{sec:model} we introduce a Hamiltonian model which includes an ideal meter as sketched in \Fig{fig:model}(c).
Integrating out the reservoirs, as in \Fig{fig:model}(b), we set up a master equation for the composite quantum state of \emph{system plus meter} and carefully discuss which momentum (counting-field) is actually conjugate to the charge registered by the meter.
Then, in \Sec{sec:ase-review} we outline the adiabatic state-evolution (\ase) approach of Sarandy and Lidar~\cite{Sarandy05,Sarandy06} to open systems
and show in \Sec{sec:meter} how it
describes the entire charge-pumping process
when applied to the composite system-plus-meter \emph{state}.
This provides the basis for resolving all mentioned concerns regarding \ar and \fcs approach:
in \Sec{sec:fcs}, we show how our \ase approach to system-plus-meter dynamics --through standard, calibrated measurements on the meter--
naturally includes the \fcs approach which focuses on charge-transfer statistics.
All features of the \fcs approach, in particular, the adiabaticity, can be clearly understood physically by considering the system plus meter.
In \Sec{sec:ar} we show how our approach
also naturally leads to the \ar approach
--with its completely opposite feature list--
by going to a reduced description in which the meter is integrated out
as in \Fig{fig:model}(a).
This allows us to explain in particular how the less familiar Landsberg geometric phase for the \emph{observable} in the \ar approach emerges \emph{physically} from the more familiar Berry-Simon geometric phase of the composite \emph{state} of system-plus-meter.
We summarize our findings in \Sec{sec:conclusion}
and indicate how the key ideas presented can be extended beyond the simple limits that we intentionally focus on here for purpose of clarity.
\section{Model, meter and master equation\label{sec:model}}

\paragraph*{Hamiltonian model.}
In \Fig{fig:model}(c) we sketch a model of an ideal meter
that we introduce following Schaller et al.\cite{Schaller09}.
In order to count the net number of electrons entering one of the reservoirs, say the left one, we extend the system plus reservoirs by a suitable Hilbert space to model a meter.
This can be any space spanned by an infinite discrete set of orthonormal vectors
$
\left \{ \ket{N} , N \in \mathds{Z} \right \}
$.
When an electron enters (leaves) the left reservoir one ``counts'' by changing the meter state from $\ket{N} \to \ket{N\pm 1}$.
The charge operator on the meter thus reads
\begin{align}
\hat{N} = \sum_{N=-\infty}^{\infty} N \ket{N}\bra{N} 
\label{eq:N}
.
\end{align}
Through this formal trick we are able to explicitly keep track of observable operators outside the system even after integrating out the reservoirs below.

We assume the meter is \emph{ideal} in the sense that it has no internal dynamics (zero Hamiltonian). As a result, the choice of the meter-space is nonunique:
it is possible to shift $\ket{N} \to \ket{N - \eta}$ by any integer $\eta \in \mathds{Z}$ without altering the construction (translational invariance).
Later we will see that there is a further nontrivial physical gauge freedom
at this point
in the choice of the meter states,
see \Eq{eq:nice}.
The charges are counted instantaneously and without backaction
on the rest of the system:
when later [\Eq{eq:ar-state}] on we discard the measurements on the meter (by tracing over the meter space)
the reduced system evolution is not modified by the presence of the meter.
This ideal coupling of the meter to the transported charge is
implemented in the dynamics by replacing the electron field operators
in the tunnel-coupling Hamiltonian of the left reservoir by a tensor product \begin{align}
	c\hphantom{^\dagger} \to c\hphantom{^\dagger} \otimes
	e^{i\hat{\phi}} ,
	\qquad
	c^\dagger \to c^\dagger \otimes
	e^{-i\hat{\phi}}
	,
	\label{eq:construction}
\end{align}
with phase operators acting on the meter space ($\eta=\pm$)
\begin{align}
	e^{i \eta \hat{\phi}} = \sum_{N=-\infty}^\infty \ket{N} \bra{N+\eta}
	= \sum_{N=-\infty}^\infty \ket{N-\eta} \bra{N}
	.
	\label{eq:kick}
\end{align}
We note that this
construction works for \emph{any} tunnel coupling model -- not just for bilinear ones as considered in \Ref{Schaller09}.
Physically, $\hat{N}$ is the position operator of the needle on an unlimited charge scale of an ideal meter 
and $\hat{\phi}$ is its conjugate momentum ($\hbar=1$):
\begin{align}
	[\hat{N}, \hat{\phi}] = i
	.
	\label{eq:Nphicomm}
\end{align}
Thus, $\hat{N} \to \partial_{i\phi}$ when acting on phase eigenkets\footnote
	{Note carefully: what is needed here is the action on kets
	$\hat{N}\ket{\phi}
	 = \sum_N N e^{i\phi N} \ket{N} =\partial_{i\phi} \ket{\phi}
	$.
	The action on a wavefunction
	$\bra{\phi}\hat{N}\ket{\psi}
	= \sum_N N e^{-i\phi N} \braket{\phi|\psi}
	=-\partial_{i\phi} \braket{\phi|\psi}
	$ has the opposite sign.
	The latter is needed to verify \Eq{eq:Nphicomm},
	$\bra{\phi} [\hat{N}, \hat{\phi}] \ket{\psi}
	=-\partial_{i\phi} \phi \braket{\phi|\psi} + \phi \partial_{i\phi} \braket{\phi|\psi}  =
	i \braket{\phi|\psi}
	$, but is avoided otherwise.
	}
$\ket{\phi}:=\sum_N e^{i N\phi}\ket{N}$.
This phase operator is well known in superconductivity~\cite{Nieto68,Carruthers68,Tsui93} and in the $P(E)$-theory of electromagnetic fluctuations of electric circuits coupled to quantum dots~\cite{Grabert91,IngoldNazarov92,Flensberg92}.
The operator \eq{eq:kick} ``kicks'' the meter by one unit $-\eta=\pm$ for every charge that tunnels to / from the left reservoir (out of / into the system).
Charge transport is measured outside the system by projective measurements of the meter observable $\hat{N}$, producing a statistics of outcomes $N$ from which the expectation values of powers $\braket{\hat{N}}$, $\braket{\hat{N}^2}$, etc. can be computed.
By measuring these at two different times, all moments and cumulants of the charge transfer (full counting statistics) can be determined,
see \App{app:cumulants} and \App{app:reservoir-charge}.

Calling this a ``meter'' emphasizes its \emph{ideal} detection aspects
which allow us to incorporate it \emph{inside} an open system,
i.e., even when integrating out the reservoirs.
This is sufficient to clarify the issues at the focus of the paper.
We stress that it is not intended as a realistic model of measurements,
but simply makes explicit what one assumes when using any of the approaches discussed in the paper (\ase, \fcs, \ar) to compute pumping effects.
For concreteness we assume that the model of the total system in \Fig{fig:model}(c) takes the form
\begin{align}
	H^\text{tot} =& H \otimes \mathds{1}+ \sum_{r=\text{L}, \text{R}} H^r \otimes \mathds{1}  \notag \\
	&+ H^{\text{T,R}} \otimes \mathds{1} + \sum_{\eta = \pm} H_\eta^{\text{T,L}}
	\otimes e^{i \eta \hat{\phi}}
	,
	\label{eq:Hamiltonian}
\end{align}
where $H$ is the system Hamiltonian.
Reservoir $r$ is described by the noninteracting Hamiltonian $H^r$ and has temperature $T$, electrochemical potential $\mu^r$ and density of states $\nu_r$.
The last line assumes bilinear coupling and for the left reservoir includes the counting of electrons entering the dot ($\eta=+$) and leaving the dot ($\eta=-$).

\paragraph*{Example.}
For a single-orbital quantum dot
one can consider, for example, the Anderson model
$H=\epsilon \sum_{\sigma} d^\dag_{\sigma}d_{\sigma} + U \prod_{\sigma} d^\dag_{\sigma}d_{\sigma}$
with $\sigma=\uparrow,\downarrow$
and bilinear tunnel coupling $H^{r \text{T}}_\eta =
\sum_{k\sigma} \sqrt{\Gamma_{r}/(2\pi \nu_r)} c_{-\eta r k\sigma} d_{\eta \sigma}$ where
$d_{\eta \sigma}=d_{\sigma}^\dag,\, d_{\sigma}$ for $\eta = \pm$,
and similarly,
$c_{\eta rk\sigma}=c_{rk\sigma}^\dag,\, c_{rk\sigma}$,
and reservoirs $H^r=\sum_{k\sigma} \epsilon_{k} c_{r k\sigma}^\dag c_{r k\sigma}$.
For simplicity we assume wide bands with constant density of states $\nu_r$.
This model was used to compute the motivating example of charge pumping  shown in \Fig{fig:pump}
with pumping parameters $V_\text{bias}=\mu_\text{L}-\mu_\text{R}$ and $V_\text{gate}=- \epsilon$,
see also \Eq{eq:ar-A} ff.

\paragraph*{Master-equation.}

By explicitly including the meter, we go back to the way \fcs of transport was originally conceived\cite{Levitov93,Levitov96}.
However, the following is not a mere rederivation of \fcs: we go beyond \Ref{Schaller09} by introducing charge and phase \emph{super}operators
in order to relate the geometry of the pumping process to meter calibrations.
Moreover, we consider driving of the Hamiltonian \eq{eq:Hamiltonian}
which is slow in the sense that the velocity of the dimensionless parameters $\vec{R}(t)$
is small relative to the transport rates ($\dot{R}:=|\dot{\vec{R}}|\ll \Gamma$).
For simplicity we also make
the weak-coupling approximation ($\Gamma \ll T$)
and
the frozen-parameter approximation (memory kernel $K$ below is independent of $\dot{\vec{R}}$)
which is sufficient for computing the linear-response in $\dot{{R}}$.
See \Ref{Pluecker17a} for a detailed discussion of these steps and also \Sec{sec:conclusion}.
After tracing out only the reservoirs\cite{Schaller09} we obtain a Born-Markov master equation
for the density operator $\Lket{\rho(t)}$ for system \emph{plus meter}:
\begin{align}
	\partial_t \Lket{\rho(t)} = K[\vec{R}(t)] \,\,  \Lket{\rho(t)}
	.
	\label{eq:MasterEquation}
\end{align}
Since it proves useful, we often consider operators as vectors in Liouville space, i.e, we write
\begin{align}
	\Lket{A}:=\hat{A},
	\qquad
	\Lbra{A}\bullet :=\tr \hat{A}^\dag \bullet
	\label{eq:superbraket}
	.
\end{align}
Here $\bullet$ denotes the operator argument of $\Lbra{B}$,
which is a covector dual to $\Lket{A}$
through the operator scalar product $\Lbraket{A|B}:=\tr \hat{A}^\dag \hat{B}$.
The kernel is a superoperator that reflects the form of the Hamiltonian~\eq{eq:Hamiltonian}
\begin{align}
	K =& K^0 \otimes \mathcal{I} + K^\text{R} \otimes \mathcal{I} 
	+ \sum_\eta K^{\text{L},\eta} \otimes 
	e^{ i \eta \Phi }
	,
	\label{eq:K}
\end{align}
where $\mathcal{I}$ is the  identity superoperator on the meter.
All time dependence is parametric and is not explicitly indicated for simplicity.
We now discuss the various terms.
\paragraph*{Example of \Fig{fig:pump}.}
For our concrete example of the Anderson model, the relevant\footnote
	{We only need the probabilities for an empty, singly ($\sigma = \uparrow$, $\downarrow$) or doubly occupied dot
	due to charge and spin conservation in the total system~\cite{Schulenborg16a}:
	in the eigenbasis of $H$ the diagonal density matrix elements decouple from the off-diagonal ones.}
part of the density matrix contains the occupation probabilities of the four charge states. 
In this case,
\begin{align}
	K^0 = 
	 - \scalemath{0.8}{
			\begin{pmatrix}  
				W_{\uparrow,0}+W_{\downarrow,0} & 0 & 0 & 0 \\
					 0 & W_{2,\uparrow}+W_{0,\uparrow} & 0 & 0 \\
					 0 & 0 & W_{2,\downarrow}+W_{0,\downarrow} & 0 \\
					 0 & 0 &  0 & W_{\uparrow,2}+W_{\downarrow,2}
			\end{pmatrix}
			}
			,
\end{align}
is the diagonal part of the rate matrix. Golden-Rule transition rates $W_{i,j}=\sum_r W_{i,j}^r$ appear in the matrix elements, where
\begin{align}
W^r_{\sigma,0} &= \Gamma_r f^+_r(\epsilon_\sigma) , &		W^r_{2,\overline{\sigma}} &= \Gamma_r f^+_r(\epsilon_\sigma + U) , \notag \\
W^r_{0,\sigma} &= \Gamma_r f^-_r(\epsilon_\sigma) ,  &		W^r_{\overline{\sigma},2} &= \Gamma_r f^-_r(\epsilon_\sigma + U) ,
\end{align}
with the Fermi functions $f^\pm_r(\omega) = (1 + e^{\pm (\omega-\mu_r)/T})^{-1}$
and the tunnel rate $\Gamma=\sum_{r=\text{L},\text{R}} \Gamma^r$.
Next, $K^\text{R}$ contains the transition rates due to the coupling of the system to the right reservoir:
\begin{align}
K^\text{R} = 
\scalemath{0.8}{
	\begin{pmatrix}  
	0 & W^\text{R}_{0,\uparrow} & W^\text{R}_{0,\downarrow} & 0 \\
	W^\text{R}_{\uparrow,0} & 0 & 0 &  W^\text{R}_{\uparrow,2} \\
	W^\text{R}_{\downarrow,0} & 0 & 0 & W^\text{R}_{\downarrow,2} \\
	0 & W^\text{R}_{2,\uparrow} &  W^\text{R}_{2,\downarrow} & 0 
	\end{pmatrix}
}
.
\end{align}
Finally, the transitions induced by an electron tunneling from / to ($\eta =\pm$) the left reservoir
are described by the superoperators $K^{\text{L},\eta}$,
in our example,
\begin{align}
K^{\text{L},+} = 
\scalemath{0.8}{
	\begin{pmatrix}  
	0 & 0 & 0 & 0 \\
	W^\text{L}_{\uparrow,0} & 0 & 0 &  0 \\
	W^\text{L}_{\downarrow,0} & 0 & 0 & 0 \\
	0 & W^\text{L}_{2,\uparrow} &  W^\text{L}_{2,\downarrow} & 0 
	\end{pmatrix}
}
,
\end{align}
\begin{align}
K^{\text{L},-} = 
\scalemath{0.8}{
	\begin{pmatrix}  
	0 & W^\text{L}_{0,\uparrow} & W^\text{L}_{0,\downarrow} & 0 \\
	0 & 0 & 0 &  W^\text{L}_{\uparrow,2} \\
	0 & 0 & 0 & W^\text{L}_{\downarrow,2} \\
	0 & 0 &  0 & 0 
	\end{pmatrix}
}
.
\end{align}

\paragraph*{Charge and phase superoperator.}
Because the reservoirs which we eliminated are normal metals (charge-diagonal state),  
in the last term of \eq{eq:K} the system superoperators $K^{\text{L},\eta}$ are combined with the phase \emph{super}operator $\Phi$:
\begin{subequations}
	\begin{align}
	e^{i \eta \Phi } \bullet
	& :=
	e^{i\eta \hat{\phi}}\bullet e^{-i\eta \hat{\phi}}
	\label{eq:superphasesym}
	\\&=  
	\sum_{N=-\infty}^\infty
	\sum_{k=-\infty}^\infty
	\Lket{N,k}
	\Lbra{N+\eta,k} \bullet
	\label{eq:superphaseshift}
	.
\end{align}
\end{subequations}
This superoperator ``kicks'' the meter by $-\eta=\pm 1$ and thereby counts the charge on the reservoir.
Due to its central importance below it deserves some comments.
We have written (functions of) pure-state projections as super(co)vectors:
\begin{subequations}
	\begin{align}
	\Lket{N}
	& := \ket{N}\bra{N}
	,
	\\
	\Lbra{N} \bullet
	& := \Tr{\text{M} } \, \left ( \ket{N}\bra{N}  \bullet \right )
	= \braket{N|\bullet | N}
	,
\end{align}\end{subequations}
where $\tr_{\text{M}}$ denotes the trace of an operator acting on the meter space.
(Only for $\Lket{N}$, and below $\Lket{N,k}$, $\Lket{\phi}$ and $\Lket{\phi,k}$
we will deviate from the convention \Eq{eq:superbraket}, thus, e.g., $\Lket{N} \neq \hat{N}$.)
From these we constructed the $k$-offdiagonal versions
\begin{align}
	\Lket{N,k}
	 := \ket{N}\bra{N+k}
	,\quad
	\Lbra{N,k} \bullet
	 := \Tr{\text{M} } \ket{N}\bra{N+k}^\dag  \bullet  
.
\end{align}
Importantly, the expectation value of any integer power ($q=1,2,\ldots$)
of the meter charge operator,
\begin{align}
	\braket{\hat{N}^q}(t) = \Tr{\text{M}} \Tr{\S} \hat{N}^q(t) \rho
	=
	\Tr{\text{M}} \Tr{\S}
	\mathcal{N}^q
	\rho(t)
	\label{eq:Nexp}
	,
\end{align}
can be expressed entirely in terms of the following charge superoperator, the anticommutator
\begin{subequations}
	\begin{align}
	\mathcal{N}
	&:=
	\tfrac{1}{2} ( \hat{N} \bullet + \bullet \hat{N})
	\label{eq:anticom}
	\\&=
	\sum_{N}\sum_{k} (N+ \tfrac{1}{2}k) \,  \Lket{N,k}\Lbra{N,k}
	\label{eq:supercharge}
	.
	\end{align}
\end{subequations}
Moreover, writing $k=N'-N$ in \Eq{eq:supercharge} one recognizes the eigenvalue $N+\frac{1}{2}k=\frac{1}{2}(N+N')$ as the  classical symmetrized coordinate of the meter's needle.
This splitting into the sum and difference of $N$ and $N'$ is well-known from the Moyal gradient expansion~\cite{Moyal49} around the semiclassical limit using Wigner~\cite{Wigner32,Bondar13} or Green's functions~\cite{Rammer}.

The relevant phase \emph{superoperator} $\Phi$
that is conjugate to the charge anticommutator $\mathcal{N} \propto \hat{N} \bullet + \bullet \hat{N} $
in the sense
\begin{align}
	[ \mathcal{N},\Phi ] = i
	\label{eq:Nphicommsuper}
\end{align}
is given by the commutator, the phase difference
\begin{align}
	\Phi
	:= \hat{\phi} \bullet - \bullet \hat{\phi}
	\label{eq:Phi}
	,
\end{align}
Thus, $\Phi$ is the momentum canonically conjugate to the classical charge-meter needle position $\mathcal{N}$:
it implements the unitary symmetry transformations \eq{eq:superphasesym} on the meter's \emph{mixed state},
i.e., the translations of the charge meter's needle.
As we will see in \Sec{sec:fcs},
the counting field in the density-operator \fcs formalism
is the eigenvalue $\phi$ of \emph{this} phase \emph{super}operator $\Phi$.
It should not be confused with  the phase operator $\hat{\phi}$ in the initial Hamiltonian formulation which is conjugate to the charge operator $\hat{N}$ [\Eq{eq:Nphicomm}].
(We note that from $\hat{N}$ and $\hat{\phi}$ one can construct another pair of conjugate superoperators for charge and phase:
$\mathcal{N}'=\hat{N}\bullet - \bullet \hat{N}$
and 
$\Phi'=\tfrac{1}{2}( \hat{\phi}\bullet + \bullet \hat{\phi} )$.
These do not enter anywhere here: these are required only when considering a \emph{phase} meter --observable $\Phi'$-- whose needle is ``kicked`` by symmetry generator $\mathcal{N}'$ for charge.)

\paragraph*{Diagonalization of the kernel in the meter space.}
In the following [\Eq{eq:densitySolution0}] we will need the diagonal form of the kernel  superoperator \eq{eq:K}.
This amounts to going to the eigenbasis of the phase superoperator:
\begin{subequations}\begin{align}
	e^{i \eta \Phi}
	& =
	\sum_{N} \sum_k \Lket{N,k} \Lbra{N+\eta,k}
	\\&= \int_{-\pi}^{\pi} \tfrac{d\phi}{2\pi} e^{i \eta \phi} \sum_k  \Lket{\phi,k}\Lbra{\phi,k}
	.
\end{align}\label{eq:kicksuper}\end{subequations}
Its eigenvectors are Fourier transforms of the pure-state meter operators,
the phase superkets\footnote
	{Note that the superkets $\Lket{\phi}:=\sum_N e^{iN\phi} \ket{N}\bra{N}$ are distinct from
	the projectors $\ket{\phi}\bra{\phi}=\sum_{N,N'} e^{i(N-N')\phi}\ket{N}\bra{N'}$
	onto the eigenvectors $\ket{\phi}$ of the phase-operator $\hat{\phi}$ which play no role here.}
\begin{align}
	\Lket{\phi}
	:= e^{i\hat{N}\phi } 
	= \sum_{N} e^{i\phi N}\Lket{N}
	,\qquad \phi \in [-\pi,\pi]
	\label{eq:ketchi}
	,
\end{align}
(noting $\Lket{\phi} \neq \hat{\phi}$, the other deviation from convention \eq{eq:superbraket})
and similar operators that are charge-off-diagonal by $k=0,\pm 1, \pm 2, \ldots$
\begin{align}
	\Lket{\phi,k}
	:= e^{i\hat{N}\phi } e^{ i\hat{\phi} k } 
	= \sum_{N} e^{i\phi N} \ket{N}\bra{N+k}
	\label{eq:ketchik}
	.
\end{align}
In agreement with \Eq{eq:Nphicommsuper},
$\mathcal{N} \to \partial_{i\phi}$
when acting on the phase superkets\footnote
	{Distinguish the action on superkets
	$\mathcal{N}\Lket{\phi}
	=\partial_{i\phi} \Lket{\phi}
	$
	from the action on a mixed state component
	$\Lbra{\phi}\mathcal{N}\Lket{\rho}
	=-\partial_{i\phi} \Lbraket{\phi|\rho}
	$
	which is needed to verify \Eq{eq:Nphicommsuper}:
	$\Lbra{\phi} [\mathcal{N}, \Phi] \Lket{\rho}
	=-\partial_{i\phi} \phi  \Lbraket{\phi|\rho}
	+\phi \partial_{i\phi}   \Lbraket{\phi|\rho}
	=
	i \Lbraket{\phi|\rho}
	$.
	}
$\Lket{\phi}=\Lket{\phi,0}$ and $\Lket{\phi,k}$ for $k \neq 0$.
In \Eq{eq:K} we thus see that after integrating out the reservoirs, the ideal meter is coupled to the transported charge through the meter's needle momentum superoperator $\Phi$.
Importantly, inserting \Eq{eq:kicksuper} into the kernel \eq{eq:K} we see that the tensor product structure
\begin{subequations}
	\begin{gather}
	K(t) =
	\int_{-\pi}^{\pi} \tfrac{d\phi}{2\pi}
	\left (
	K_0(t) + K^\text{R}(t)
	+ \sum_\eta K^{\text{L},\eta}(t) e^{i\eta \phi }
	\right ) 
	\notag
	\\
	\otimes  \sum_k \Lket{\phi,k}\Lbra{\phi,k}  
	\\
	 =\int_{-\pi}^{\pi} \tfrac{d\phi}{2\pi} W^\phi(t)
	\otimes  \sum_k \Lket{\phi,k}\Lbra{\phi,k}
	\label{eq:chiKernel}
	.
\end{gather}\label{eq:Kdiag}\end{subequations}
reflects the ideality of the meter model \eq{eq:Hamiltonian}.
The important $\phi$-dependent superoperator $W^\phi$ --acting only on the system space-- emerges naturally.

Finally, we note that in the following we can drop all $k$ sums in the above expressions
since we will restrict our attention to physical meter states
that are $\hat{N}$-diagonal\footnote
	{Acting on $\hat{N}$-diagonal meter states the $k\neq 0$ terms in the superoperators give zero
	and can be dropped.
	They contain no terms that couple the $k=0$ and $k\neq 0$ components of the state
	since the difference $\hat{N} - \hat{n}$ between the meter and the left reservoir charge is conserved
	by the construction \eq{eq:construction}.
	(This is no longer true if superconducting electrodes are considered.)},
i.e., mixed states of the form $\sum_N G^N \ket{N}\bra{N}$ with some probability distribution $G^N$.
We then only need to deal with pure meter charge-states $\Lket{N}:=\Lket{N,0}=\ket{N}\bra{N}$ and their Fourier transforms, the phase superkets $\Lket{\phi}:=\Lket{\phi,0}$.
However, the $k$ sums were important above to relate the charge ($\hat{N}$) and phase operators ($\hat{\phi}$) to the relevant superoperators ($\mathcal{N}$ and $\Phi$, respectively).

\section{Geometric phases of adiabatically evolving mixed states\label{sec:ase-review}}

To keep the paper self-contained, this section collects the key steps of the adiabatic state evolution (\ase) approach to slowly-driven open quantum systems
guided by our points of interest in \Tab{tab:compare}.
This approach due to Sarandy and Lidar~\cite{Sarandy05,Sarandy06}
--in principle not related to transport--
generalizes the adiabatic closed-system approach:
starting from a time-local \emph{master equation} of the form \eq{eq:MasterEquation} it leads to a Berry-Simon type geometric phase for the \emph{mixed-state operator}.
In \Sec{sec:meter}, we will apply this approach to the pumping model which explicitly includes the meter.
Here we first discuss the less specific problem of the driven state-evolution
governed by \Eq{eq:MasterEquation} [i.e., without assuming \Eq{eq:Kdiag}].

\paragraph*{Mixed states, modes and decay rates.}
First we consider fixed parameters $\vec{R}$ (not indicated), for which the time-evolved state reads
\begin{align}
	\Lket{\rho(t)}
	=
	\Lket{k_0} + \sum_{m=1,2,\ldots} e^{k_m t} \C_m(0) \, \Lket{k_m}
	.
	\label{eq:rhot-ase}
\end{align}
Here, the \emph{modes} are the right eigenvectors which satisfy $K\Lket{k_m}=k_{m}\Lket{k_m}$. These are operators $\Lket{k_m}=\hat{k}_m$
labeled $m=1,2,\ldots$.
These nonstationary modes account for the dissipative decay\cite{Schulenborg16a} since their eigenvalues have $\text{Re}\, k_m < 0$.
We label by $m=0$ the zero-mode with eigenvalue $k_0=0$, i.e., $\Lket{k_0}=\hat{k}_0$ is the
stationary density operator.
We assume for simplicity that  $K$ has a completely nondegenerate spectrum for all parameter values\footnote
	{See \Ref{Pluecker17a} [Eq. (G 10)] for a discussion of the gap condition for adiabatic decoupling in the steady-state limit,
	cf. \Refs{Sarandy05,Sarandy06}.},
in particular, that the stationary state $\Lket{k_0}$ is unique.
This applies to a very large class of practically relevant cases.
Because the evolution is non-Hamiltonian,
$K$ is not a normal  operator ($K^\dagger K \neq K K^\dagger$)
and the left eigenvectors (covectors) are \emph{not} the hermitian adjoints of the right ones
and need to be determined separately\cite{Schulenborg16a}.
For each eigenvalue $k_m$ the corresponding left supereigenvector $\Lbra{\bar{k}_m}$
is thus specified by an operator $\hat{\bar{k}}_m$ different from $\hat{k}_m$,
as indicated by the overbar.
The covector plays a different role: 
given the initial state for \Eq{eq:rhot-ase},
$\Lket{\rho(0)}=\sum_m \C_m(0) \Lket{k_m}$,
it determines the amplitude for mode $\Lket{k_m}$ being excited:
\begin{align}
	\C_m(0)
	=\Lbraket{\bar{k}_m|\rho(0)}
	= 	\tr \Big( \hat{\bar{k}}_m^\dag \rho(0) \Big)
	.
	\label{eq:cm0}
\end{align}
Furthermore, the zero mode $\Lket{k_0}$ always exists since the trace-preservation $\tr K \bullet \propto \Lbra{\bar{k}_0}K\bullet = 0$ requires that there is a corresponding covector with $\hat{\bar{k}}_0 \propto \unit$.

\paragraph*{Adiabatic mixed-state and decoupling.}
When slowly driving the parameters,
it can be shown~\cite{Sarandy05,Sarandy06,Pluecker17a} that in the leading approximation
the mixed state maintains the form \eq{eq:rhot-ase}:
\begin{align}
	\Lket{{\rho}(t)} =
	\sum_{m=0,1,\ldots}  e^{z_m(t)} \C_m(0) 
	\, \Lket{k_m[\vec{R}(t)]}
	\label{eq:ase}
	.
\end{align}
It is therefore meaningful to refer to the $m=0$ ($m\neq0$) contributions as its \emph{parametrically (non)stationary} components.
A key question addressed in this paper is to physically understand how this adiabatic procedure manages to (correctly) capture \emph{non}adiabatic effects when applied to a \emph{suitably modified} open system (including a meter).
In the adiabatic decoupling approximation \eq{eq:ase}, the modes $\Lket{k_m}$ evolve independently since any crosstalk is negligible due to the slow driving. 
Their coefficients have evolution phases
$z_m(t) = z_{\n,m}(t)+z_{\g,m}(t)$ which are no longer given by \Eq{eq:rhot-ase}.
After one period, at $t=\T$, the modes all return to their initial values\footnote
	{We assume that the $\Lket{k_m[\vec{R}]}$ are chosen as continuous and smooth functions of $\vec{R}$ in all accessed parameter regimes.}
since the parameters have traversed a closed curve $\curve$ in the parameter space, $\vec{R}(\T)=\vec{R}(0)$.
However, the state $\Lket{\rho(\T)}$ has changed relative to $\Lket{\rho(0)}$ due to all the geometric and nongeometric phases, respectively, 
\begin{subequations}
\begin{align}
z_{\g,m}(\T) &:=
-
\oint_\curve d\vec{R} \Lbraket{\bar{k}_m|\delR k_m}
\label{eq:ase-zg}
\\
z_{\n,m}(\T)
&=\int^{\T}_0 dt \, k_m[\vec{R}(t)]
\label{eq:ase-zn}
,
\end{align}\label{eq:ase-z}\end{subequations}
one for \emph{each} mode $m$
contributing to the state \eq{eq:ase}.

Thus, the \ase approach formally resembles the Berry-Simon approach to closed systems,
except for distinguishing left from right eigenvectors.
In closed systems one is essentially only concerned with the right eigenvectors of the time-evolution
because the closed system Hamiltonian $H$ is hermitian (implying the left eigenvectors are simply their adjoints).
For such Hamiltonian dynamics, we have $K(t)=-i [H(t),\bullet]$ and the master equation  \eq{eq:MasterEquation} reduces to the usual Liouville-von Neumann equation.
One verifies that in this case one exactly recovers the standard closed-system approach for ket vectors
including Berry's result for a spin in a rotating field~\cite{Pluecker_phd}.

\begin{figure*}
	\includegraphics[width=0.9\linewidth]{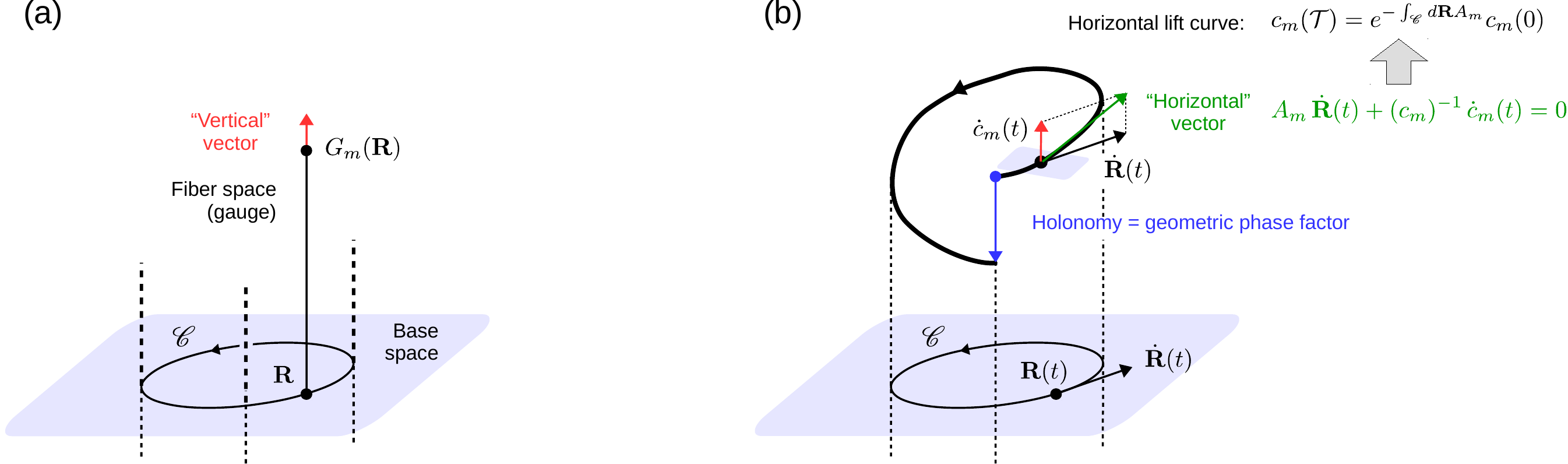}
	\caption{
		What is geometric about the adiabatic solution factor?
		(a) Space of driving parameters ($\vec{R}$) $\times$ gauge values ($g_m$).
		(b) The geometric factor of the adiabatic evolution coefficient of decay mode $\Lket{k_m}$ corresponds to a curve in the space of (a). Along this curve, the parallel transport condition  \eqref{eq:mode-partrans} holds,
		or, equivalently, the curve minimizes the Fock condition \eqref{eq:fock}.
		Importantly, the parallel transport of the adiabatic mixed \emph{state} $\Lket{\rho}$ is defined by the parallel transport of its components, each represented by such a curves in its own space.
	}
	\label{fig:geometry}  
\end{figure*}

\paragraph*{Geometric phases and nonstationarity.}

However, there is an important difference in the way physical restrictions enter for open systems.
In particular, the kernel needs to preserve the trace and hermiticity of the density operator. This requires the adiabatically decoupled state to take the more specific form
\begin{align}
	\Lket{\rho}
	= \C_0(t) \Lket{k_0}+ \sum_{m \geq 1} \Big[ \C_m(t) \Lket{k_m}+\C_m^{*}(t) \Lket{k_m^\dag} \Big]
	\label{eq:rho-restrict}
	,
\end{align}
where the coefficients are $\C_m(t)=\Lbraket{\bar{k}_m|\rho(t)}$
and $\Lket{k_m}$ [$\Lket{k_m^\dag}$] is the eigenoperator for the complex(-conjugate)
eigenvalue $k_m$ ($k_m^{*}$).
In view of the following discussion we assume that for $m \neq 0$ there all eigenvalues are complex.
This implies that \emph{steady-state geometric} phases arise only if the mixed state has a \emph{parametric nonstationary} component, similar to closed systems, see \App{app:nonstationary}.
We need to distinguish the two cases:

First, the parametrically stationary mode $m=0$ in \Eq{eq:rho-restrict} has a real eigenvalue, so its operator can be chosen hermitian,
$\hat{k}_0=\hat{k}_0^\dag$.
If we trace-normalize it to be a physical state (stationary state),
$\tr \hat{k}_0=1$,
then $\C_0=1$ is fixed for all times as in \Eq{eq:rhot-ase}.
These physical restrictions force both
the geometric and the non-geometric phase for the zero mode to vanish:
$z_0 =z_{\n,0}=z_{\g,0}=0$
since $k_0=0$ in \Eq{eq:ase-zn} and $\Lbraket{\bar{k}_0|k_0}=\Lbraket{\unit|k_0}=1$ in \Eq{eq:ase-zg}.
That $\Lket{k_0}$ can be trace-normalized reflects that it is the only term in the expansion on the right-hand side of \Eq{eq:ase} that is a valid physical state
[consistent with $\Lket{\rho(t)} \to \Lket{k_0}$ in the stationary limit for fixed parameters].
This should be contrasted with closed systems, where each term in a superposition of pure-state vectors similar to \Eq{eq:ase} represents a physical state.
Since in the slowly driven steady-state limit the unique zero-mode contribution dominates,
the time-dependent steady-state \eq{eq:ase}
exhibits \emph{no geometric phase}
as a result of trace-normalization.

Second, the parametrically nonstationary modes $m\geq 1$ in \Eq{eq:rho-restrict} are
pairs of non-hermitian operators with complex-conjugate coefficients
(complex eigenvalue $k_m$).
The operators for these nonstationary components are necessarily
traceless, $\tr \hat{k}_m=\tr \hat{k}_m^\dag=0$,
and therefore are not quantum states by themselves or in any combination.
They can only be added to the stationary state as in \Eq{eq:rho-restrict}
to form the physical state $\Lket{\rho}$.
This lack of trace-normalization also allows the nonstationary modes $\Lket{k_m}$ ---and only these---  to accumulate nonzero (non)geometric phases
\eq{eq:ase-z}.

We note that the positivity of the density operator $\hat{\rho}(t)$
imposes a further nonlinear constraint on the evolution.
Whereas in closed systems this is automatically ensured,
in open systems this is not the case.
This has received little attention in the context of pumping and geometric phases
and we will return to it in \Sec{sec:meter} [\Eq{eq:nice}].

\paragraph*{Gauge freedom and physical restrictions.}
The appearance of a geometric phase such as \Eq{eq:ase-zg} in general requires some gauge freedom.
Here, similar to closed systems, the gauge freedom  resides in the possibility of choosing a different normalization
of the eigenvectors in the expansion of the state while leaving the state invariant:
each pair of right and left eigenvectors can be modified to gauged eigenvectors
\begin{align}
	\Lbra{\bar{k}_m}_{G_m} = (G_m)^{-1} \Lbra{\bar{k}_m}
	,\qquad
	\Lket{k_m}_{G_m} = G_m \Lket{k_m}
	,
	\label{eq:ase-gauge}
\end{align}
if the coefficients are simultaneously adjusted to $\C_{m,G_m} = (G_m)^{-1} \C_{m}$ [\Eq{eq:cm0}].
A gauge for mode $m$ is thus \emph{any} choice of a continuous nowhere-vanishing complex function of the parameters, $G_m[\vec{R}]$.
In contrast to closed systems,
in open systems the physical requirements for the gauges are less restrictive:
Although preservation of the form \eq{eq:rho-restrict},
written compactly as $\Lket{\rho}= \sum_{m \in \mathds{Z}} \C_m(t) \, \Lket{k_m}$ with $\tr \hat{k}_0 = 1$,
$\Lket{k_{-m}}:=\Lket{k_{m}^\dag}$ and $\C_{-m}:=\C_m^{*}$, requires
\begin{align}
	G_{0}=1,
	\qquad
	G_m^{*}= G_{-m}
	\label{eq:ase-gauge-restrict}
	,
\end{align}
this still allows
the modes $\Lket{k_m}$ to accumulate a normalization factor\footnote
	{Note that the gauge transformations do maintain the normalization between left- and right eigenvectors
	$\Lbraket{\bar{k}_m|k_m}=1$.}
with real phases under non-unitary evolution,
in addition to the usual imaginary phase factors.

\paragraph*{Parallel transport of nonstationary components.}
Finally, to see what is precisely geometric about the adiabatic evolution
 [\Eqs{eq:ase}-\eq{eq:ase-z}]
 we can closely follow the Berry-Simon approach to closed quantum systems.
In particular, in \Eq{eq:ase} the evolution factor $e^{z_{\g,m}(\T)}$ for each parametrically-nonstationary mode ($m\geq1$)
is equivalent to a geometric parallel transport in the space of parameters ($\vec{R}$) and nonzero complex gauge values ($G_m$).
This means for each $m$ separately that if we try to gauge away this evolution factor we will in general fail
as illustrated in \Fig{fig:geometry}.
Suppose we look for $\C_m(t)$ such that
$\Lket{k'_m(t)}:=\C_m(t) \, \Lket{k_m(t)}$ together with $\Lbra{\bar{k}'_m} =  \Lbra{\bar{k}_m} \, / \C_m(t)$
satisfy the parallel transport condition
\begin{align}
	\Lbraket{\bar{k}'_m| \tfrac{d}{dt} k'_m} = 0
	,
	\label{eq:mode-partrans}
\end{align}
for $t\in [0,\T]$,
such that the geometric phase \eq{eq:ase-zg} is zero
in this new ``rotating frame''.
\Eq{eq:mode-partrans} is equivalent to
\begin{align}
A_{m,\C_m} \dot{\vec{R}} := A_m \dot{\vec{R}} + (\C_m)^{-1} \tfrac{d}{dt} \C_m = 0
\label{eq:Amc}
,
\end{align}
where $A_{m}= \Lbraket{\bar{k}_m |\delR k_m}$ is the connection appearing in the result \eq{eq:ase-zg}.
Solving this for $\C_m(t)$ one obtains (only) the nontrivial geometric evolution factor
$
\C_m(\T)=e^{
	-\oint_\curve d\vec{R} A_m
}\C_m(0)
= e^{z_{\g,m}(\T)} \C_m(0)
$
in \Eq{eq:ase},
explaining\footnote
	{The left hand side of \Eq{eq:Amc} is one of several ways of defining a geometric connection.
	The definition employed here leads to a clear physical picture later on [\Eqs{eq:AphiG-partrans}, \eq{eq:fcs-cumcur-partrans}, \eq{eq:ar-cur-partrans}].}
its geometric origin.
The geometric phase cannot be gauged away: for a small driving curve $\curve$ around a point $\vec{R}$ with nonzero curvature $B_m=\delR \times A_m$,
the solution $\C_m(t)$ will be discontinuous and fails to be a valid gauge.
The phases $z_{\g,m}$ are thus unavoidable and physically relevant.
We stress that the adiabatic evolution of a single mixed state involves the parallel transport of \emph{all} its nonstationary modes $m=1,2,\ldots$.
Only the family of geometric pictures, each as in \Fig{fig:geometry},is physically meaningful,
in contrast to closed systems where often consideration of the geometric phase of a single vector suffices.

\paragraph*{Discussion: difference with Uhlmann's phase.}
This completes our summary of the \ase approach as applied for the following.
It is, however, relevant to point out that the parallel transport condition \eq{eq:mode-partrans} can be formulated also as a Fock-type condition~\cite{Fock1928}:
\begin{align}
\Lbraket{ \tfrac{d}{dt} \bar{ k}' | \tfrac{d}{dt} k' }
\quad \text{is stationary}
\label{eq:fock}
\end{align}
for all possible curves $\C_m$ parametrized by time.
Here the overbar in the covector indicates that the distance measure used
depends on the particular nonhermitian evolution kernel $K$ in \Eq{eq:MasterEquation},
see \App{app:fock} for details.
This should be contrasted with parallel transport in closed systems, where due to the hermicity of the Hamiltonian, the Fock condition ``$\braket{ \frac{d}{dt}\psi | \frac{d}{dt} \psi }$ is stationary'' involves the standard hermitian metric, \emph{independent} of the Hamiltonian.
Thereby,
\Eq{eq:fock} also differs from the definition of the \emph{Uhlmann geometric phase}~\cite{Uhlmann86} for density operators
which is also defined using the Fock condition for a closed ``purified'' system\footnote
	{The \emph{hermitian} inner product is taken between pure states which purify two density operators.
	In terms of operators, this is the Hilbert-Schmidt scalar product between polar decompositions $\sqrt{\rho} U$ with a unitary $U$.}.
This difference is relevant in view of the recent interest~\cite{Huang14,Viyuela15,Budich15a,Budich15b} in finding measurable quantities that are related to the Uhlmann phase.
The inverse route taken in the present paper --starting from physical evolution and measurable quantities and then identifying the relevant geometric structures--
leads to a different type of geometric parallel transport condition,
one that depends on the particular open-system dynamics from which it was derived.
Of course, for \emph{topological} quantities the connection used to compute them does not matter by the Chern-Weil theorem,
but our discussion indicates that considerations of measurements are tied to different types of connections.

\section{Adiabatic state evolution of system + meter\label{sec:meter}}

In the following we derive geometric pumping
in terms of the established geometric approach to adiabatic mixed states [\Sec{sec:ase-review}]
by exploiting the special structure \eq{eq:Kdiag} of the evolution \eq{eq:MasterEquation} of an open system extended by an ideal meter [\Sec{sec:model}].
We purposefully postpone comparison with the \fcs and \ar approach to Secs.~\sec{sec:fcs}  and \sec{sec:ar}, respectively, to systematically answer the questions that they raised.

\subsection{Adiabatic, nonstationary composite-state evolution.}
Since we integrate out only the reservoirs,
while explicitly keeping the meter as part of the open system,
the master equation \eq{eq:MasterEquation} treats the system-meter coupling as instantaneous.
As a result, the \emph{adiabatic} evolution of the mixed \emph{composite} state correctly describes pumping effects.
Thus, in this description there is no visible ``lag'' between driven system and charge measurement,
cf. Secs.\sec{sec:fcs}-\sec{sec:ar}.
The adiabatic solution of the master equation \eq{eq:ase} involves the decoupling of the kernel eigenmodes.
We start from \Eq{eq:Kdiag}
\begin{align}
	K(t) 
	=\int_{-\pi}^{\pi} \tfrac{d\phi}{2\pi} W^\phi(t)
	\otimes  \Lket{\phi}\Lbra{\phi}
	\label{eq:chiKernel2}
\end{align}
and further diagonalize the kernel also in the system space.
The resulting eigenvectors inherit the special tensor-product form of \eq{eq:Kdiag}
\begin{subequations}
	\begin{align}
	\Lket{k_{\p,\phi}(t)} &= \Lket{w^\phi_\p(t)} \otimes \Lket{\phi}
	\label{eq:eigenvector}
	, \\
	\Lbra{\bar{k}_{\p,\phi}(t)} &= \Lbra{\bar{w}^\phi_\p(t)} \otimes \Lbra{\phi}
	\label{eq:eigencovector}
	,
\end{align}
\label{eq:eigenvectors}\end{subequations}
involving the meter-momentum eigenvectors $\Lket{\phi}$ given by \Eq{eq:ketchi}.
For each value of $\phi$ we denoted by $\{ \Lket{w^\phi_\p(t)} \}_{\p=0,1,\ldots}$ the discrete set of eigenvectors of the superoperator $W^\phi$ in \Eq{eq:chiKernel2} whose eigenvalues coincide with those of $K(t)$, i.e.,
$k_{\p,\phi}(t) = w_{\p}^\phi(t)$.
Compared to the previous section,
the eigenmodes are thus labeled by a composite mode index, $m \to (\p,\phi)$:
a discrete system index $\p=0,1,2,\ldots$  and the continuous meter momentum $\phi \in [-\pi,\pi]$.

As a consequence of the ideal coupling, the meter kets $\Lket{\phi}$ are time- and parameter independent.
The modes $\Lket{k_{\p,\phi}}$ with different $\phi$ are thus already decoupled before the adiabatic approximation.
Denoting the meter density operator by $\rho^\M(0)$,
we find that starting from any classical meter state,
a statistical mixture $\rho^\M(0)=\sum_N \ket{N}\bra{N} \rho^\M \ket{N} \bra{N}$
with well-defined charge $[\hat{N},\rho^\M(0)]=0$,
the exact evolution of the ideal meter model takes the form
\begin{align}
	\Lket{\rho(t)}
	&=
	\int_{-\pi}^{\pi} \frac{d\phi}{2\pi}
	\Lket{\rho^\phi(t)} \otimes \Lket{\phi}
	\label{eq:rhoform}
	.
\end{align}
The tensor product expresses the absence of backaction [\Eq{eq:Nphicomm} ff.].
The adiabatic approximation [\Eq{eq:ase}] thus only involves the decoupling of the eigenmodes with the same $\phi$ but different $\p =0,1,\ldots$:
\begin{align}
	\Lket{\rho(t)} & \approx
	\sum_{l=0,1,\ldots}
	\int_{-\pi}^{\pi} \frac{d\phi}{2\pi}
	\, e^{z_\p^\phi(t)}
	\,
	\C_\p^\phi(0)
	\, 
	\Lket{w^\phi_\p(t)} \otimes \Lket{\phi}
	\label{eq:densitySolution0}
	.
\end{align}
where we assume that the eigenvalues $w^\phi_{\p}$ are gapped\footnote
	{Crossings of the eigenvalues may actually occur and lead to interesting effects as noted only recently in related \fcs studies\cite{Ivanov10,Ren13}, see also the discussion in \Sec{sec:conclusion}.
	In our formulation crossings lead to \emph{non}adiabatic evolution of \emph{system-plus meter}
	(not to be confused with nonadiabaticity of the system only, cf. \Sec{sec:ar}.}
for all $\phi$.

Next, we take the steady-state limit of the driven evolution:
for each $\phi$, only the $\p=0$ term survives\footnote
	{In \Eq{eq:densitySolution0} only the $\p=0$ eigenspace contributes in the long time limit.
	The $\p \neq 0$ modes decay on the time scale $\Gamma^{-1}$, see App. D of \Ref{Pluecker17a}.}
\begin{align}
	\Lket{\rho(\T)} &
	\approx
	\int_{-\pi}^{\pi} \frac{d\phi}{2\pi}
	\,
	Z^\phi
	\, 
	\Lket{w^\phi_0(\T)} \otimes \Lket{\phi}
	\label{eq:densitySolution}
	,
\end{align}
and we denote the remaining $\p=0$ coefficient at $t=\T$ by $Z^\phi = e^{z^\phi}$.
We stress that this involves two steps:
to extract the steady-state component of $\rho(t)$ one must ``reset'' the initial condition of the system plus meter, $c_0^\phi(0)=1$ for all $\phi$.
Importantly, before this resetting the function $c_0^\phi(0)$ cannot be properly gauged away (i.e., in a continuous fashion).
We discuss this in detail in \App{app:steady} since this is easily overlooked~\cite{Nakajima15}.
After this resetting, we are thus free to fix a reference gauge for the eigenvectors $\Lket{k_{\p,\phi}(t)}$ by requiring
\begin{align}
	\Lbraket{\unit|w^\phi_0(t)}=1
	\quad\text{for \emph{all} $\phi$ and $t$.}
	\label{eq:gauge-ref}
\end{align}
This gauge ensures that the meter readout statistics is entirely contained in the coefficient $Z^\phi$ and no additional information remains in the eigenvector normalization:
the reduced density operator of the meter obtained by additionally tracing out the system is then
\begin{align}
\rho^\M
:= \tr_{\S} \rho
=\int \tfrac{d\phi}{2\pi} Z^\phi \Lket{\phi}
.
\label{eq:rhoM}
\end{align}

The quantum state \Eq{eq:densitySolution} describes the system plus meter in an ideal adiabatic steady-state pumping experiment.
The contributions to $\Lket{\rho}$ from all the nonzero meter momenta $\phi \neq 0$  are parametrically-\emph{nonstationary} [\Eq{eq:rho-restrict} ff.]:
for fixed parameters these terms account for the fact that the charge meter keeps running,
even when (the current through) the system --without meter-- has become \emph{stationary}.
(This latter point is made explicit in the \ar approach discussed in \Sec{sec:ar}, cf. \Eq{eq:ar-rho-exp}.)

\subsection{Continuum of Berry-Simon phases\newline and pumped charge.\label{sec:meter-pump}}
As we have seen [\Eq{eq:rho-restrict} and \App{app:nonstationary}],
parameter driving in general leads to geometric phase accumulation
\emph{only} in the parametrically nonstationary components of the time-dependent state.
Indeed, the coefficients 
\begin{align}
Z^\phi = e^{z^\phi}
,\quad
z^\phi = z_\n^\phi+z_\g^\phi
.
\end{align}
of the system-plus-meter steady state pick up a \emph{continuum} of phases for $\phi \neq 0$.
The nongeometric part,
\begin{align}
	z_\n^{\phi} & := \int_0^{\T} dt \, k_{0,\phi}(t)
		= \int_0^{\T} dt \, w_0^\phi(t)
	\label{eq:zn}
	,
\end{align}
is a ``sum of snapshots'' of the eigenvalues [\Eq{eq:ase-zn}].
The geometric part,
\begin{align}
	z_\g^{\phi} := 
	-
	\oint_\curve d\vec{R} A^{\phi}(\vec{R})
	\label{eq:zg}
	,
\end{align}
is determined by the eigenvectors through the geometric connection of the \emph{Berry-Simon type}
[\Eq{eq:ase-zg}]
\begin{align}
	A^{\phi}
	:=	
	\Lbraket{ \bar{k}_{0,\phi} | \delR | k_{0,\phi} }
	=
	\Lbraket{ \bar{w}_0^\phi | \delR | w_{0}^\phi }
	\label{eq:Aphi}.
\end{align}
Since it relies on the eigenvectors, it arises even when the eigenvalues $w_0^\phi$ in \Eq{eq:zn} were all zero along the path $\curve$ of accessed parameters, i.e., if there was no transport at all (including no fluctuations). It thus accounts for \emph{pumping} contributions to the entire, slowly driven transport process.

In our physical model \eq{eq:Hamiltonian}, it is clear from the beginning that the transport statistics of charge is exactly registered by the meter incorporated in the quantum state $\rho(\T)$ --and thus in $Z^\phi$ or $z^\phi$--
because every passing electron ``kicks'' the meter [\Eq{eq:kick} and \eqref{eq:superphaseshift}].
More precisely, the projective measurements of the reservoir charge at two times
 -- with outcomes $n$ at $t=0$ and $n'$ at $t=\T$--
 determine the statistics for the \emph{change} of the reservoir electron number through
the 2-point moments $\mom^{(k)}$ of order $k=1,2,\ldots$:
\begin{align}
\mom^{(k)}:=\sum_{n'n} (n'-n)^k p_{n'n}
.
\label{eq:moments}
\end{align}
Here $p_{n'n}$ is the joint 2-point probability distribution for this system-reservoir process.
At this point it is convenient to initialize the meter in the pure state with zero average and no dispersion, $\Lket{0}=\ket{0}\bra{0}$.
Then, the 2-point moments of charge transport appear as 1-point measurements of powers of the charge $\hat{N}$ indicated by the meter,
\begin{align}
	\mom^{(k)}=\braket{\hat{N}^k }(\T)
	\label{eq:deltaNk}
	,
\end{align}
which is verified in \App{app:cumulants}.
This is possible because the meter only registers  \emph{changes} of the charge by its ideal coupling to transport through the phase superoperator $\Phi$ [\Eq{eq:superphaseshift}].
Since we include the meter in the open system,
we simply compute these averages in the standard way from the meter reduced density operator:
\begin{align}
\mom^{(k)} & 
= \Tr{\M} \hat{N}^k  \rho^\M(\T)
=
\left . (-\partial_{i\phi})^k Z^\phi
\right |_{\phi=0}
\label{eq:momentDerivative}
.
\end{align}
Although the last formula is familiar from formal \fcs considerations to be discussed in \Sec{sec:fcs},
here we can physically understand all of its aspects as shown by the steps of its derivation:
\begin{subequations}
	\begin{align}
	\Tr{\M} \hat{N}^k \rho^\M(\T)
	=
	\Tr{\M} \mathcal{N}^k \rho^\M(\T)
	\label{eq:deltaN-asea}
	\\=
	\int_{-\pi}^{\pi} \tfrac{d\phi}{2\pi}
	Z^\phi
	 \Lbraket{\unit|w^\phi_0(t)}
	\cdot (\partial_{i\phi})^k 2\pi \delta(\phi)
	,
	\label{eq:deltaN-ase}
	\end{align}
\end{subequations}
The first equality \eq{eq:deltaN-asea} was discussed in \Eq{eq:Nexp}.
The second step highlights the physical meaning of the quantity $\phi$,
the counting field of the \fcs approach discussed in \Sec{sec:fcs}:
it is the momentum conjugate to the relevant ``classical'' charge \emph{super}operator $\mathcal{N}$,
the position of the needle of the charge meter
[\Eq{eq:Nphicomm} and \eq{eq:Nphicommsuper} ff.]:
\begin{align}
	\mathcal{N} \to \partial_{i\phi} \quad \text{ when acting on $\Lket{\phi}$}
	 \label{eq:conjugation}
	.
\end{align}
The phase should \emph{not} be confused with the conjugate to the quantum charge operator $\hat{N}$ [\Eq{eq:Phi} ff.].
Third, the phase derivative in \eq{eq:deltaN-ase} is made to act on the system factor $Z^\phi \, \Lket{w^\phi_0(\T)}$ of the composite state $\Lket{\rho(\T)}$ [\Eq{eq:densitySolution}]
by partial integration using the gauge condition $\Lbraket{\unit|w^\phi_0(t)}=1$ [\Eq{eq:gauge-ref}].
We see that the extra sign in $-\partial_{i\phi}$ that appears in the formula \eq{eq:momentDerivative} relative to \Eq{eq:conjugation}
simply reflects that the charge added to the system is counted negative by the meter.
Finally, this derivative is taken \emph{only} at $\phi=0$
due the partial trace which traces out the meter:
\begin{align}
\tr_{\M}  \mathcal{N} \Lket{\phi}
= \tr_{\M} \partial_{i\phi}  \Lket{\phi}
= \partial_{i\phi} 2\pi \delta(\phi)
\label{eq:phizerotrace}
\end{align}
This shows that setting $\phi=0$ corresponds to discarding information about further measurement outcomes on the meter.
Thus, the formal phase-derivative at zero phase in \Eq{eq:deltaN-ase} emerges naturally
from a measurement of the charge observable on the meter, after which the meter is discarded.
Relations \eq{eq:deltaNk}, \eq{eq:conjugation} and \eq{eq:phizerotrace} will also clarify the structure of the \ar equations in \Sec{sec:ar}.

The formula \eq{eq:momentDerivative} furthermore underlines that
the (non)geometric part of the transported charge ($\mom^{(1)})$ is \emph{never equal} to a \emph{single} (non)geometric phase of some quantum \emph{state}:
Physically this is clear since an ideal charge meter has
an infinite number of discrete needle positions
with necessarily a continuum of conjugate momenta $\phi$
(kinematics).
All these are required to describe the charge detection process,
resulting in a corresponding \emph{continuum} of (non)geometric phases
(dynamics).
A particular measurement $\braket{ \hat{N}^k } (\T) = \mom^{(k)}$ only accesses a certain Taylor coefficient [\Eq{eq:momentDerivative}]
but not the entire function $Z^\phi$ which is essentially the state \eq{eq:densitySolution} or \eq{eq:rhoM}.\footnote
	{Note carefully: in \Sec{sec:ar} the pumped charge is nevertheless expressed
	directly as \emph{single} geometric (Landsberg) phase
	but this is not a (Berry-Simon) phase of a \emph{state}, but a (Landsberg) phase of the observable.}

\subsection{Gauge transformations -- meter recalibration\label{sec:meter-gauge}}
It remains to clarify why geometric quantities emerge in the transport process as a whole
by identifying the \emph{physical} origin of the gauge freedom.
(In the general \ase approach of \Sec{sec:ase-review} the freedom to re-normalize eigenvectors was merely formal.)
This origin is simply the possibility of calibrating the meter
and geometric parallel transport is the (failed) attempt to calibrate away any effect of the transport process.
To develop both ideas we need to take a few steps.

First, the gauge freedom \eqref{eq:ase-gauge}
in the present problem translates to multiplying the system and meter tensor-factor in \Eq{eq:rhoform}
by opposite gauge functions such that the state $\Lket{\rho}$ remains invariant:
\begin{subequations}
	\begin{align}
	\Lket{\rho^\phi} & \to
	\Lket{\rho^\phi}_{G}
	:= (G^\phi)^{-1} \Lket{\rho^\phi}
	\label{eq:rhochi-gauge}
	\\
	\Lket{\phi} & \to
	\Lket{\phi}_{G}:= \, G^\phi \, \Lket{\phi}	
	\label{eq:chi-gauge}
	.
\end{align}\label{eq:gauge}\end{subequations}
Because there is a continuum of modes, the gauge  $G^\phi$ is a function of the meter-momentum $\phi$.
As function of time it is convenient and without loss of generality to fix the initial condition to
$G^\phi[\vec{R}(0)]=1$ for all $\phi$.
Preservation of the trace and hermiticity  of the composite state $\rho$ require that we maintain [\Eq{eq:ase-gauge-restrict}]
\begin{align}
G^{0}=1, \qquad (G^\phi)^{*}= G^{-\phi}
.
\label{eq:asem-gauge-restrict}
\end{align}
Analogous to other problems with a gauge structure, the gauge freedom can be exploited as follows:
A transformation \eq{eq:chi-gauge} to an arbitrary gauge $G^\phi$ [relative to the reference~\eq{eq:gauge-ref}]
amounts to writing
\begin{align}
\Lket{\rho}
= \int_{-\pi}^{\pi} \tfrac{d\phi}{2\pi}
S^\phi \Lket{ w^\phi_0 } \otimes \Big[ G^\phi \Lket{\phi} \Big]
.
\label{eq:rhoformSG}
\end{align}
and correspondingly for the meter density operator \eq{eq:rhoM}
\begin{align}
\rho^\M
= \int \tfrac{d\phi}{2\pi} S^\phi \, \Lket{\phi}_{G}
\label{eq:rhoM-gauged}
.
\end{align}
with new coefficients $S^\phi=Z^\phi (G^\phi)^{-1}$.
Expectation values on the meter are gauge-invariant:
\begin{align}
	\mom^{(k)} & 
	= \Tr{M} \hat{N}^k  \rho^\M(\T)
	=
	\left . (-\partial_{i\phi})^k S^\phi G^\phi
	\right |_{\phi=0}
	.
\end{align}
One thus splits up the problem of computing the $Z^\phi = G^\phi S^\phi$ in \Eq{eq:momentDerivative}
by decomposing it into a simple or convenient gauge function $G^\phi$ (absorbed into the meter ket) and an unknown complicated part $S^\phi$,
the solution to be computed in this gauge.
For example, by a choice of $G^\phi$ one can eliminate~\cite{Calvo12a} the parametrically time-dependent capacitive screening charges from the calculation of the pumped charge (first moment).

Next, we note that gauge transformations can only be combined with the geometric factor of the solution
\begin{align}
Z^\phi = Z_\g^\phi Z_\n^\phi
,
\label{eq:ZSD}
\end{align}
which is a functional $Z_\g^\phi=e^{z_\g^\phi}$ of the parameters $\vec{R}(t)$ only.
(The nongeometric factor $Z_\n^\phi=e^{z_\n^\phi}$ additionally has a functional dependence on the velocities $\dot{\vec{R}}(t)$.)
Thus, when working in some gauge one rather considers the split
\begin{align}
Z_\g^\phi = G^\phi S^\phi
.
\label{eq:SGH}
\end{align}
rather than splitting up $Z^\phi$.

Finally, to clearly see how physical restrictions affect gauge transformations we need to consider the charge representation
(in contrast to geometric considerations which are most evident in the phase representation).
The special form of the composite state \eq{eq:rhoform} implies
\begin{align}
	\Lket{\rho}
	= \int_{-\pi}^{\pi} \tfrac{d\phi}{2\pi} \Lket{ \rho^\phi } \otimes \Lket{\phi}
	= \sum_{N} \Lket{\rho^{-N}} \otimes \Lket{N}
	\label{eq:rhoformn}
	.
\end{align}
This reflects that by our construction \eq{eq:construction} of the ideal meter model
the classical information about the charge exchanged between system and reservoir ($-N$, i.e., lost by system)
is ``copied'' to the ideal meter ($N$) without any backaction.
The charge representation in \Eq{eq:rhoformn} shows that the gauge transformation \eq{eq:chi-gauge} leading to \Eq{eq:rhoformSG}
corresponds to changing the \emph{pure} meter states to
\begin{align}
	\Lket{N} & \to \Lket{N}_{G}
	\label{eq:nice}
	\\
	& :=\int \frac{d\phi}{2\pi} e^{-i\phi N} G^\phi \Lket{\phi}
	= \sum_{N'} G^{N-N'} \ket{N'}\bra{N'}
	\notag
	.
\end{align}
These are physical \emph{mixed states} provided $G^N \geq 0$ for all $N$
in addition to the restrictions \eq{eq:ase-gauge-restrict} which translate to $\sum_N G^{N}=1$ and $G^N \in \mathbb{R}$.
In this case, the gauge transformation in the charge representation $G^{N-N'}$ is the classical probability of randomly finding the pure state $\ket{N'}\bra{N'}$  when the meter is in the gauged state $\Lket{N}_G$.
We have thus found that working in such a physical gauge, part of the transport statistics has been absorbed into parametrically evolving \emph{mixed} meter states
$\Lket{N}_G$ that are used to count charge:
the meter has been \emph{recalibrated} by making the meter states ``noisy'' with a probability distribution $G^{N-N'}$ (assumed to be known).
This is what the formal freedom of re-normalizing eigen-supervectors of the evolution superoperator $K$ [\Eq{eq:gauge}] physically amounts to
for an ideal pumping process.

\subsection{Bochner's constraints of positivity.\label{sec:bochner}}

Before we can develop a physical picture of geometric parallel transport,
we need to discuss whether a gauge $G^N$ always has (and should have)  \emph{positive} values and thus make sense as probabilities~\cite{Levitov96}.
Although for closed systems this is always true,
for open systems negative values are unavoidable,
but, on the other hand, never lead to incorrect results.

We first note that in any case $Z^N$ must be positive
because the reduced density operator of the meter \eq{eq:rhoM},
$\rho^\M
= \sum_{N} Z^{-N} \ket{N}\bra{N}$,
is a valid quantum state.
In the following we simply call $G^\phi$ ``positive''
if it Fourier transforms to a positive distribution $G^N$.
Importantly, this property can be checked with Bochner's criterion~\cite{BochnerLecture} directly on $G^\phi$ without actually performing the Fourier transform, as discussed in \App{app:bochner}.
This criterion makes explicit that in the original $\phi$-representation
there is a serious constraint on the class of allowed gauge functions $G^\phi$
if they are to have physical meaning as the Fourier transform of the statistical mixing coefficients of meter states [\Eq{eq:nice}].

\begin{figure*}[t]  
	\centering\includegraphics[width=0.75\linewidth]{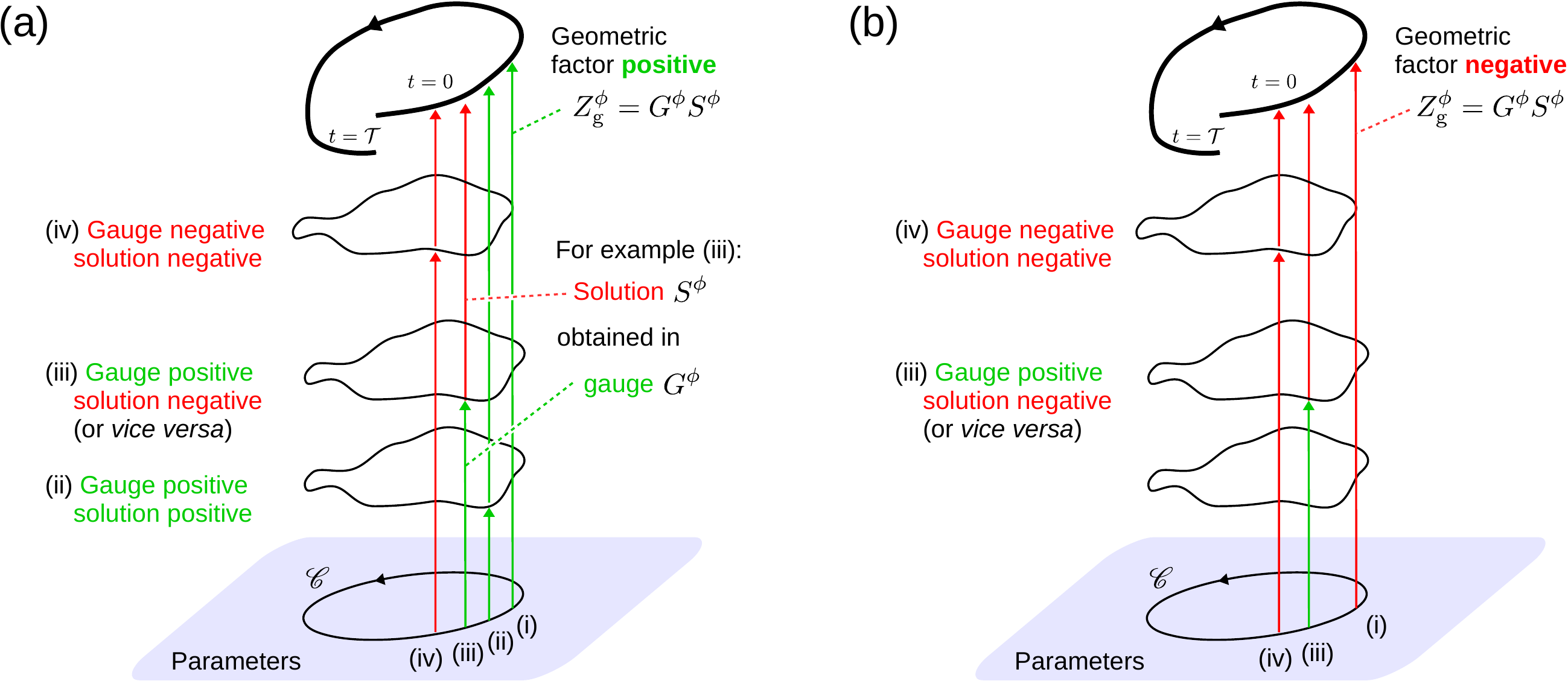}
	\\ \vspace{10pt}
	\centering\includegraphics[width=0.75\linewidth]{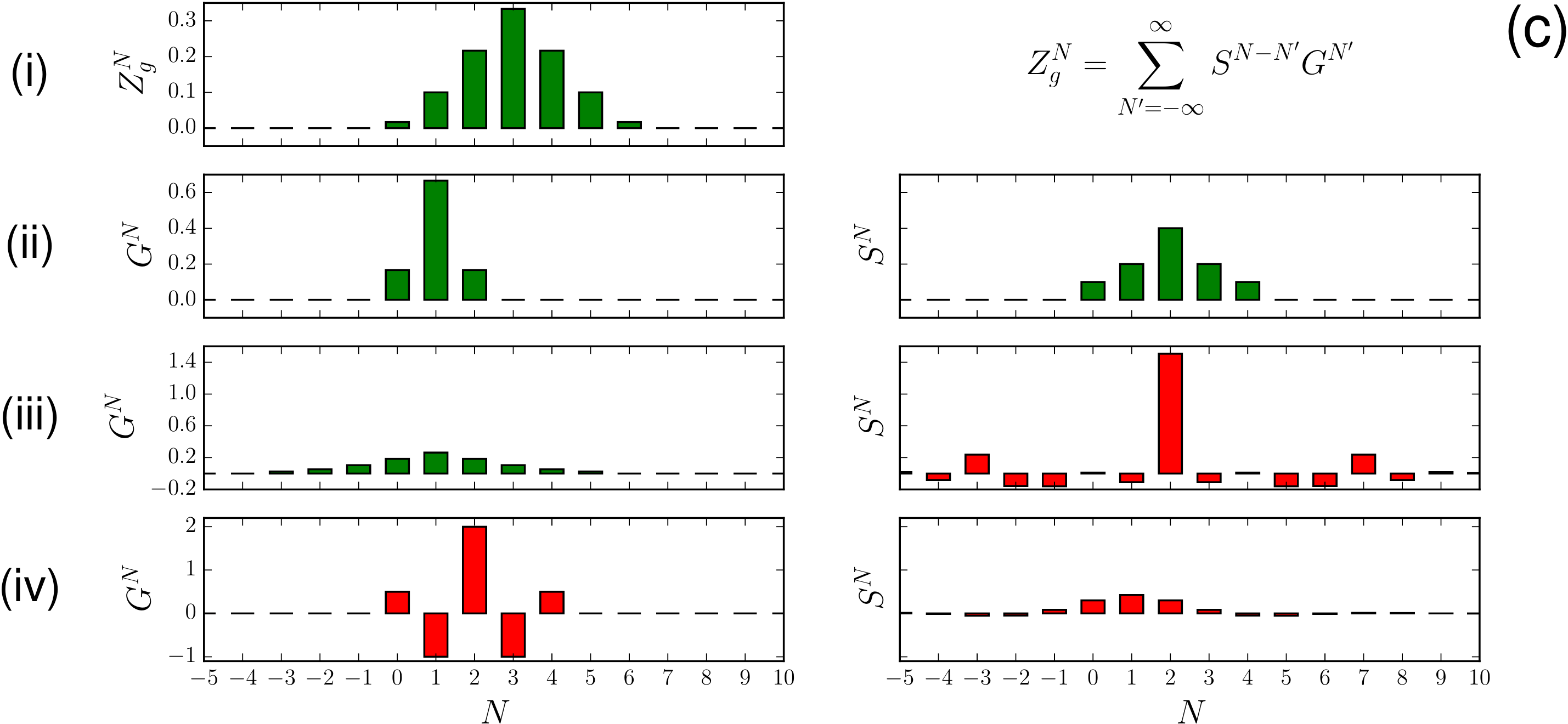}			
	\caption{
		Positivity restrictions and gauge transformations.
		Panels (a) and (b) sketch the \emph{geometric} factor $Z_\g^\phi$ of $Z^\phi=Z_\n^\phi Z_\g^\phi$
		as a curve in the total space of parameter values $\times$ gauge coordinates.
		The various possible cases of splitting $Z_\g^\phi = G^\phi S^\phi$ into a continuous gauge $G^\phi$ and a discontinuous solution $S^\phi$ (obtained in this gauge) are sketched.
		Panel (c) shows examples for these cases in the $N$-representation.	
		(a) First consider $Z_\g^\phi \approx Z^\phi$ implying that $ Z_\g^\phi$ is positive.
		(i) The factor $Z_\g^\phi$ is obtained in the reference gauge, corresponding to the plane in the sketch.
		However, one can always choose some other gauge $G^\phi$ to determine the solution $S^\phi$. In (ii)-(iii) we choose $G^\phi$ positive.		
		(ii) 
		If $S^\phi$ satisfies Bochner's stringent criterion,
		it is positive.
		In fact, the only positive gauges for which $S^\phi$ is \emph{guaranteed} to also be positive are
		$G^\phi = e^{i \phi k}$ with integer $k$,
		corresponding to the trivial shift $G^N=\delta_{N-k}$ of probability distributions.
		(iii)
		It may thus be that $S^\phi$ fails to be positive
		even if $G^\phi$ (by choice) and $Z_\g^\phi$ (necessarily) are positive.
		In this case, the solution $S^N$ is \emph{not} a probability distribution.
		(iv)
		It is even possible to choose a nonpositive gauge $G^\phi$ to compute a solution $S^\phi$,
		which can again be both positive or nonpositive.
		(b)
		In general $Z_\g^\phi$ may well be nonpositive, in contrast to (a): only $Z^\phi = Z_\n^\phi Z_\g^\phi$ is required to be positive.
		In this case, at least one of $G^\phi$ and $S^\phi$ is nonpositive, i.e., 
		case (ii) cannot occur here.
		(c)
		Examples of the splitting $Z_\g^\phi=G^\phi S^\phi$ in the $N$ representation.
		(i) ``Target'' probability distribution $Z_\g^N$.
		(ii) A positive $G^N$ that is shifted and narrowed relative to the target distribution $Z_\g^\phi$ leads to a positive solution $S^\phi$.
		(iii) Broadening the gauge function in (ii) at some point forces the solution $S^\phi$ to take on negative values to narrow down the convolution and produce the target $Z_\g^N$.
		(iv) Even when choosing a nonpositive gauge, sharply varying in sign, still allows to achieve the positive target by an also negative solution $S^\phi$.
	}
	\label{fig:gauges}  
\end{figure*}

Performing gauge transformations in an open system is thus not an innocent procedure as it is in closed quantum system:
in general, two probability distributions $Z^N$ and $G^N$ cannot be related by a convolution
$Z^N = \sum_{N'} S^{N-N'}G^{N'}$
with a third function $S^N$ that is \emph{also a probability distribution}.
Although the function $S^\phi$ in \Eq{eq:rhoM-gauged} can be found, given a positive $G^\phi$,
Bochner's criterion imposes nontrivial constraints on $S^\phi$ to be positive as well.
We stress, that the correctness of the final result $Z^\phi$ to be computed is never at stake.
This should be contrasted with closed systems where physical restrictions and geometric considerations can be cleanly separated.
To develop some intuition how such negative values arise
we consider some simple examples in \Fig{fig:gauges}.

In \Fig{fig:gauges}(a) three possible scenarios are sketched when assuming for simplicity that
the geometric factor dominates, $Z^\phi \approx Z_\g^\phi$, and thus is positive.
This figure shows that a positive function can be split into two positive functions, a positive and a negative one, or even into two negative ones.
In these scenarios, the correct physical, positive $Z_\g^\phi$ is obtained even though either the \emph{chosen} gauge ($G^\phi$) or the computed solution ($S^\phi$) or both may be nonpositive and lack a classical probabilistic interpretation.
The final answer is obtained by \Eq{eq:SGH}
corresponding to a charge-convolution $Z_\g^N = \sum_{N'} S^{N-N'} G^{N'}$.

In \Fig{fig:gauges}(b) we show additional scenarios that arise in the case where $Z_\g^\phi$ does not dominate, i.e.,
the coefficient 
$Z^\phi = Z_\g^\phi Z_\n^\phi$ consists of both a nontrivial geometric factor $Z_\g^\phi$ and nongeometric factor $Z_\n^\phi$.
Since it is only required by \Eq{eq:ZSD} that the convolution
$Z^N = \sum_{N'} Z_\n^{N-N'}Z_\g^{N'}$ is positive,
there seems to be no reason that the individual functions $Z_\n^{N}$ and $Z_\g^{N}$ must be positive.
Therefore scenarios arise where $Z_\g$ is negative and can be split into a negative and a positive or even two negative functions.
This means that geometric and nongeometric \emph{contributions} generally do not correspond to well-defined
classical \emph{processes}, i.e., with probability distributions.
For example, if nongeometric and pumping effects compete, it should be possible that their contributions partially cancel (e.g., a reduction of noise due to pumping), which requires a \emph{nonpositive} $Z_\g^\phi$ as we illustrated in \Fig{fig:gauges}(b).

Thus, generally,
considerations of the geometry / gauge structure and of the probabilistic structure of open quantum systems are nontrivially intertwined.
However, in no case one makes a mistake by applying gauge transformation:
one is only working in an intermediate picture in which the gauge and/or the solution computed in that gauge may have no physical significance.
	
\subsection{Parallel transport / horizontal lift.}

Aware of the positivity constraints
we can now express geometric parallel transport in terms of the pumping process
following the geometrical picture in \Fig{fig:gauges}.

First, as explained in  \Sec{sec:ase-review} [\Eq{eq:mode-partrans} ff.]
suppose we try to gauge away the geometric factor of the solution, $Z_\g^\phi$,
by a gauge transformation $G^\phi=Z_\g^\phi$ [\Eq{eq:rhoformSG}]
\begin{align}
	\Lket{\rho(t)} = \int_{-\pi}^{\pi} \tfrac{d\phi}{2\pi}
	Z_\n^\phi \,
	\Lket{ w^\phi_0 } \otimes \left [ Z_\g^\phi \Lket{\phi} \right ]
	.
	\label{eq:rhoformZnZg}
\end{align}
We are thus going to a ``rotating frame'' of mixed meter states \eq{eq:nice} with parametric time dependence.
As in \Sec{sec:ase-review}, this corresponds to maintaining the parallel transport condition
\begin{align}
	A^{\phi}_{G} \dot{\vec{R}}
	=
	A^{\phi} \,	\dot{\vec{R}}
	+ (G^\phi)^{-1} \tfrac{d}{dt} G^\phi
	=
	0
	\label{eq:AphiG-partrans}
\end{align}
with the gauged connection [\Eq{eq:gauge-ref}]
\begin{align}
	A^{\phi}_{G}:=
	{}_{G}\Lbraket{ \bar{w}_0^\phi | \delR | w_{0}^{\phi} }_{G}
	=
	A^{\phi}
	+ (G^\phi)^{-1} \delR G^\phi
	.
	\label{eq:AphiG}
\end{align}
Indeed, the geometric factor of the solution for
$G^\phi(t) = Z_\g^\phi(t)=e^{
	-
	\int_0^t d\tau \dot{\vec{R}}(\tau) A^{\phi} 
}$
satisfies
\Eq{eq:AphiG-partrans}, i.e., it is a horizontal lift
and a nonzero value of the geometric curvature
$B^\phi : =\delR \times A^{\phi}$
will force this curve to be discontinuous.
This failure to find a proper gauge (continuous) means that the  \emph{entire} quantum state --system plus meter--
exhibits a geometric effect:
not just the average charge, but the entire transport process (all fluctuations).

Whenever $Z_\g^\phi$ is positive,
the geometric parallel transport thus has the physical meaning of a meter recalibration
to proper physical \emph{mixed} states [\Eq{eq:nice}]
\begin{align}
	\Lket{N}_{Z_\g}
	= \sum_{N'} Z^{N-N'}_\g \ket{N'}\bra{N'}
	\label{eq:nice-Zg}
\end{align}
This holds for the ideal case of ``pure pumping'' where
$Z_\g^\phi$ dominates $Z^\phi$ and both are necessarily positive.
In this case, the information about the entire transport process can be completely incorporated in a \emph{recalibration of the meter},
which does not evolve anymore (``rotating frame'').
Parallel transport \eq{eq:AphiG-partrans} is thus the (impossible) attempt to literally calibrate away the pumping effect on \emph{all} charge transport quantities (current, noise, etc.) in \emph{continuous} fashion by an adjustment of the ideal meter.

One should, however, keep in mind [\Sec{sec:bochner}] that the splitting $Z^\phi=Z^\phi_\g Z^\phi_\n$ does not in general guarantee that $Z_\g$ is positive, i.e., pumping is not a separate process.
With this provision, the above arguments can be reversed:
\emph{because} an ideal meter can be literally recalibrated
down to the level of its classical fluctuations,
slowly driven transport measurements should in general pick up geometric contributions
in \emph{all} moments of the measurement statistics.
One overlooks this simple physical origin of geometric effects
if one assumes that the states \Eq{eq:N} in the ideal meter are pure, as we did initially [\Eq{eq:N} ff.] following Schaller et. al~\cite{Schaller09}.
Instead, one should in general allow for measurements with ``noisy'' meter states with \emph{known} gauged statistics $G^N$
(i.e., which are part of the meter design).

The equivalence of ``trivial charge transport'' meter-calibration and
geometric parallel transport was demonstrated for the average pumped in \Ref{Pluecker17a}.
Here we have generalized this natural physical idea to the \emph{entire} transport process (all moments) [cf. also  \Sec{sec:fcs} and \Sec{sec:ar}]
noting the nontrivial intertwining with the physical constraint of positivity.
In \Sec{sec:ar} we will show that this complication goes unnoticed when one considers only the first moment (= first cumulant) of pumping as in most studies of pumping.
Only in that case one
can always consider a given gauge transformation of the average charge as realized by \emph{some} meter-state calibration.

\subsection{Summary}

In this main section of this paper
we have shown that charge pumping results from the continuum of geometric phases accumulated in the adiabatic dynamics of the \emph{composite} system-plus-meter state.
In particular, these phases are picked up by
the components of the state with nonzero meter-momentum ($\phi \neq 0$)
which account for the nonstationarity of the charge meter (``the meter keeps running'').
There is necessarily a \emph{continuum} of geometric phases,
one for each value of the momentum of the meter's needle.
The gauge freedom underlying these geometric phases emerges from the simple physical freedom to recalibrate the meter
which includes \emph{making the meter ``noisy''} in a \emph{known way}.
However, the requirement of positivity nontrivially constrains the \emph{physical} meaning of \emph{geometric} quantities in open quantum systems,
in strong contrast to closed quantum systems:
only a subset of gauge transformations correspond to a probability distribution in the charge representation.
Nevertheless, we showed that the Berry-Simon type parallel transport condition of adiabatic open-system time evolution
can be understood as trying to maintaining a ``trivial charge transport \emph{process}'' on the meter.

\section{Full counting statistics (FCS)\newline of adiabatic pumping\label{sec:fcs}}

Having discussed our comprehensive approach [\Sec{sec:meter}] in terms of mixed quantum-state evolution,
we show in the remainder of the paper
how two prominent density-operator approaches to pumping,
\fcs\cite{\FCS} and \ar\cite{\ARNAIVE}, can be elegantly derived from it,
thereby reconciling their seemingly conflicting features in \Tab{tab:compare}.

In the present section, we first discuss the \fcs approach
which is most closely related to the \ase approach for an open system with meter \emph{inside},
as illustrated in \Fig{fig:model}(b).
The expressions derived in the main part of this paper [\Sec{sec:meter}] were written such that one obtains the \fcs approach by simply dropping the meter ket $\Lket{\phi}$ as we now show.
Importantly, this does \emph{not} mean that the meter is physically discarded or eliminated -- as in the case of the \ar approach discussed in the next \Sec{sec:ar}.

\paragraph*{State evolution, generating operator and counting field.}

In order to compute the statistics of charge transfer to the reservoir,
the \fcs approach in its usual formulation introduces a generating operator $\rho^\phi$ which is defined only on the system~\cite{Esposito09rev}.
This auxiliary object is constructed such that the generating function $Z^\phi = \tr_{\S} \rho^\phi$ produces the desired $k$th moment and cumulant, respectively, through\footnote
	{Our sign convention is opposite to the usual one in \fcs but is physically motivated. It is fixed by letting the ideal meter indicate the same charge as counted in reservoir, \emph{outside} the system, and by defining $\phi$ to be the momentum conjugate to the meter-needle position, see \Sec{sec:model}.
	This moreover ensures that there are no signs in the phase representation $\Lket{\rho} = \int \frac{d \phi}{2\pi} \Lket{\rho^\phi} \otimes \Lket{\phi}$
	in which the adiabatic decoupling is made.}
\begin{align}
	\mom^{(k)}=(-\partial_{i\phi})^k Z^\phi |_{\phi=0}
	,\quad
	\cum^{(k)}=(-\partial_{i\phi})^k z^\phi |_{\phi=0}
	\label{eq:fcsCumulants}
	.
\end{align}
for all $k=1,2,\ldots$.
The \emph{cumulant generating function} $z^\phi$ is the exponent:
\begin{align}
	Z^\phi = e^{z^\phi}
	,\quad
	z^\phi
	= \sum_{k=0}^\infty \tfrac{1}{k!}(-i\phi)^k \cum^{(k)}
	\label{eq:cumulants}
	.
\end{align}
Here, no reference is made to a composite system-meter \emph{state} as in \Sec{sec:meter},
the meter momentum $\phi$ is treated as an auxiliary ``counting field'' variable:
taking derivatives and setting $\phi=0$ is just a way of generating the desired expressions for the moments / cumulants \eq{eq:fcsCumulants} based on measurements in the reservoir (see \Eq{eq:app-cumulants-mk}-\eq{eq:app-pnn}).

\paragraph*{Derivation of \fcs equations from \ase.}
The \fcs approach thus effectively accounts for the measurements performed outside the system
through the $\phi$ dependence of the quantity $\rho^\phi$ defined only on the system.
In contrast, in our \ase approach to pumping \Eq{eq:fcsCumulants} arises naturally from the ideal \emph{meter inside} the open system
in line with the original motivation of the \fcs by a meter model~\cite{Levitov93},
see our discussion of \Eq{eq:momentDerivative}.
There we identified the meter phase $\phi$ as the meter-momentum [eigenvalue of the superoperator $\Phi$, \Eq{eq:Phi}]
that is conjugate to  the classical charge [superoperator $\mathcal{N}$, \Eq{eq:supercharge}].
Phase derivatives $\partial_{i\phi}$ relate to meter-charge measurements and setting $\phi=0$ discards the meter afterwards.
Finally, in our approach the generating function $Z^\phi=e^{z^\phi}$ naturally appeared when expanding the mixed meter \emph{state} (after integrating out both system and reservoirs) in the meter-momentum basis $\Lket{\phi}$.

The central equation of the \fcs approach describing the evolution of $\rho^\phi$
is derived quite simply from the adiabatic  state evolution of the system-meter model [\eq{eq:MasterEquation}, \eq{eq:chiKernel}]
by formally leaving out all reference to the meter space in \Eq{eq:densitySolution}
and treating $\phi$ as a parameter
\begin{align}
	\tfrac{d}{dt} \Lket{\rho^\phi} = W^\phi \Lket{\rho^\phi}
	\label{eq:fcs-mastereq}
	.
\end{align}
This follows ultimately from the ideality of the meter model
[tensor product structure in \Eq{eq:chiKernel2} and \eq{eq:rhoform}],
\emph{not} by discarding the meter (which we do not).
We stress that after leaving out $\int d\phi/(2\pi) \ldots  \Lket{\phi}$
the remaining quantity $\rho^\phi$ by itself is \emph{not} a state of some system,
even though \Eq{eq:fcs-mastereq} formally resembles the evolution of a quantum state.
First, it is not the state of the system,
the reduced density operator $\rho^\S$,
even though it includes that state.\footnote
	{If one does integrate out the meter [\Sec{sec:ar}]
	we indeed have
	$\rho^\phi|_{\phi=0}=\rho^\S$
	and from \Eq{eq:fcs-mastereq}
	$
	\tfrac{d}{dt} \Lket{\rho^\S} = W \Lket{\rho^\S}
	$
	with $W:=W^\phi|_{\phi=0}$.}
This difference between $\rho^\phi$ and $\rho^\S$  is crucial since physical restrictions (normalization, hermiticity) do not quench~\cite{Sinitsyn09} the gauge freedom for $\rho^\phi$ as they do for $\rho^\S$ [\Sec{sec:ar}], allowing $\rho^\phi$ to accumulate a nonzero geometric phase for $\phi\neq 0$.
Second, $\Lket{\rho^\phi}$ is also not the state of the system plus meter:
only the full system-meter expression \Eq{eq:rhoform} has this direct physical meaning.

\paragraph*{Nonstationary \fcs and Berry-Simon phase.}
For slowly driven parameters, the \ase and \fcs approach account for pumping of transport quantities
in the same way by performing an adiabatic decoupling approximation.
Taking account of the steady-state limit as explained in \Sec{sec:meter},
in particular, ``resetting'' $\C^\phi_0=1$, and defining the gauge by $\Lbraket{\unit|w^\phi_0}=1$ [\Eq{eq:densitySolution0} ff.] leads to
\begin{align}
	\Lket{\rho^\phi} \approx Z^\phi \Lket{w^\phi_0}
	\label{eq:fcs-rhochi-adia}
	.
\end{align}
Apart from parametrically following of the eigenvector $\Lket{w^\phi_0}$,
the \fcs generating operator $\rho^\phi$ picks up the factor $Z^\phi=e^{z^\phi}$.
In our approach this parametric nonstationarity\footnote
	{For fixed parameters $\vec{R}$, the eigenvalue of $W^\phi$ with the smallest negative real part determines the cumulant generating function at long times, $dz^\phi/dt \approx w^\phi_0$,
	giving a nonstationary $\Lket{\rho^\phi} \approx e^{w^\phi_0 t} \Lket{w^\phi_0}$ for $\phi \neq 0$.}
is seen to correspond to the meter needle ``running'' (net current) and fluctuating (higher cumulants).
The geometric part of the exponent $z^\phi$, the second term in 
\begin{subequations}
	\begin{align}
	z^\phi
	& = \int_0^{\T} dt w^\phi_0 - \oint_\curve d\vec{R} A^\phi
	,
	\label{eq:fcs-zphi}
	\\
	A^\phi &= \Lbraket{w^\phi_0| \delR | w^\phi_0}
	.
	\label{eq:fcs-Achi}
\end{align}
\end{subequations}
is of the Berry-Simon type due to the formal similarity
of \Eq{eq:fcs-mastereq} to the closed-system Hamiltonian evolution [\Sec{sec:ase-review}].
Because in our approach we keep $\Lket{\phi}$, we see that this
phase appears in the nonstationary component of the physical \emph{state} of system plus meter,
which is always the case [\Sec{sec:ase-review}, \Eqs{eq:rho-restrict}, \eq{eq:rhoM} and \App{app:nonstationary}].
Thus, even though $\rho^\phi$ itself is not a state, its geometric phase is an adiabatic Berry-Simon phase \eq{eq:zg} of a mixed state.

\paragraph*{Gauge freedom.}
As discussed in \Sec{sec:meter-gauge}, in our \ase approach, the gauge transformations
\begin{align}
	G^\phi = e^{g^\phi}
	,\quad
	g^\phi = \sum_{k=0}^\infty \tfrac{1}{k!}(-i\phi)^k g^{(k)}
	,
	\label{eq:fcs-gtaylor}
\end{align}
can be naturally regarded as changes of the meter factor $\Lket{\phi}$ in the system-meter state:
gauge freedom is thus a consequence of physical recalibration of the meter.
If one instead sticks to the \fcs equation \eq{eq:fcs-mastereq} in which all reference to the meter space remains implicit, gauge transformations are not easily related to meter recalibration.
Instead of \eq{eq:gauge}, gauge transformations are then equivalently introduced as a simultaneous rescaling of the eigenvectors and of the generating function
\begin{subequations}
	\begin{gather}
	\Lbra{\bar{w}^\phi_0}   \to (G^\phi)^{-1} \Lbra{\bar{w}^\phi_0}
	,\qquad
	\Lket{w^\phi_0}   \to G^\phi \Lket{w^\phi_0}
	\\
	Z^\phi   \to (G^\phi)^{-1} Z^\phi := S^\phi	
	\end{gather}
\end{subequations}
such that the generating operator $\Lket{\rho^\phi}$ in \Eq{eq:fcs-rhochi-adia} stays invariant.
As in \Sec{sec:meter}, $S^\phi$ denotes the solution sought working in the gauge $G^\phi$.
This absorption of a factor of the generating function into the eigenvector suggests no relation to physical meter calibrations.

\paragraph*{Geometric parallel transport $=$ ``trivial charge transport''.}
In the \fcs approach the parallel transport condition \eq{eq:AphiG-partrans} of the system-plus-meter state,
a condition holding for all values of $\phi$,
can be expressed in terms of geometric parts\footnote
	{The geometric part of a cumulant is in general not separately measurable (except for the first moment / cumulant~\cite{Pluecker17a}).
	See the corresponding discussion regarding the splitting $Z^\phi = Z^\phi_\n Z^\phi_\g$ in \Sec{sec:bochner}.}
of the cumulants.
To see this, consider the adiabatic evolution equation of the cumulant generating function that leads to \Eq{eq:fcs-zphi},
\begin{align}
	\tfrac{d}{dt} z^\phi_{G} 
	&= w^\phi_0 - A^\phi \dot{\vec{R}} - \tfrac{d}{dt} g^\phi
	\label{eq:fcs-dzdt}
\end{align}
in an arbitrary gauge $G^\phi=e^{g^\phi}$ [\Eq{eq:fcs-gtaylor}]. By \Eq{eq:fcsCumulants} this determines the $k$th \emph{cumulant current}
	\begin{align}
	\tfrac{d}{dt} \cum^{(k)}_{G} 
	&= w^{(k)}_0 - A^{(k)} \dot{\vec{R}} - \tfrac{d}{dt} g^{(k)}
	\label{eq:fcs-dcdt}
	\end{align}
where, similar to the real-valued cumulants $\cum^{(k)}$ \eq{eq:cumulants}, the terms on the right hand side of \Eq{eq:fcs-dzdt} have been expanded in powers of $\phi$,
\begin{align}
	w^\phi_0
	= \sum_{k=0}^\infty \tfrac{1}{k!}(-i\phi)^k w^{(k)}_0
	,\quad
	A^\phi
	= \sum_{k=0}^\infty \tfrac{1}{k!}(-i\phi)^k A^{(k)}
	.
	\label{eq:wAphi}
\end{align}
Geometric parallel transport \eq{eq:AphiG-partrans} of the system-meter state due to adiabatic evolution is then equivalent to maintaining zero for the \emph{geometric} part of the cumulant current for all $k=0,1,2,\ldots$:
\begin{align}
	A^{(k)}_G
	\dot{\vec{R}}
	= A^{(k)} \dot{\vec{R}} + \tfrac{d}{dt} g^{(k)}
	=0
	,
	\label{eq:fcs-cumcur-partrans}
\end{align}
i.e., the second, \emph{pumping} contribution to
\begin{align}
\cum^{(k)}=
\int_0^{\T} d t w_0^{(k)} 
- \oint_{\curve} d\vec{R} A^{(k)} 
\label{eq:DeltaCk-fcs}
.
\end{align}
Geometrically, \Eq{eq:fcs-cumcur-partrans} defines a horizontal lift curve in the space of parameters $\vec{R}$ $\times$ \emph{cumulant gauges} $g^{(k)}$.
However, one should note that by our discussion in \Sec{sec:bochner},
maintaining a ``trivial transport process'' is physically realizable as a recalibration of the meter state only if the \emph{geometric} factor of $Z^\phi$ has a positive inverse Fourier transform (guaranteed by Bochner's criterion).

\paragraph*{Adiabatic decoupling produces nonadiabatic current?}

Equation \eq{eq:fcs-cumcur-partrans}
generalizes the parallel transport condition derived in \Ref{Pluecker17a} for the first cumulant / moment ($k=1$) to arbitrary cumulants ($k \geq 2$).
However, in \Ref{Pluecker17a} the considerations were based on the explicitly \emph{nonadiabatic} \ar approach.
In contrast, the \fcs / \ase result \eq{eq:fcs-cumcur-partrans}
was derived using the adiabatic decoupling  procedure \eq{eq:fcs-rhochi-adia}.
In this last subsection we address this apparent conflict from the \fcs side
by borrowing a few simple considerations from the \ar approach that are further discussed in the next section.

The geometric phase in \Eq{eq:fcs-Achi} is generated by a driving-velocity dependent argument in the contribution
$
- \oint_\curve d\vec{R} A^\phi
=
- \int_0^{\T} dt \dot{\vec{R}} A^\phi
$
which is well-known from analogous closed-system state-evolution expressions.
However, one may equally well consider the $\dot{\vec{R}}$-dependence
in this expression to indicate a nonadiabatic effect as one finds in response calculations~\cite{Berry93,Landsberg92,Landsberg93}.
This point naturally arises when one inquires into the \emph{currents} defined by \Eq{eq:fcs-dzdt} that flow through the system-reservoir boundary,
e.g., in the gauge $G^\phi=1$, $g^\phi=0$:
\begin{align}
I^{(k)}_\cum
:= \tfrac{d}{dt} \cum^{(k)} = w^{(k)}_0 -  A^{(k)} \tfrac{d}{dt} \vec{R}
.
\label{eq:fcs-Ik}
\end{align}
We now compare this with the form one would expect in a response calculation,
i.e., the expansion of the exact cumulant \emph{currents}
(with a functional dependence on the driving)
to linear order in the driving velocity:
$I^{(k)}_{\cum}
\approx I^{(k)}_{\cum}[\vec{R}] + \delta I^{(k)}_{\cum}(\vec{R},\dot{\vec{R}})+\ldots$
We see that the adiabatic cumulant currents
are just the real-valued coefficients of the
parametric eigenvalue [\Eq{eq:wAphi}]
$I^{(k)}_{\cum}[\vec{R}]=w^{(k)}_0[\vec{R}]$.
The geometric cumulant current $\delta  I^{(k)}_{\cum}( \vec{R}, \dot{\vec{R}} ) = - A^{(k)} [\vec{R}] \,  \dot{\vec{R}}$ is linear in the driving velocity $\dot{\vec{R}}$
and therefore generates the \emph{pumping} contribution to the cumulant $\cum^{(k)}=\int_{_0}^{\T} I^{(k)}_{\cum} dt$.
The real-valued geometric connection $A^{(k)}$ has the direct physical meaning
of an adiabatic-response coefficient:
\begin{align}
- A^{(k)}[\vec{R}]
= \left. \frac{
	\delta  I^{(k)}_{\cum}
	}{ \delta \dot{\vec{R}} } \right|_{\dot{\vec{R}}=0}
.
\end{align}
This is the first \emph{nonadiabatic} correction to the adiabatic cumulant current.
The parallel transport \eq{eq:fcs-cumcur-partrans} is thus equivalent to (trying to find) gauges $g^{(k)}$ that maintain \emph{zero nonadiabatic cumulant current}
$ \delta  I^{(k)}_{\cum,G}
:= \tfrac{d}{dt} \cum^{(k)}_{G}
=
\delta  I^{(k)}_{\cum} + \tfrac{d}{dt} g^{(k)}$
for all $k=0,1,2,\ldots$:
\begin{align}
\delta  I^{(k)}_{\cum,G}( \vec{R}, \dot{\vec{R}} )
=0
.
\label{eq:fcs-cumcur-partrans2}
\end{align}
Thus, once one turns to a response formulation,
geometry and \emph{non}adiabaticity are seen to be closely linked in open systems, even ``from within'' the \fcs / \ase approach based on adiabatic decoupling.

The nonadiabaticity responsible for $\delta  I^{(k)}_{\cum}$ is due to the finite ``time-lag'' between system and measurement of the charge outside the system [see \Sec{sec:intro} and \Sec{sec:ar}].
This is confirmed already by considering the simplest case of the first moment / cumulant $k=1$ for which \Eq{eq:fcs-dcdt} with $G^{(1)}=1$ gives
\begin{subequations}\begin{align}
	\cum^{(1)} &= \braket{\hat{N}(\T)} - \braket{\hat{N}(0)}
	\\
	&=
	\int_0^{\T} d t (-\partial_{i\phi}) w_0^{\phi} |_{\phi=0}
	+ \oint_{\curve} d\vec{R} \, \partial_{i\phi} A^\phi |_{\phi=0}
	.
\end{align}\label{eq:DeltaN-fcs}\end{subequations}
When one actually evaluates the expressions for
$A^{(1)}=-\partial_{i\phi} A^\phi |_{\phi=0}$ within the \ar approach,
i.e., by really integrating out the meter as discussed in \Sec{sec:ar},
one finds the result explicitly contains a \emph{finite} ``lag'' time
and thus cannot be an adiabatic quantity \emph{as seen by the system only}.
This second geometric part \eq{eq:DeltaN-fcs} was plotted for our example system in \Fig{fig:pump}.

One can thus equally well trace the physical origin of the $\dot{\vec{R}}$-dependence
that turns $
- \oint_\curve d\vec{R} A^\phi
$ into a geometric curve integral to the ``lag'' or retardation between the system and the measured observable.
In the \ase and \fcs approach this explicit expression of the ``lag'' is never manifest
since it accounts for a meter that is ''running'' \emph{inside} the open system.
The lag is correctly kept track of but in a fragmented and implicit way.\footnote
	{The ``lag'' effect is distributed over various pieces that only come together in the final step of the calculation.
	This is already the case for the pumping current ($k=1$) as explicitly worked out in Sec.~V~C.4 of \Ref{Pluecker17a}.
	More generally, to compute $A^{(k)}= -\partial_{i\phi}^k A^\phi |_{\phi=0}$ from
	$A^\phi = \Lbraket{w^\phi_0|\delR w^\phi_0}$
	[\Eq{eq:fcs-Achi}] requires contributions from both $\Lbra{\bar{w}^\phi_0}$ and $\Lket{w^\phi_0}$ to \emph{all} lower orders $l=0,\ldots,k$ in $\phi$.}

\paragraph*{Summary.}
In open quantum systems there is an \emph{additional} caveat in the term ``adiabatic'':
the confusion whether the $\dot{\vec{R}}$-dependence indicates adiabaticity or nonadiabaticity is \emph{not}\footnote
	{In \Ref{Pluecker17a} we checked that \ase and \ar approach when applied to the \emph{steady} state of the \emph{system only}, 	are fully consistent:
	what is (non)adiabatic in one approach, is thus also (non)adiabatic in the other.
	In particular, the pumping curent is nonadiabatic in both the \ase and \ar approach applied to the \emph{system only}.
	Note carefully: in \Sec{sec:meter} and effectively also \Sec{sec:fcs} we applies the \ase approach to the \emph{system-plus-meter},
	shifting the boundary of the open system to include the meter.}
 a matter of defining ``adiabatic'' differently (i.e., which energy scales bound the driving frequency).
Although in all discussed approaches (\ase, \fcs, \ar) it is required~\cite{Pluecker17a} that $\dot{R} \ll \Gamma$ and the same results are obtained~\cite{Nakajima15,Pluecker17a},
the \fcs is adiabatic \emph{relative} to the system \emph{including the meter},
whereas the \ar is nonadiabatic \emph{relative} to the system \emph{without the meter}.
It is revealing that the analogy to closed systems --characteristic of the \fcs / \ase approach--
breaks down precisely when one inquires in the spirit of the \ar approach into quantities [\Eq{eq:fcs-cumcur-partrans2}] characteristic of an open system (currents).
\Tab{tab:compare} outlines how the question of adiabaticity is closely tied to many other difficulties that depend on whether the system boundary includes / excludes the meter.
That this is not merely a formal choice becomes even more evident in the next section,
where we explicitly place the meter outside the system, changing all the features listed in \Tab{tab:compare}.

\section{Nonadiabatic pumping current: AR approach\label{sec:ar}}

Finally, we show how the \ar approach
can be derived from the adiabatically evolving state of system plus meter [\Sec{sec:meter}],
thereby clarifying the remaining unsettled questions raised in the introduction.
The \ar approach is obtained from the \ase approach by \emph{physically discarding} the meter (measurement outcomes).
Similar to the \fcs approach, the \ar approach thus eliminates $\Lket{\phi}$ from the equations in \Sec{sec:meter},
but this time one additionally sets $\phi=0$, which corresponds to tracing out the meter [\Sec{sec:meter-pump}].
This leads to a very different physical and geometric picture.

\paragraph*{Derivation of \ar equations from \ase.}
To achieve a description referring only to the system,
both for the state and for measurements done \emph{outside} the system,
we need to trace out the meter relative to \Sec{sec:meter}.
We first do this for the state $\rho(t)$:
tracing over the master equation \eq{eq:MasterEquation}, accounting for
\Eq{eq:chiKernel2} and \eq{eq:rhoform}, gives a master equation for the reduced state $\rho^\S(t) = \tr_{\M} \rho(t)$:
\begin{subequations}
	\begin{align}
	\tfrac{d}{dt}\Lket{\rho^\S}
	= \Tr{\M}  \tfrac{d}{dt} \Lket{\rho}
	=
	W^\phi \big|_{\phi=0} \, \Lket{ \rho^\S}
	= W\Lket{\rho^\S} 
	\label{eq:ar-state}
	.
\end{align}
Here the key step is to use $\tr_{\M} \Lket{\phi} = 2\pi \delta(\phi)$, i.e., setting $\phi=0$ integrates out the meter [cf. \Eq{eq:phizerotrace}].
Preservation of hermiticity and normalization for the composite system implies that these are also preserved by the master equation \eq{eq:ar-state}.
As we discussed in \Sec{sec:ase-review}, these properties restrict the Berry-Simon geometric phase of $\rho^\S$ to be zero in the driven steady state~\cite{Pluecker17a}.
Thus, once we physically discard the meter (by setting $\phi=0$) the geometric phase of the system state is lost.

To recover the pumping effect as a geometric phase we need to consider the \emph{transport} of an \emph{observable} outside the system,
i.e., keep track of additional information about the reservoirs.
Using the reduced state $\rho^\S$ we cannot describe measurements of the charge $\hat{N}$ registered by the meter.
This is reflected by the system kernel $W$ being the $\phi=0$ component [cf. \Eq{eq:phizerotrace}]  of system-meter kernel $K$ [\Eq{eq:chiKernel2}], the sum over all possible charge transfers:
\begin{align}
	W:=W^\phi|_{\phi=0} = \sum_N W^N
	\label{eq:ar-W}
	.
\end{align}
To recover the charge as computed in the \ar approach we need to consider the transport \emph{current} into the reservoir
as registered by the meter in the \ase approach,
$I_N:=\tfrac{d}{dt}\braket{\hat{N}}$
where $\braket{\hat{N}}=\tr_{\M} \hat{N}\rho^\M
=\tr_{\M} (\mathcal{N}\rho^\M)$
[\Eq{eq:Nexp}].
Using \Eq{eq:phizerotrace}, i.e.,
tracing out the meter \emph{after} the measurement,
we obtain
\begin{align}
	\tfrac{d}{dt}\braket{N}
	&
	=
	\Tr{\M}  \Tr{\S} \mathcal{N} \tfrac{d}{dt} \Lket{\rho}
	\\
	&
	=
	\Tr{\S}
	\int_{-\pi}^{\pi} \tfrac{d\phi}{2\pi}
	W^\phi \Lket{ \rho^\phi }
	\otimes \Tr{\M}   \mathcal{N}\Lket{\phi}
	\\
	&
	=
	\Tr{\S} (-\partial_{i\phi}  W^\phi)
	\, 
	\Lket{ \rho^\phi } \Big|_{\phi=0}
	+
	\Tr{\S} W^\phi
	\, 
	(-\partial_{i\phi})  \Lket{ \rho^\phi } \Big|_{\phi=0}
	\notag
	\\
	&
	=
	\Lbra{\unit}
	W_{I_N} \Lket{ \rho^\S}
	\label{eq:ar-transport} 
	.
\end{align}
The second term vanishes by the trace preservation of the master equation \eq{eq:ar-state},
$\tr_\S W =\tr_\S W^\phi \big|_{\phi=0}
= 0$.
Thus, by integrating out the meter from our \ase approach we obtained\footnote
	{Usually, these equations --including the explicit expressions for the kernels $W$ and $W_{I_N}$-- are derived more economically~\cite{Splettstoesser06} without any reference to \fcs, counting fields or the meter model,
	making it a very practical approach.} 
the coupled set of \ar equations \eq{eq:ar-eq},
in which the observable current is ``enslaved'' to the driven state~\cite{\ARNAIVE}.
The \emph{current-kernel}
\begin{align}
	W_{I_N} & :=
	- \partial_{i\phi}  W^\phi  |_{\phi=0}
	\label{eq:ar-WIN}
\end{align}\label{eq:ar-eq}\end{subequations}
accounts for the additional system-reservoir correlations that are needed to compute the current of charge measured outside the system.
We point out the difference between the current
$\tfrac{d}{dt}\braket{\hat{N}}$,
namely the average \emph{velocity} of the needle of the charge-meter,
and the needle's canonical \emph{momentum} $\phi$.
Our result \eq{eq:ar-transport}-\eq{eq:ar-WIN}
relates these as
$\tfrac{d}{dt}\braket{\hat{N}}=
\tfrac{\partial}{\partial \phi} \Lbra{\unit}
i W^\phi \Lket{\rho^\S}  |_{\phi=0}$,
which interestingly resembles a classical Hamilton equation for the velocity $\tfrac{d}{d t} x=\tfrac{\partial}{\partial p}H$,
while the phase $\phi$ plays the role of momentum
and the meter is again discarded after the measurement by setting $\phi=0$ [\Eq{eq:phizerotrace}].

\paragraph*{Nonadiabatic, driven stationary state.}
We now follow the \ar approach
and compute the current coupled to the state evolution by \Eq{eq:ar-eq}.
Notably, to obtain any pumping contribution at all for the first moment $\mom^{(1)}=\braket{N}(\T)-\braket{N}(0)$
we need to treat the parameter driving in equations \Eq{eq:ar-eq} in a way that \emph{differs} from both the \ase{}  / \fcs approach.
By putting the meter outside the system in the \ar approach we have to deal with the finite physical ``lag'' between system and meter: we need a \emph{nonadiabatic} approximation
relative to the adiabatic system state which obeys the instantaneous equation (``i'')
$W[\vec{R}] \,  \Lket{\rho^\i [\vec{R}]}=0$.
In the notation of \Sec{sec:meter}-\sec{sec:fcs} this is the parametric stationary state, $\Lket{\rho^\i}=\Lket{w_0}$. 
The required nonadiabatic correction or \emph{response} $\Lket{\rho^{\S,\a}}$ (``r'')
\begin{subequations}\begin{align}
	\Lket{\rho^\S} 
	&\approx
	\Lket{\rho^{\S,\i}} 
	+ \Lket{\rho^{\S,\a}}
	,
	\label{eq:ar-rho}
\end{align}
is easily found to depend on $\vec{R}$ \emph{and} linearly on $\dot{\vec{R}}$:
\begin{align}
	\Lket{\rho^{\S,\a} }
	&
	\approx \frac{1}{W}  \dot{\vec{R}} \, \delR  \Lket{\rho^{\S,\i}}
	\label{eq:ar-lag}
	.
\end{align}\label{eq:ar-rho-exp}\end{subequations}
Since we have discarded the meter
the steady system state $\rho^\S(t)$ given by \Eq{eq:ar-rho} does \emph{not} exhibit a geometric phase,
even though it includes nonadiabatic effects\footnote
	{Although $\Lket{\rho^{\S,\i}}$, and the nonadiabatic correction $\Lket{\rho^{\S,\a}}$ \emph{can both} acquire geometric phases,
	these phases were shown to be zero due to trace preservation in \Ref{Pluecker17a} [App. G and H], see also \Sec{sec:ase-review}:
	the steady state $\Lket{\rho^\S(t)}$ of the \emph{system only} is geometrically trivial.
}.
This closely correlates with the fact that for fixed parameters the system (without the \emph{running} meter) has a well-defined \emph{stationary} state.
In contrast, the system-meter state $\rho(t)$ in the \ase approach does have such nonstationary components with geometric phases.
Our analysis shows that this opposite state of affairs
[\Tab{tab:compare}] is a consequence only\footnote
	{In our \ase approach we do not put in / leave out anything by hand relative to the \fcs and \ar approach,
	as explicitly verified in \Refs{Nakajima15,Pluecker17a}.
	See \App{app:steady} regarding a superfluous term in \Ref{Nakajima15}.}
of having shifted the boundary defining the open system to exclude the meter [$\phi=0$] as in \Fig{fig:model}.
This is a characteristic of geometric effects in open systems:
shifting the boundary can drastically change both the physical and the geometric description without changing any prediction for measurements.

\paragraph*{Landsberg's geometric phase and transported charge.}

The \ar equations \eq{eq:ar-state}-\eq{eq:ar-WIN}
by their ``enslaved'' structure generate a dissipative geometric phase
which was considered in particular by Landsberg~\cite{Landsberg92,Landsberg93,Andersson03thesis,Andersson05}, cf. also \Ref{Sinitsyn09}.
The splitting \eq{eq:ar-rho} of the geometrically trivial state
translates into
a nongeometric and geometric contribution to the transported charge
\begin{subequations}
\begin{align}
	\mom^{(1)}
	&= \braket{\hat{N}(\T)} - \braket{\hat{N}(0)}
	\\& \approx \int_0^{\T} d tI_N[\vec{R}(t)] + \oint_{\curve} d\vec{R} A
	.
\end{align}\label{eq:DeltaN-ar}\end{subequations}
The adiabatic part $\Lket{\rho^{\S,\i}}$ leads to the nongeometric adiabatic current
$I_N[\vec{R}] = \Lbra{\unit} W_{I_N} \Lket{\rho^{\S,\i}}$
that survives even at zero driving velocity ($\dot{\vec{R}}=0$).
The nonadiabatic response correction $\Lket{\rho^{\S,\a}}$ gives rise to a nonadiabatic current $\delta I_N(\vec{R},\dot{\vec{R}} ) = \Lbra{\unit} W_{I_N} \Lket{\rho^{\S,\a}}$ 
that naturally defines a geometric connection
\begin{align}
	\left. \frac{ \delta  I_N	 }{ \delta \dot{\vec{R}} } \right|_{\dot{\vec{R}}=0}
	=
	A[\vec{R}]
	=
	\Lbra{\unit} W_{I_N} \frac{1}{W} \nabla_{\vec{R}} \Lket{\rho^{\S,\i}}
	\label{eq:ar-A}
	.
\end{align}
Physically, this Landsberg connection~\cite{Pluecker17a} is just the leading current-response to the driving velocity, a \emph{nonadiabatic} quantity:
it depends on the inverse relaxation kernel $W^{-1}$, i.e., a finite ``lag'' or retardation [\Eq{eq:ar-lag}].
The pumped charge contribution is equal to the \emph{single} geometric phase $\oint_{\curve} d\vec{R} A$ of Landsberg,
in contrast to the \ase and \fcs approach
where exactly the same result is obtained from an adiabatic approximation and a continuum of Berry-Simon phases.
In \Ref{Nakajima15,Pluecker17a} it was technically verified that
\Eq{eq:DeltaN-fcs} and \eq{eq:DeltaN-ar} are equivalent\footnote
	{The opposite signs are in a way unavoidable because in the \ase and \fcs approach leading to $A^{(1)}$ the natural convention is to require the expression for the connection [\Eq{eq:fcs-Achi}] to resemble that of Berry [\Sec{sec:ase-review}].
	Landsberg's connection $A$, however, is naturally equated [\Eq{eq:ar-A}] to the physical pumping effect without a sign.}:
$A=-A^{(1)}=\partial_{i\phi} A^{\phi}|_{\phi=0}$ and
$I_N^{\i} = -\partial_{i\phi} w_0^{\phi}|_{\phi=0}$.
Here, however, we have \emph{physically} related these two very different phases:
the Landsberg pumping phase appears when one traces out an ideal meter from a composite system that accumulates Berry-Simon phases.
Notably, it does not appear by simply taking the partial trace of the state, but additionally requires the partial trace of a \emph{transport equation} [\Eq{eq:ar-transport}].
Moreover, in the previous section [\Eq{eq:fcs-Ik}] we indicated that
higher \fcs cumulants $\cum^{(k)}$ similarly can be understood as nonadiabatic Landsberg phases.

Practically, the Landsberg curvature $B=\delR\times A$ obtained from \Eq{eq:ar-A} simplifies calculations~\cite{Reckermann10a,Calvo12a,Pluecker17a}
and it directly corresponds to the physical pumped charge per unit area in driving-parameter space.
Our illustration in \Fig{fig:pump} of the charge pumping effect for our example quantum dot of \Sec{sec:model} was computed this way~\cite{Pluecker17a},
coinciding with the \fcs expressions.

\paragraph*{Gauge transformations -- geometric phase accumulated by observable?}
In the \ar approach, the gauge freedom that allows the Landsberg geometric phase to accumulate lies in the possibility of recalibrating the meter observable~\cite{Sinitsyn09,Pluecker17a} as $\hat{N} \to \hat{N}+g \mathds{1}$ [\Eq{eq:GaugeFreedom}].
Our considerations in \Sec{sec:meter} readily reveal this freedom:
the gauge transformation \eq{eq:nice} of the \emph{meter states}  $\Lket{N} \to \Lket{N}_G$ defines a new gauged \emph{meter observable} operator\footnote
	{Note in \Eq{eq:ar-Nshift} that only for the charge and phase superkets
	we do not use the Liouville bra-ket convection \eq{eq:superbraket},
	i.e., $\Lket{N}\neq \hat{N}$ and $\Lket{\phi}\neq \hat{\phi}$}:
\begin{align}
	\hat{N} = \sum_N N \Lket{N} \to \hat{N}_G = \sum_N N \Lket{N}_G = \hat{N} + g \unit
	\label{eq:ar-Nshift}
	,
\end{align}
where $
g=\sum_{N} N G^N
= - \partial_{i\phi} G^\phi |_{\phi=0}
= - \partial_{i\phi} g^\phi |_{\phi=0}
=g^{(1)}
$
is the real linear coefficient in $G^\phi$ or $g^\phi$ [\Eq{eq:fcs-gtaylor}].
This provides a direct physical understanding how this gauge freedom in the \emph{observable} --uncommon in quantum physics--
emerges from the gauge freedom in the system-meter \emph{state} in the familiar Berry-Simon setting.
In the usual formulation of the \ar approach~\cite{Splettstoesser06}
this requires establishing the gauge covariance of the \ar equations \eq{eq:ar-eq}
through a careful derivation of the \emph{current kernel}~\cite{Pluecker17a}.

Another insight going beyond \Ref{Pluecker17a} is that for the first moment $\mom^{(1)}$ the gauge transformation of the observable can always be understood physically as arising from \emph{some} recalibration of the meter states (mixed instead of pure ones),
rather than recalibrating the meter observable \eq{eq:ar-Nshift}.
In \App{app:observable-gauge} we show that the positivity restriction discussed in \Sec{sec:bochner}
does not prohibit the construction of a meter-state gauges corresponding to any given real value of $g$.
Since for pumping of the first moment, the gauge freedom is exhausted by the observable recalibration \eq{eq:ar-Nshift},
this implies that Bochner's positivity criterion \Sec{sec:bochner} imposes no constraint on pumping of the average charge
unlike the situation for higher moments.

\paragraph*{Parallel transport = zero nonadiabatic current.}
Finally, by discarding the meter,
the ``trivial charge-transport condition'' of the \ar approach,
\begin{align}
	\delta  I_G ( \vec{R}, \dot{\vec{R}} )
	=
	A_G [\vec{R}]\dot{\vec{R}}
	=
	A [\vec{R}] \dot{\vec{R}} + \tfrac{d}{dt} g
	=0
	\label{eq:ar-cur-partrans}
	,
\end{align}
is found to be just the first of a family of parallel transport conditions [\Eq{eq:AphiG-partrans}]
that defines the parallel transport of the adiabatically evolving system-meter state (the entire transport process),
see \Eq{eq:fcs-cumcur-partrans} for $k=1$.
The condition \eq{eq:ar-cur-partrans} was related in \Ref{Pluecker17a} to attempting to gauge away [by the recalibration \eq{eq:ar-Nshift}] the nonadiabatic pumping current (as seen by the system),
i.e., Landsberg's connection \emph{equals} the pumping current.

\paragraph*{Summary.}
Our approach developed in \Sec{sec:meter} allowed us to show how the \ar approach
--with all its unfamiliar aspects listed in \Tab{tab:compare} -- 
emerges naturally from an adiabatic state-evolution approach similar to closed systems
by just a single physical operation:
discarding the meter, i.e., shifting the boundary from \Fig{fig:model}(b) to \Fig{fig:model}(a).

\section{Conclusion and Outlook\label{sec:conclusion}}

Starting from an explicit model of system, reservoirs and an ideal meter
we have shown that geometric effects in driven transport have their direct origin
in the physically obvious freedom to calibrate the meter.
In contrast to \Ref{Pluecker17a} where only the first moment / cumulant was addressed,
this applies to the \emph{entire} transport process, i.e., all possible charge measurements performed on the meter,
or, equivalently, all moments / cumulants of the transported charge.
The recalibration allows for a ``noisy'' meter  (with known noise)
that counts charges using \emph{mixed} quantum states instead of the pure ones assumed in \Ref{Schaller09}.
Notably, physical recalibrations form only a semigroup in the group of all possible geometric gauges
since they must satisfy the nontrivial restrictions of Bochner's criterion for positive probability distributions.
Nevertheless, one can make use of the full geometric gauge group to apply
standard considerations of connections on a fiber bundle.

We showed that two widely used approaches to pumping
--full counting statistics (\fcs) and adiabatic response (\ar)--
transparently follow from our approach.
We showed their many crucial differences in \Tab{tab:compare}
all go back to the choice of the open system boundary, either including or excluding the meter.
In particular, the central issue of the (non)adiabaticity of pumping transport was fully clarified this way.
We also noted that by the computational choice of working either with moments (\ar) or cumulants (\fcs)
one decides \emph{where one places the open-system boundary}
and commits oneself to one of the two very different physical and geometric pictures.
If one entirely focuses on the average pumped charge -- as often done-- this is easily overlooked
since the first moment equals the first cumulant.

Finally, we showed that when going beyond the average charge, the probability distributions that appear both in the \ase and \fcs approach may turn into nonpositive functions when working in certain gauges.
As we explained, this is a generic situation but it is not problematic since the computed results are gauge invariant:
unlike closed systems the ``intermediate'' picture obtained in some gauges need not make physical sense.
That the physical constraint of positivity is incompatible with geometric gauge structure,
is important if one wants to use gauge invariance as a principle for \emph{constructing} models within the open system approach (Liouville space).
	
Finally, we comment on the few simplifying assumptions that we made:

(i)
We limited our attention to the \ar approach as formulated for the first moment $\mom^{(1)}$.
However, an extension of the \ar approach to \emph{all} moments $\mom^{(k)}$ is possible by calculating additional \emph{observable} memory kernels.
This leads to additional Landsberg geometric phases, one for each $\mom^{(k)}$, which are nontrivially related [\Eq{eq:cummom} ff.]
to the cumulants $\cum^{(k)}$ obtained by \Eq{eq:DeltaCk-fcs} from the Berry-Simon phases of \ase / \fcs approach.

(ii)
Although for clarity we have focused on charge transport, our considerations can be extended to
spintronics (spin counting~\cite{Lopez12a})
quantum thermodynamics (energy / heat transport~\cite{Sanchez12a,Sanchez13a,Yuge13}
and simultaneously measurable quantities (multi-counting statistics~\cite{Nakajima15}).

(iii) For simplicity, we assumed that the eigenvalues $w_0^\phi$ are gapped for all $\phi$ [\Eq{eq:densitySolution0}]
but we noted that crossings of the eigenvalues may in principle occur.
This leads to interesting \emph{topological} effects studied recently in \cite{Ivanov10,Ren13}
and our considerations can be adapted to this.
	
(iv)
Importantly, even when maintaining the strong restrictions of the ideal meter model
--underlying  all discussed approaches (\ase, \fcs, \ar)--
many of the presented considerations can be extended further,
in particular, to general \emph{non-Markovian} open-system dynamics.
	
(vi) Finally, we raise the interesting question how our relation between meter calibration
(gauge freedom) and pumping effects (geometric phases) is modified when extending the model beyond \Ref{Schaller09} to \emph{non-ideal} measurements.
It seems promising to combine our approach with related insights from quantum-information and measurement theory (``quantum instruments''). 

\acknowledgments
We acknowledge stimulating discussions with
M. Ansari, G. Catelani, S. Das, M. Martin-Delgado, Y. Mokrousov, M. Pletyukhov, V. Reimer, R.-P. Riwar and B. Terhal.
T. P.  was  supported  by  the  Deutsche Forschungsgemeinschaft  (RTG  1995)
and  J.  S.  by  the Knut and Alice Wallenberg Foundation and the Swedish VR.

\appendix
\section{Moments and cumulants\label{app:cumulants}}

In this appendix we discuss the different moments and cumulants used in Sects.~\sec{sec:meter}-\sec{sec:fcs} and verify \Eq{eq:deltaNk},
\begin{align}
	\braket{\hat{N}^k }({\T})
	= \mom^{(k)}
	.
	\label{eq:mkNkapp}
\end{align}
We distinguish between measurements on the reservoir (charge operator $\hat{n}$) as considered in most works~\cite{Esposito09rev}
and the measurements on the ideal meter (meter position $\hat{N}$).

\paragraph*{Reservoir charge $\hat{n}$: 2-point moments}
For measurements of the reservoir charge
at two times, with outcome $n'$ at $t={\T}$ and $n$ at $t=0$,
one can compute the statistical 2-point moments
\begin{align}
	\mom^{(k)}:=\sum_{n n'} (n'-n)^k p^{n'n}
	= \sum_{\Delta n=-\infty}^\infty (\Delta n)^k p^{\Delta n}
	\label{eq:app-cumulants-mk}
,
\end{align}
from the distribution
$p^{\Delta n}:=\sum_{n=0}^\infty p^{n+\Delta n,n}$
for changes $\Delta n=n'-n$.
The joint distribution $p_{n',n}$ can be computed~\cite{Esposito09rev} by the standard quantum rules:
expanding $\hat{n}=\sum_n n \hat{O}^n$ with projective measurement operators $\hat{O}^n$,
assuming that $[\hat{n},\rho^\R]=0$,
and tracing out the system (S) and reservoirs (R):
\begin{align}
	p^{n'n} =
	\Tr{\S} \Tr{\R} \hat{O}^{n'}
	U
	\hat{O}^{n} (\rho^\R \otimes \rho^\S) \hat{O}^{n}
	U^\dag
	\hat{O}^{n'}
	\label{eq:app-pnn}
	,
\end{align}
where $U$ is the evolution operator from time $0$ to ${\T}$.

For \App{app:reservoir-charge} we note that
the $k$-th moment is simply the average of the $k$th-power of the change of the Heisenberg charge operator $\hat{n}(t)$,
\begin{align}
	\mom^{(k)}= \braket{ \hat{T} [\hat{n}(T)-\hat{n}(0)]^{k} }
	\label{eq:mom-heis}
\end{align}
provided one time-orders the expression by $\hat{T}$.
By definition, the generating function 
has Taylor coefficients proportional to the moments:
\begin{align}
	Z^\phi = \sum_{k=0}^\infty \frac{(-i\phi)^k}{k!} \mom^{(k)} 
	,
	\quad
	\mom^{(k)}= \left. (-\partial_{i\phi})^k Z^\phi \right|_{\phi=0}
	\label{eq:mkdef}
\end{align}
Summing the series with \Eq{eq:mom-heis}, one obtains~\cite{Belzig01,Nazarov03b,Bagrets03}:
\begin{subequations}
	\begin{align}
Z^\phi
&=
\braket{ e^{-i\phi \hat{n}({\T})} e^{i\phi \hat{n}(0)} }
\\&=\braket{\hat{T} e^{-i\phi [\hat{n}({\T}) - \hat{n}(0)]}}	
=
\braket{\hat{T} e^{-i\phi \int_{0}^{\T} dt \hat{I_n}(t)} }
.
\end{align}\label{eq:Zchi-formal}\end{subequations}

\paragraph*{Meter charge $\hat{N}$: 1-point moments.}

By construction of the ideal meter model, changes of the reservoir charge $\hat{n}$ are exactly copied to the meter position $\hat{N}$:
the probabilities for measuring the same change $\Delta n = \Delta N$
between times $0$ and $\T$ are thus equal:
\begin{align}
	p^{\Delta N} = P^{\Delta N}
	.
	\label{eq:app-cumulants-same}
\end{align}
Thus, from the time-propagation of the first $r$ moments $\braket{\hat{N}}, \braket{\hat{N}^2}, \ldots, \braket{\hat{N}^r}$ of the meter one should be able to extract the first $r$ moments of interest $\mom^{(1)},\ldots, \mom^{(r)}$ (or the first $r$ cumulants).
Below we establish this recursive relation [\Eq{eq:binom}].
Moreover, when initializing the meter to $\ket{0}\bra{0}$, this simplifies to \Eq{eq:mkNkapp}.
Thus the formulas of \Sec{sec:meter} --based on 1-point meter measurements-- indeed produce the exact 2-point statistics \eq{eq:mkNkapp} of charge transport to the left reservoir. There are three steps:

\noindent(i) Without the adiabatic / steady-state limit the system-meter state has the form \eq{eq:rhoform}.
Tracing out the system, the meter density operator \eq{eq:rhoM}
at time $t=\T$ reads\footnote
	{The negative sign in $Z^{-N}$ indicates that charge $N$
	counted by the meter and entering the reservoir is lost on the system ($-N$).}
$
\Lket{\rho^\M({\T})}
= \sum_{N'} Z^{-N'} \Lket{N'}
$.
Therefore, the charge-diagonal elements $P^N(t) =\bra{N}\rho^\M(t) \ket{N}=\Lbraket{N|\rho^\M(t)}$ evolve as
\begin{align}
	P^{N'}({\T}) = \sum_N Z^{-(N'-N)} P^N(0)
\end{align}
where we restored for clarity a general initial condition, instead of $P^N(0)=\delta_{N,0}$ used in the main text.

\noindent(ii)
The conditional distribution defined by $P^{N'}(\T) = \sum_N P^{N'|N} P^{N}(0)$ is thus translation invariant:
\begin{align}
P^{N'|N}=Z^{-(N'-N)}=P^{\Delta N}\big|_{\Delta N=N'-N}
\label{eq:cond-prob}
.
\end{align}
Then the expectation value of $\hat{N}^k$ at time $\T$ reads
\begin{subequations}
	\begin{gather}
\braket{\hat{N}^k}({\T})
= \sum_{N'} (N')^k P^{N'}
\\
= \sum_{N'N} (N'-N+N)^k P^{N'|N} P^{N}
\\=
\sum_{q=0}^k \tbinom{k}{q}
\sum_{\Delta N} (\Delta N)^{k-q} P^{\Delta N} 
\sum_{N} N^q P^{N}
\\
\overset{\text{\Eq{eq:app-cumulants-same}}}{=}
\sum_{q=0}^k \tbinom{k}{q}
\sum_{\Delta n} (\Delta n)^{k-q} p^{\Delta n} 
\sum_{N} N^q P^{N}
\\=
\sum_{q=0}^k
\tbinom{k}{q}
\, \mom^{(k-q)}
\, 
\braket{\hat{N}^q}(0)
.
\end{gather}\label{eq:binom}\end{subequations}
\noindent(iii)
When in the main text the meter is initialized in a pure state,
$P^N(0)=\braket{N|0}\braket{0|N}=\delta_{N,0}$,
we have $\braket{\hat{N}^q}(0)=\delta_{q,0}$ in \Eq{eq:binom} which gives the result \eq{eq:mkNkapp}.

\paragraph*{Reservoir charge $\hat{n}$: 2-point cumulants.}
Finally, in \Sec{sec:fcs}
we used that the transport process as a whole is characterized either by all cumulants or all moments.
This relies on the explicit relation
\begin{align}
	\cum^{(k)}:= \mom^{(k)} - \sum_{l=1}^{k} \tbinom{k-1}{l-1} \, \cum^{(l)} \mom^{(k-l)}
	\label{eq:cummom}
	,
\end{align}
which is obtained by inserting the series \Eq{eq:mkdef} and
$z^\phi=\sum_{k=0}^\infty \tfrac{1}{k!}\cum^{(k)} (-i\phi)^k$
into
$z^\phi = \ln Z^\phi$.
The relation \eq{eq:cummom} is nonlinear for $k \geq 2$ which complicates the comparison of geometric approaches that compute moments (\ar) on the one hand, and cumulants (\fcs) on the other.
For this reason the present paper focused on comparing $\cum^{(1)}= \mom^{(1)}$.

\section{Recalibrating the reservoir charge $\hat{n}$? \label{app:reservoir-charge}}

Here we point out how gauge transformations enter in formulations
that do not use an open-system (density operator) formulation
and provide some comments.
When expression \eq{eq:Zchi-formal} is used as a starting point,
then the gauge transformations
$G^\phi = e^{g^\phi}=e^{-i\phi \gamma^\phi}$
discussed in the main text correspond to the formal replacements~\cite{Levitov96}
\begin{align}
\hat{n}(t) \to \hat{n}(t) + \gamma^\phi \unit
\quad
\text{or}
\quad
 \hat{I_n}(t) \to   \hat{I_n}(t) + \partial_{t} \gamma^\phi \unit
 \label{eq:formal-gauge}
\end{align}
with $\gamma^\phi(t)=\gamma^\phi[\vec{R}(t)]$
Also, $\gamma^\phi|_{\phi=0}
=-\partial_{i\phi}g^\phi|_{\phi=0} = g^{(1)}$
and $\gamma^\phi=-(\gamma^{-\phi})^{*}$
since $g^\phi|_{\phi=0}=0$ and $g^\phi=(g^{-\phi})^{*}$.

For the $\phi$-independent case $\gamma^\phi=g$
the gauge  transformation \eq{eq:formal-gauge} physically corresponds to parametrically changing the charge operator in time,
$\hat{n}(t) \to \hat{n}(t) + g[\vec{R}(t)] \unit$,
which will cancel out in the change $\hat{n}({\T}) - \hat{n}(0)$.
This \emph{only} captures the gauge freedom of the observable \eq{eq:ar-Nshift},
responsible for geometric pumping of the first moment $\mom^{(1)}$ / cumulant $\cum^{(1)}$ as discussed in \Sec{sec:ar}.
However, the gauge freedom responsible for the geometric higher moments / cumulants discussed in the main text
arises when $\gamma^\phi$ is $\phi$-dependent,
in which case it is less clear what \Eq{eq:formal-gauge} physically means.

Another problem is that one cannot really consider the counting field $\phi$ in \Eq{eq:formal-gauge} as the conjugate to the reservoir charge observable $\hat{n}$.
First, the replacement \eqref{eq:formal-gauge}, would amount to shifting the operator $\hat{n}$ by the eigenvalue of its \emph{noncommuting} conjugate $\hat{\phi}$, which is difficult to understand.
Moreover, it is known in quantum measurement and estimation theory~\cite{Holevobook11}
that due to the lower bound $n=0$ of the spectrum there does not exist \emph{any} phase observable operator\footnote
	{Such a phase does have description as a POVM-measurement (projection-valued operator measure).}
that is conjugate to a Fock-occupation operator (such as $\hat{n}$).

None of these complications arise in the ideal meter model \eq{eq:Hamiltonian} that we used following \Ref{Schaller09}:
the charge operator $\hat{N}$ has a well-defined conjugate observable $\hat{\phi}$
because $\hat{N}$ is the \emph{position}-operator for the meter needle running from $-\infty$ to $\infty$.
It is an excess charge operator that registers discrete \emph{changes} of $\hat{n}$ in the reservoirs,
similar to the situation in, e.g., superconductivity\cite{Nieto68,Carruthers68,Tsui93}
and the $P(E)$-theory of electromagnetic circuits\cite{Grabert91,IngoldNazarov92,Flensberg92}.\section{Measurable geometric phase and parametrically nonstationary state.\label{app:nonstationary}}

In \Sec{sec:ase-review} and \Sec{sec:meter} we found that in open systems the geometric phase is accumulated in the \emph{parametrically nonstationary} part of the quantum \emph{state}
and remarked that the same is true for closed systems.
The latter point easily goes unnoticed when working with state vectors as usual:
from the parametric eigenkets
$H\ket{k_m}= k_{m} \ket{k_m}$
one constructs the adiabatic approximation to the solution of
$\tfrac{d}{dt}\ket{\psi(t)} = -i H(t)\ket{\psi(t)}$,
starting from an arbitrary superposition 
$\ket{\psi(0)} = \sum_m c_m(0)\ket{k_m(0)}$
of parametric eigenstates of $H(0)$.
Decoupling of the eigenspaces gives
$\ket{\psi(t)} = \sum_m e^{z_m(t)} c_m(0)  \ket{k_m(t)}$ 
where \emph{each} term seems to have an imaginary phase
$
z_m(t)=
-i\int_0^\T dt k_m(t)
- \int_0^\T dt \bra{k_m} \tfrac{d}{dt}\ket{k_m(t)}
$.

However, consider the corresponding density operator:
\begin{subequations}
\begin{align}
& 
 \rho(t) = \ket{\psi(t)}\bra{\psi(t)}
 \notag
 \\
&= \sum_{m=1}^d |c_m(t)|^2 \, \ket{k_m(t)} \bra{k_{m}(t)}
\label{eq:rho-purea}
 \\
&
+\sum_{m=1}^d \sum_{m'\neq m} c_m^{*}(t) c_{m'}(t) e^{z_m(t)-z_{m'}(t)} \ket{k_m(t)} \bra{k_{m'}(t)}
\label{eq:rho-pureb}
\end{align}\label{eq:rho-pure}\end{subequations}
In \Eq{eq:rho-purea} all closed-system phases have canceled out.
This is the parametrically stationary part which would be time-independent if the parameters were fixed.
The measurable phases appear only in the parametrically nonstationary part \eq{eq:rho-pureb} of the state.
This expresses that single state-vector geometric phases are not measurable (like any absolute phase in quantum physics).

A second point made by \Eq{eq:rho-pure} is that consideration of measurements is also crucial for understanding geometric phases in closed systems.
To observe a geometric phase one needs some interference measurement~\cite{Chruscinskibook04}.
(i) Temporal: One prepares a superposition of states leading to interference terms \eq{eq:rho-pureb} with observable phase difference.
(ii) Spatial: One adjoins an auxiliary system with simple reference phases and lets the two components interfere.
In the latter case, one verifies that when \emph{actually including} this auxiliary system into the pure state description, the phases of interest appear as observable phase differences in the parametrically \emph{nonstationary} part \eq{eq:rho-pureb} of the composite state.
Whereas for a closed system this is all clear, the main text of the paper discusses this in the more complicated setting of open systems,
cf. also \Ref{Sjoeqvist00}.\section{Fock variational formulation of parallel transport for mixed-state adiabatic evolution\label{app:fock}}

Here we show how the Fock variational condition,
that we generalized to mixed-state evolution in \Eq{eq:fock},
\begin{align}
	\Lbraket{ \tfrac{d}{dt} \bar{\rho} | \tfrac{d}{dt} \rho }
	\quad \text{is stationary}
	\label{eq:fock-fock}
\end{align}
when varied over all operators $\rho$,
is equivalent to the parallel transport condition \Eq{eq:Amc} as it arises from adiabatic evolution.
First, in \Eq{eq:fock-fock} the bar is understood as follows in terms of the left- and right eigenvectors:
\begin{align}
	\Lbra{\bar{\rho}}
	:= \sum_m \C_m^{-1} \Lbra{\bar{k}_m}
	,\qquad
	\Lket{\rho}
	:= \sum_m \C_m \Lket{k_m}
	,
\end{align}
The inverses of the coefficients $\C_m$ in the dual vector $\Lbra{\bar{\rho}}$ guarantee that in the innerproduct
\begin{align}
	\Lbraket{\bar{\rho}|\rho}
	=\sum_m \Lbraket{\bar{k}_m|k_m} = \sum_m 1
	\label{eq:fock-constant}
\end{align}
\emph{each term} is time-independent: $\Lbraket{\bar{k}_m|k_m}=1$.
Functional variation of the expression \eq{eq:fock-fock} with respect to the coefficients is effected by replacing
$\C_m \to \C_m e^{\epsilon g_m}$,
maintaining the constancy of each term \Eq{eq:fock-constant},
expanding to $\text{O}(\epsilon)$,
and noting that the gauge exponent $g_m(\vec{R}(t))$ is a parametric function of time.
Using
\begin{subequations}
\begin{align}
\C_m^{-1}\Lbraket{\bar{k}_m| \tfrac{d}{dt} \rho}
&=
\C_m^{-1}\Lbraket{\bar{k}_m}  \tfrac{d}{dt} \big[ \Lket{ k_m}\C_m \big]
\\
\Lbraket{\tfrac{d}{dt} \bar{\rho} | k_m} \C_m
&=
\big[ \tfrac{d}{dt} \C_m^{-1} \Lbra{ \bar{k}_m } \big] \, \Lket{k_m} \C_m
\\
\tfrac{d}{dt} [ \C_m^{-1} \Lbraket{ \bar{k}_m | k_m} \C_m ]
&=\tfrac{d}{dt} \Lbraket{ \bar{k}_m | k_m} =0
\end{align}\end{subequations}
we obtain to O($\epsilon$) the stationarity condition
\begin{subequations}
	\begin{align}
	&
	{}_g{}\Lbraket{\tfrac{d}{dt}\bar{\rho}|\tfrac{d}{dt}\rho}{}_{g}
	-
	\Lbraket{\tfrac{d}{dt}\bar{\rho}|\tfrac{d}{dt}\rho}
	\\&
	=
	\epsilon 
	\sum_m
	\Big[
	-\tfrac{d g_m}{dt} \C_m^{-1}
	\Lbraket{\bar{k}_m| \tfrac{d}{dt} \rho}
	+\Lbraket{\tfrac{d}{dt} \bar{\rho} | k_m} \C_m
	\tfrac{d g_m}{dt}
	\Big]
	\\&
	=
	2 \epsilon 
	\sum_m
	\tfrac{d g_m}{dt}
	\C_m^{-1}\Lbra{\bar{k}_m} \tfrac{d}{dt} \C_m \Lket{k_m}
	\\&
	=
	2 \epsilon 
	\sum_m
	\tfrac{d g_m}{dt} A_{m,\C_m} \tfrac{d}{dt}\vec{R}
	=
	0
\end{align}\end{subequations}
Because the variations $g_m$ are independent
we have for \emph{each} mode  $m$ separately
\begin{align}
	A_{m,\C_m}
	=\Lbra{\bar{k}_m} \delR \Lket{k_m}
	+\C_m^{-1} \delR \C_m
	=0,
\end{align}
This shows \Eq{eq:fock} and \Eq{eq:Amc} are equivalent.
\section{Steady-state limit\label{app:steady}}

In this appendix we discuss the transition to the steady state in
\Eqs{eq:densitySolution0} and \eq{eq:densitySolution}.
Our expressions in Secs.~\sec{sec:meter}-\sec{sec:fcs} are consistent with those of \Ref{Sinitsyn07EPL}
and we comment on a different expression obtained in \Ref{Nakajima15}.
To obtain \Eq{eq:densitySolution},
\begin{subequations}
	\begin{align}
	\Lket{\rho(t)} &\approx
	\sum_{\p=0,1,\ldots}
	\int_{-\pi}^{\pi} \frac{d\phi}{2\pi}
	e^{z_\p^\phi(t)}
	\,
	\C_\p^\phi(0)
	\, 
	\Lket{w^\phi_\p(t)} \otimes \Lket{\phi}
	\label{eq:app-steady-decoupling}
	\\
	&
	\approx
	\int_{-\pi}^{\pi} \frac{d\phi}{2\pi}
	\,
	e^{z^\phi_0(t)}
	\C_0^\phi(0)
	\, 
	\Lket{w^\phi_0(t)} \otimes \Lket{\phi}
	\label{eq:app-steady-limit}
	\end{align}\end{subequations}
we have \emph{set} $\C_0^\phi(0)=1$ and identified $Z^\phi:=e^{z_0^\phi(\T)}$ after one driving period.
Since the adiabatic decoupling has already been performed in \Eq{eq:app-steady-decoupling},
the step \eq{eq:app-steady-limit} only concerns the steady-state limit.
This can be understood clearly by considering measurable quantities:
since we consider slow driving, $\T \gg \Gamma^{-1} \propto \dot{R}$,
the cumulant current \eq{eq:fcs-Ik} reaches a time-dependent steady-state already at the beginning of the first driving period, at times $t \ll \T$
and the charge-transfer statistics becomes steady.
As explained after \Eq{eq:fcs-Ik} the steady-state cumulant currents $I_{\cum}^{(k)}$ are functions of both $\vec{R}$ and $\dot{\vec{R}}$
and determine the cumulant-generating current $I^\phi:=\tfrac{d}{dt} z^\phi=\sum_k \tfrac{1}{k!} I_{\cum}^{(k)}(-i\phi)^k$ [\Eq{eq:cumulants}]. Thus
\begin{align}
	z^\phi(\T) -z^\phi(0)
	& =
	\int_0^{\T} dt \ I^{\phi}(\vec{R}(t), \dot{\vec{R}}(t))
	\\
	& =
	\int_0^{\T} dt \ \left( I^{\phi}[\vec{R}(t)]	
	-
	\dot{\vec{R}}(t)) A^{\phi}[\vec{R}(t)] \right)
\end{align}
is periodic: $z^\phi(n\T)=n\cdot z^\phi(\T)$ for integer $n$
since the first term, the sum of ``snapshots'', takes the same ``snapshots`` every period and the second term is geometric and depends on the traversed parameter curve, which is also the same for every period. Thus, in \Eq{eq:app-steady-limit}
$
	e^{z^\phi_0(n\T)} \C_0^\phi(0)
	=
	\big( e^{z^\phi_0(\T)}\big)^n \C_0^\phi(0)	
$
showing that $Z^\phi= e^{z^\phi_0(\T)}$ is the steady-state generating function for one-period \fcs.
It is extracted formally by setting  $\C_0^\phi=1$
as we did in the main text after which we denoted $z^\phi:=z_0^\phi$.

\paragraph*{Extra contribution in \Ref{Nakajima15}.}
Notably, \Ref{Nakajima15} does not set $\C_0^\phi=1$ and obtains a modified \fcs phase $-\oint_{\curve} A^\phi + \ln \C_0^\phi$
relative to \Ref{Sinitsyn07EPL} and the present paper.
The authors of \Ref{Nakajima15} rationalize this difference by noting that the extra term does not contribute to the first cumulant $\cum^{(1)}$ (pumped charge).
Although this last observation is correct\footnote
	{The extra term in \Ref{Nakajima15}
	$\ln \C_0^\phi$ in the geometric phase $z^\phi$
	in principle gives an extra contribution to
	$\mom^{(1)} = \cum^{(1)} =
	 -\partial_{i\phi}
	\exp({ z^\phi })
	|_{\phi=0}
	$
	equal to
	$
	-\partial_{i\phi}
	\C^\phi_0
	|_{\phi=0}
	=0
	$
	calculated as follows.
	We already assumed
	$\Lket{\rho(0)}=\Lket{ \rho^\S(0) } \otimes \Lket{ \rho^\M(0) }$
	and initialized the meter in the pure state $\rho^\M(0)=\Lket{0}=\ket{0}\bra{0}$
	and we now assume also that the system is initially stationary, $\Lket{\rho^\S(0)}=\Lket{w_0}$.
	This gives
	$\C^\phi_0 =
	\Lbraket{ \bar{k}_{0,\phi} | \rho(0) } 
		= \Lbraket{\bar{w}^\phi_0|w_0}
	$
	by \Eq{eq:eigencovector} for $l=0$ and using $\Lbraket{\phi|0}=1$.
	Using Eq. (126) of \Ref{Pluecker17a}, we obtain
	$\C^\phi_0
	=\Lbraket{\unit|w_0}
	-i \phi \Lbra{\unit}W_{I_N}W^{-1}\Lket{w_0} = 1+\text{O}(\phi^2)$
	since the pseudo-inverse $W^{-1}$ projects out $\Lket{w_0}$.
	This gives $
	-\partial_{i\phi}
	\C^\phi_0
	|_{\phi=0}=0
	$.}
provided one assumes that $\rho(0)$ is initialized as the parametric stationary state $\Lket{w_0}$,
this does not take away the fact that higher moments really do depend on the initial condition through $c^\phi_0$:
the modified result does \emph{not} apply to the \emph{steady-state} \fcs.
For the above reason this did not lead to any inconsistencies in \Ref{Nakajima15} when their \fcs result was compared with the \emph{steady-state} \ar result for the first moment only.
However, when comparing with other \fcs works, \Ref{Pluecker17a} or the present paper one should ignore the extra term in \Ref{Nakajima15}.
In the present paper we consistently treat each of the compared approaches (\ase, \fcs, and \ar) in the steady-state limit.\section{Bochner's criterion and gauge~transformations\label{app:bochner}}

\paragraph*{Bochner's criterion.}
Bochner's criterion\footnote
	{\Eq{eq:app-bochner-criterion} is actually Herglotz result, a special case of Bochner's more general criterion.}
states that the inverse Fourier transform $G^N$ of a smooth function $G^\phi$ of $\phi \in [-\pi,\pi]$ is a positive (semidefinite) function, $G^N \geq 0$ for all $N \in \mathbb{Z}$, 
if and only if the associated quadratic form
\begin{align}
	\sum_{k k'} v^{*}_k \,  G^{\phi_k-\phi_{k'}} \, v_{k'} \geq 0
	\label{eq:app-bochner-criterion}
\end{align}
is positive (semidefinite) for an arbitrary finite complex vector with components $v_k$
and a corresponding arbitrary discrete sampling of Fourier frequencies $\phi_k \in [-\pi,\pi]$.
The ``if'' part is easily verified
and a constructive proof of the converse can be found in \Ref{lukacsBook}.

\paragraph*{Examples of \Fig{fig:gauges}(c).}
The criterion \eq{eq:app-bochner-criterion} is easily verified for  phase factors  $G^\phi=e^{i N \phi}$ with a fixed discrete $N \in \mathbb{Z}$.
It can easily be made to fail as illustrated for finite Fourier sums $G^\phi = \sum_N G^N e^{iN\phi}$ in \Fig{fig:gauges}(c) of the main text.
These examples indicate that the splitting of a function into a convolution of two positive ones is rather rare unless it involves transformations close to shifts.
For the figure we constructed the target distribution $Z_\g^\phi$ as a product $Z_\g^\phi := G^\phi_\text{(ii)} S^\phi_\text{(ii)}$ of two positive functions
\begin{align}
	G^\phi_\text{(ii)} &= \tfrac{1}{6}+\tfrac{2}{3} e^{i \phi}+\tfrac{1}{6} e^{2 i \phi} \\
	S^\phi_\text{(ii)} &= \tfrac{1}{10}+\tfrac{2}{10} e^{i \phi}+\tfrac{4}{10} e^{2 i \phi}+\tfrac{2}{10} e^{3 i \phi}+\tfrac{1}{10} e^{4 i \phi}
	.
\end{align}
We also split the target function in another way, $Z_\g^\phi = G^\phi_\text{(iii)} S^\phi_\text{(iii)}$ using a different, still positive gauge function
\begin{align}
  G^\phi_\text{(iii)} =& \frac{1}{3.8} \left ( 0.1 e^{-3 i \phi}+0.2 e^{-2 i \phi}+0.4 e^{-1 i \phi}+0.7 e^{0 i \phi} \right .
	\notag \\
	+&\left . 1 e^{1 i \phi}+0.7 e^{2 i \phi}+0.4 e^{3 i \phi}+0.2 e^{4 i \phi}+0.1 e^{5 i \phi} \right )
	.
\end{align}
whose width exceeds the width of the target distribution $Z_\g^N$.
Considering the convolution graphically in \Fig{fig:gauges}(c) one sees that negative coefficients in $S^N_\text{(iii)}$ are needed to narrow down the chosen $G_\text{(iii)}^N$ to match the target $Z_\g^N$.
The final example of a negative gauge and a resulting negative solution $S^\phi_\text{(iv)} = Z_\g^\phi / G^\phi_\text{(iv)}$ is
\begin{align}
	G^\phi_\text{(iv)}= \tfrac{1}{2} e^{0 \phi}-1 e^{1 i \phi}+2 e^{2 i \phi}-1 e^{3 i \phi}+\tfrac{1}{2} e^{4 i \phi}
	.
\end{align}
These examples indicate given an arbitrary positive gauge $G^N$, the corresponding solution $S^N$ may require negative values
to adjust the gauge distribution (e.g., its width) to the target $Z^N_\g$.
\section{Calibrating meter~observables \newline and meter~states\label{app:observable-gauge}}

Here we explain the point made at the end of \Sec{sec:ar} that for the first moment $\mom^{(1)}$ the gauge transformation of the observable $\hat{N}$ can always be understood physically as arising from \emph{some} recalibration of the meter states $\Lket{N}=\ket{N}\bra{N}$ to mixed states.
As shown there,
a \emph{meter-gauge} $G^\phi=\sum_N G^N e^{i\phi N}$ determines
a real-valued \emph{observable-gauge} $\hat{N} \to \hat{N} + g$
through 
$
-\partial_{i\phi} G^\phi|_{\phi=0}=\sum_N N G^N = g
$.
Here, the geometric gauge freedom --unconstrained by physics-- allows $G^\phi$ to be any function
(not necessarily with positive Fourier transform $G^N$.)
We now show conversely that for a given observable gauge $g$, one can find many state gauges $G^N$ that produce this value since their differences cancel out in $\sum_N N G^N=g$.
Moreover, these functions always include positive ones which correspond to a physical mixing of the meter states [\Sec{sec:bochner}].
One simple construction of a class of positive meter gauges is
\begin{align}
	{G}^{N}=
	(1-|\tfrac{g}{k}|) \, \delta_{N,0}
	+ |\tfrac{g}{k}| \, \delta_{N,k}
	.
	\label{eq:example1}
\end{align}
The weighted sum $\sum_N N G^{N}$ achieves the value $g$ by a single contribution at integer value $k$ with $\text{sign} \ k = \text{sign} \ g$.
Both nonzero values of $G^N$ are positive if $k$ is taken sufficiently large ($|k| \geq g$) but other than this $k$ is arbitrary.
When $g$ is parameter dependent, $k$ can always be chosen large enough such that $G^N \geq 0$ for all $\vec{R}$ values accessed during the driving cycle.
One can thus always consider the observable gauge $g$ as arising from some physical mixing of meter states during a pumping cycle.

Whereas example \eq{eq:example1} is only piecewise differentiable for $|g| >0$ (which is sufficient) one can also construct a differentiable gauge:
\begin{align}
G^N =
\tfrac{1}{2}(1-\tfrac{g}{k}) \, \delta_{N,-k}
+
\tfrac{1}{2}(1+\tfrac{g}{k}) \, \delta_{N,k}
\end{align}
Now the weighted sum achieves the given value of $g$  by two unequal contributions with positive weights if $|k| \geq g$.
This has the side effect that the trivial observable gauge $g=0$ corresponds to a nontrivial (``noisy'') meter-state $G^{N}=
(\delta_{N,-k}
+
\delta_{N,k})/2$,
which is fine if one only considers the first moment.


\begin{thebibliography}{83}
\expandafter\ifx\csname natexlab\endcsname\relax\def\natexlab#1{#1}\fi
\expandafter\ifx\csname bibnamefont\endcsname\relax
  \def\bibnamefont#1{#1}\fi
\expandafter\ifx\csname bibfnamefont\endcsname\relax
  \def\bibfnamefont#1{#1}\fi
\expandafter\ifx\csname citenamefont\endcsname\relax
  \def\citenamefont#1{#1}\fi
\expandafter\ifx\csname url\endcsname\relax
  \def\url#1{\texttt{#1}}\fi
\expandafter\ifx\csname urlprefix\endcsname\relax\def\urlprefix{URL }\fi
\providecommand{\bibinfo}[2]{#2}
\providecommand{\eprint}[2][]{\url{#2}}

\bibitem[{\citenamefont{Altland and Zirnbauer}(1997)}]{Altland97}
\bibinfo{author}{\bibfnamefont{A.}~\bibnamefont{Altland}} \bibnamefont{and}
  \bibinfo{author}{\bibfnamefont{M.~R.} \bibnamefont{Zirnbauer}},
  \bibinfo{journal}{Phys. Rev. B} \textbf{\bibinfo{volume}{55}},
  \bibinfo{pages}{1142} (\bibinfo{year}{1997}).

\bibitem[{\citenamefont{Read and Green}(2000)}]{Read00}
\bibinfo{author}{\bibfnamefont{N.}~\bibnamefont{Read}} \bibnamefont{and}
  \bibinfo{author}{\bibfnamefont{D.}~\bibnamefont{Green}},
  \bibinfo{journal}{Phys. Rev. B} \textbf{\bibinfo{volume}{61}},
  \bibinfo{pages}{10267} (\bibinfo{year}{2000}).

\bibitem[{\citenamefont{Snyder and Toberer}(2008)}]{Snyder08}
\bibinfo{author}{\bibfnamefont{G.~J.} \bibnamefont{Snyder}} \bibnamefont{and}
  \bibinfo{author}{\bibfnamefont{E.~S.} \bibnamefont{Toberer}},
  \bibinfo{journal}{Nat. Mater.} \textbf{\bibinfo{volume}{7}},
  \bibinfo{pages}{105} (\bibinfo{year}{2008}).

\bibitem[{\citenamefont{Nayak et~al.}(2008)\citenamefont{Nayak, Simon, Stern,
  Freedman, and Das~Sarma}}]{Nayak08}
\bibinfo{author}{\bibfnamefont{C.}~\bibnamefont{Nayak}},
  \bibinfo{author}{\bibfnamefont{S.~H.} \bibnamefont{Simon}},
  \bibinfo{author}{\bibfnamefont{A.}~\bibnamefont{Stern}},
  \bibinfo{author}{\bibfnamefont{M.}~\bibnamefont{Freedman}}, \bibnamefont{and}
  \bibinfo{author}{\bibfnamefont{S.}~\bibnamefont{Das~Sarma}},
  \bibinfo{journal}{Rev. Mod. Phys.} \textbf{\bibinfo{volume}{80}},
  \bibinfo{pages}{1083} (\bibinfo{year}{2008}).

\bibitem[{\citenamefont{Kitaev}(2009)}]{Kitaev09Conference}
\bibinfo{author}{\bibfnamefont{A.}~\bibnamefont{Kitaev}}, \bibinfo{journal}{AIP
  Conf. Proc.} \textbf{\bibinfo{volume}{1134}}, \bibinfo{pages}{22}
  (\bibinfo{year}{2009}).

\bibitem[{\citenamefont{Ryu et~al.}(2010)\citenamefont{Ryu, Schnyder, Furusaki,
  and Ludwig}}]{Ryu10}
\bibinfo{author}{\bibfnamefont{S.}~\bibnamefont{Ryu}},
  \bibinfo{author}{\bibfnamefont{A.}~\bibnamefont{Schnyder}},
  \bibinfo{author}{\bibfnamefont{A.}~\bibnamefont{Furusaki}}, \bibnamefont{and}
  \bibinfo{author}{\bibfnamefont{A.}~\bibnamefont{Ludwig}},
  \bibinfo{journal}{New. J. Phys.} \textbf{\bibinfo{volume}{12}},
  \bibinfo{pages}{065010} (\bibinfo{year}{2010}).

\bibitem[{\citenamefont{Hasan and Kane}(2010)}]{Hasan10}
\bibinfo{author}{\bibfnamefont{M.~Z.} \bibnamefont{Hasan}} \bibnamefont{and}
  \bibinfo{author}{\bibfnamefont{C.~L.} \bibnamefont{Kane}},
  \bibinfo{journal}{Rev. Mod. Phys.} \textbf{\bibinfo{volume}{82}},
  \bibinfo{pages}{3045} (\bibinfo{year}{2010}).

\bibitem[{\citenamefont{Qi and Zhang}(2011)}]{Qi11}
\bibinfo{author}{\bibfnamefont{X.-L.} \bibnamefont{Qi}} \bibnamefont{and}
  \bibinfo{author}{\bibfnamefont{S.-C.} \bibnamefont{Zhang}},
  \bibinfo{journal}{Rev. Mod. Phys.} \textbf{\bibinfo{volume}{83}},
  \bibinfo{pages}{1057} (\bibinfo{year}{2011}).

\bibitem[{\citenamefont{Budich et~al.}(2015)\citenamefont{Budich, Zoller, and
  Diehl}}]{Budich15a}
\bibinfo{author}{\bibfnamefont{J.~C.} \bibnamefont{Budich}},
  \bibinfo{author}{\bibfnamefont{P.}~\bibnamefont{Zoller}}, \bibnamefont{and}
  \bibinfo{author}{\bibfnamefont{S.}~\bibnamefont{Diehl}},
  \bibinfo{journal}{Phys. Rev. A} \textbf{\bibinfo{volume}{91}},
  \bibinfo{pages}{042117} (\bibinfo{year}{2015}).

\bibitem[{\citenamefont{Chiu et~al.}(2016)\citenamefont{Chiu, Teo, Schnyder,
  and Ryu}}]{Chiu15}
\bibinfo{author}{\bibfnamefont{C.-K.} \bibnamefont{Chiu}},
  \bibinfo{author}{\bibfnamefont{J.~C.~Y.} \bibnamefont{Teo}},
  \bibinfo{author}{\bibfnamefont{A.~P.} \bibnamefont{Schnyder}},
  \bibnamefont{and} \bibinfo{author}{\bibfnamefont{S.}~\bibnamefont{Ryu}},
  \bibinfo{journal}{Rev. Mod. Phys.} \textbf{\bibinfo{volume}{88}},
  \bibinfo{pages}{035005} (\bibinfo{year}{2016}).

\bibitem[{\citenamefont{Kennedy and Zirnbauer}(2016)}]{Kennedy16}
\bibinfo{author}{\bibfnamefont{R.}~\bibnamefont{Kennedy}} \bibnamefont{and}
  \bibinfo{author}{\bibfnamefont{M.~R.} \bibnamefont{Zirnbauer}},
  \bibinfo{journal}{Commun. Math. Phys.} \textbf{\bibinfo{volume}{342}},
  \bibinfo{pages}{909} (\bibinfo{year}{2016}).

\bibitem[{\citenamefont{Diehl et~al.}(2011)\citenamefont{Diehl, Rico, Baranov,
  and Zoller}}]{Diehl11a}
\bibinfo{author}{\bibfnamefont{S.}~\bibnamefont{Diehl}},
  \bibinfo{author}{\bibfnamefont{E.}~\bibnamefont{Rico}},
  \bibinfo{author}{\bibfnamefont{M.~A.} \bibnamefont{Baranov}},
  \bibnamefont{and} \bibinfo{author}{\bibfnamefont{P.}~\bibnamefont{Zoller}},
  \bibinfo{journal}{Nature Phys.} \textbf{\bibinfo{volume}{7}},
  \bibinfo{pages}{971} (\bibinfo{year}{2011}).

\bibitem[{\citenamefont{Bardyn et~al.}(2013)\citenamefont{Bardyn, Baranov,
  Kraus, Rico, İmamoğlu, Zoller, and Diehl}}]{Bardyn13}
\bibinfo{author}{\bibfnamefont{C.-E.} \bibnamefont{Bardyn}},
  \bibinfo{author}{\bibfnamefont{M.~A.} \bibnamefont{Baranov}},
  \bibinfo{author}{\bibfnamefont{C.~V.} \bibnamefont{Kraus}},
  \bibinfo{author}{\bibfnamefont{E.}~\bibnamefont{Rico}},
  \bibinfo{author}{\bibfnamefont{A.}~\bibnamefont{İmamoğlu}},
  \bibinfo{author}{\bibfnamefont{P.}~\bibnamefont{Zoller}}, \bibnamefont{and}
  \bibinfo{author}{\bibfnamefont{S.}~\bibnamefont{Diehl}},
  \bibinfo{journal}{New. J. Phys.} \textbf{\bibinfo{volume}{15}},
  \bibinfo{pages}{085001} (\bibinfo{year}{2013}).

\bibitem[{\citenamefont{Iemini et~al.}(2016)\citenamefont{Iemini, Rossini,
  Fazio, Diehl, and Mazza}}]{Iemini16}
\bibinfo{author}{\bibfnamefont{F.}~\bibnamefont{Iemini}},
  \bibinfo{author}{\bibfnamefont{D.}~\bibnamefont{Rossini}},
  \bibinfo{author}{\bibfnamefont{R.}~\bibnamefont{Fazio}},
  \bibinfo{author}{\bibfnamefont{S.}~\bibnamefont{Diehl}}, \bibnamefont{and}
  \bibinfo{author}{\bibfnamefont{L.}~\bibnamefont{Mazza}},
  \bibinfo{journal}{Phys. Rev. B} \textbf{\bibinfo{volume}{93}},
  \bibinfo{pages}{115113} (\bibinfo{year}{2016}).

\bibitem[{\citenamefont{Huang and Arovas}(2014)}]{Huang14}
\bibinfo{author}{\bibfnamefont{Z.}~\bibnamefont{Huang}} \bibnamefont{and}
  \bibinfo{author}{\bibfnamefont{D.~P.} \bibnamefont{Arovas}},
  \bibinfo{journal}{Phys. Rev. Lett.} \textbf{\bibinfo{volume}{113}},
  \bibinfo{pages}{076407} (\bibinfo{year}{2014}).

\bibitem[{\citenamefont{Viyuela et~al.}(2015)\citenamefont{Viyuela, Rivas, and
  Martin-Delgado}}]{Viyuela15}
\bibinfo{author}{\bibfnamefont{O.}~\bibnamefont{Viyuela}},
  \bibinfo{author}{\bibfnamefont{A.}~\bibnamefont{Rivas}}, \bibnamefont{and}
  \bibinfo{author}{\bibfnamefont{M.~A.} \bibnamefont{Martin-Delgado}},
  \bibinfo{journal}{2D Mater.} \textbf{\bibinfo{volume}{2}},
  \bibinfo{pages}{034006} (\bibinfo{year}{2015}).

\bibitem[{\citenamefont{Budich and Diehl}(2015)}]{Budich15b}
\bibinfo{author}{\bibfnamefont{J.~C.} \bibnamefont{Budich}} \bibnamefont{and}
  \bibinfo{author}{\bibfnamefont{S.}~\bibnamefont{Diehl}},
  \bibinfo{journal}{Phys. Rev. B} \textbf{\bibinfo{volume}{91}},
  \bibinfo{pages}{165140} (\bibinfo{year}{2015}).

\bibitem[{\citenamefont{Uhlmann}(1986)}]{Uhlmann86}
\bibinfo{author}{\bibfnamefont{A.}~\bibnamefont{Uhlmann}},
  \bibinfo{journal}{Rep. Math. Phys.} \textbf{\bibinfo{volume}{24}},
  \bibinfo{pages}{229} (\bibinfo{year}{1986}).

\bibitem[{\citenamefont{Sj\"oqvist et~al.}(2000)\citenamefont{Sj\"oqvist, Pati,
  Ekert, Anandan, Ericsson, Oi, and Vedral}}]{Sjoeqvist00}
\bibinfo{author}{\bibfnamefont{E.}~\bibnamefont{Sj\"oqvist}},
  \bibinfo{author}{\bibfnamefont{A.~K.} \bibnamefont{Pati}},
  \bibinfo{author}{\bibfnamefont{A.}~\bibnamefont{Ekert}},
  \bibinfo{author}{\bibfnamefont{J.~S.} \bibnamefont{Anandan}},
  \bibinfo{author}{\bibfnamefont{M.}~\bibnamefont{Ericsson}},
  \bibinfo{author}{\bibfnamefont{D.~K.~L.} \bibnamefont{Oi}}, \bibnamefont{and}
  \bibinfo{author}{\bibfnamefont{V.}~\bibnamefont{Vedral}},
  \bibinfo{journal}{Phys. Rev. Lett.} \textbf{\bibinfo{volume}{85}},
  \bibinfo{pages}{2845} (\bibinfo{year}{2000}).

\bibitem[{\citenamefont{Sarandy and Lidar}(2005)}]{Sarandy05}
\bibinfo{author}{\bibfnamefont{M.~S.} \bibnamefont{Sarandy}} \bibnamefont{and}
  \bibinfo{author}{\bibfnamefont{D.~A.} \bibnamefont{Lidar}},
  \bibinfo{journal}{Phys. Rev. A} \textbf{\bibinfo{volume}{71}},
  \bibinfo{pages}{012331} (\bibinfo{year}{2005}).

\bibitem[{\citenamefont{Sarandy and Lidar}(2006)}]{Sarandy06}
\bibinfo{author}{\bibfnamefont{M.~S.} \bibnamefont{Sarandy}} \bibnamefont{and}
  \bibinfo{author}{\bibfnamefont{D.~A.} \bibnamefont{Lidar}},
  \bibinfo{journal}{Phys. Rev. A} \textbf{\bibinfo{volume}{73}},
  \bibinfo{pages}{062101} (\bibinfo{year}{2006}).

\bibitem[{\citenamefont{Nakajima et~al.}(2015)\citenamefont{Nakajima, Taguchi,
  Kubo, and Tokura}}]{Nakajima15}
\bibinfo{author}{\bibfnamefont{S.}~\bibnamefont{Nakajima}},
  \bibinfo{author}{\bibfnamefont{M.}~\bibnamefont{Taguchi}},
  \bibinfo{author}{\bibfnamefont{T.}~\bibnamefont{Kubo}}, \bibnamefont{and}
  \bibinfo{author}{\bibfnamefont{Y.}~\bibnamefont{Tokura}},
  \bibinfo{journal}{Phys. Rev. B} \textbf{\bibinfo{volume}{92}},
  \bibinfo{pages}{195420} (\bibinfo{year}{2015}).

\bibitem[{\citenamefont{Pluecker et~al.}(2017)\citenamefont{Pluecker, Wegewijs,
  and Splettstoesser}}]{Pluecker17a}
\bibinfo{author}{\bibfnamefont{T.}~\bibnamefont{Pluecker}},
  \bibinfo{author}{\bibfnamefont{M.~R.} \bibnamefont{Wegewijs}},
  \bibnamefont{and}
  \bibinfo{author}{\bibfnamefont{J.}~\bibnamefont{Splettstoesser}},
  \bibinfo{journal}{Phys. Rev. B} \textbf{\bibinfo{volume}{95}},
  \bibinfo{pages}{155431} (\bibinfo{year}{2017}).

\bibitem[{\citenamefont{Bardyn et~al.}(2017)\citenamefont{Bardyn, Wawer,
  Altland, Fleischhauer, and Diehl}}]{Bardyn17a}
\bibinfo{author}{\bibfnamefont{C.~E.} \bibnamefont{Bardyn}},
  \bibinfo{author}{\bibfnamefont{L.}~\bibnamefont{Wawer}},
  \bibinfo{author}{\bibfnamefont{A.}~\bibnamefont{Altland}},
  \bibinfo{author}{\bibfnamefont{M.}~\bibnamefont{Fleischhauer}},
  \bibnamefont{and} \bibinfo{author}{\bibfnamefont{S.}~\bibnamefont{Diehl}}
  (\bibinfo{year}{2017}), \bibinfo{note}{arXiv:1706.02741}.

\bibitem[{\citenamefont{Splettstoesser
  et~al.}(2006)\citenamefont{Splettstoesser, Governale, K\"{o}nig, and
  Fazio}}]{Splettstoesser06}
\bibinfo{author}{\bibfnamefont{J.}~\bibnamefont{Splettstoesser}},
  \bibinfo{author}{\bibfnamefont{M.}~\bibnamefont{Governale}},
  \bibinfo{author}{\bibfnamefont{J.}~\bibnamefont{K\"{o}nig}},
  \bibnamefont{and} \bibinfo{author}{\bibfnamefont{R.}~\bibnamefont{Fazio}},
  \bibinfo{journal}{Phys. Rev. B} \textbf{\bibinfo{volume}{74}},
  \bibinfo{pages}{085305} (\bibinfo{year}{2006}).

\bibitem[{\citenamefont{Calvo et~al.}(2012)\citenamefont{Calvo, Classen,
  Splettstoesser, and Wegewijs}}]{Calvo12a}
\bibinfo{author}{\bibfnamefont{H.~L.} \bibnamefont{Calvo}},
  \bibinfo{author}{\bibfnamefont{L.}~\bibnamefont{Classen}},
  \bibinfo{author}{\bibfnamefont{J.}~\bibnamefont{Splettstoesser}},
  \bibnamefont{and} \bibinfo{author}{\bibfnamefont{M.~R.}
  \bibnamefont{Wegewijs}}, \bibinfo{journal}{Phys. Rev. B}
  \textbf{\bibinfo{volume}{86}}, \bibinfo{pages}{245308}
  (\bibinfo{year}{2012}).

\bibitem[{\citenamefont{Avron et~al.}(2011)\citenamefont{Avron, Fraas, Graf,
  and Kenneth}}]{Avron11}
\bibinfo{author}{\bibfnamefont{J.}~\bibnamefont{Avron}},
  \bibinfo{author}{\bibfnamefont{M.}~\bibnamefont{Fraas}},
  \bibinfo{author}{\bibfnamefont{G.}~\bibnamefont{Graf}}, \bibnamefont{and}
  \bibinfo{author}{\bibfnamefont{O.}~\bibnamefont{Kenneth}},
  \bibinfo{journal}{New. J. Phys.} \textbf{\bibinfo{volume}{13}},
  \bibinfo{pages}{053042} (\bibinfo{year}{2011}).

\bibitem[{\citenamefont{Sinitsyn and Nemenman}(2007)}]{Sinitsyn07EPL}
\bibinfo{author}{\bibfnamefont{N.}~\bibnamefont{Sinitsyn}} \bibnamefont{and}
  \bibinfo{author}{\bibfnamefont{I.}~\bibnamefont{Nemenman}},
  \bibinfo{journal}{Eur. Phys. Lett.} \textbf{\bibinfo{volume}{77}},
  \bibinfo{pages}{58001} (\bibinfo{year}{2007}).

\bibitem[{\citenamefont{Thouless}(1983)}]{Thouless83}
\bibinfo{author}{\bibfnamefont{D.~J.} \bibnamefont{Thouless}},
  \bibinfo{journal}{Phys. Rev. B} \textbf{\bibinfo{volume}{27}},
  \bibinfo{pages}{6083} (\bibinfo{year}{1983}).

\bibitem[{\citenamefont{Cohen}(2003)}]{Cohen03}
\bibinfo{author}{\bibfnamefont{D.}~\bibnamefont{Cohen}},
  \bibinfo{journal}{Phys. Rev. B} \textbf{\bibinfo{volume}{68}},
  \bibinfo{pages}{201303} (\bibinfo{year}{2003}).

\bibitem[{\citenamefont{Splettstoesser
  et~al.}(2008)\citenamefont{Splettstoesser, Governale, and
  K\"onig}}]{Splettstoesser08a}
\bibinfo{author}{\bibfnamefont{J.}~\bibnamefont{Splettstoesser}},
  \bibinfo{author}{\bibfnamefont{M.}~\bibnamefont{Governale}},
  \bibnamefont{and} \bibinfo{author}{\bibfnamefont{J.}~\bibnamefont{K\"onig}},
  \bibinfo{journal}{Phys. Rev. B} \textbf{\bibinfo{volume}{77}},
  \bibinfo{pages}{195320} (\bibinfo{year}{2008}).

\bibitem[{\citenamefont{Winkler et~al.}(2009)\citenamefont{Winkler, Governale,
  and K\"onig}}]{Winkler09}
\bibinfo{author}{\bibfnamefont{N.}~\bibnamefont{Winkler}},
  \bibinfo{author}{\bibfnamefont{M.}~\bibnamefont{Governale}},
  \bibnamefont{and} \bibinfo{author}{\bibfnamefont{J.}~\bibnamefont{K\"onig}},
  \bibinfo{journal}{Phys. Rev. B} \textbf{\bibinfo{volume}{79}},
  \bibinfo{pages}{235309} (\bibinfo{year}{2009}).

\bibitem[{\citenamefont{Reckermann et~al.}(2010)\citenamefont{Reckermann,
  Splettstoesser, and Wegewijs}}]{Reckermann10a}
\bibinfo{author}{\bibfnamefont{F.}~\bibnamefont{Reckermann}},
  \bibinfo{author}{\bibfnamefont{J.}~\bibnamefont{Splettstoesser}},
  \bibnamefont{and} \bibinfo{author}{\bibfnamefont{M.~R.}
  \bibnamefont{Wegewijs}}, \bibinfo{journal}{Phys. Rev. Lett.}
  \textbf{\bibinfo{volume}{104}}, \bibinfo{pages}{226803}
  (\bibinfo{year}{2010}).

\bibitem[{\citenamefont{Avron et~al.}(2012)\citenamefont{Avron, Fraas, and
  Graf}}]{Avron12}
\bibinfo{author}{\bibfnamefont{J.}~\bibnamefont{Avron}},
  \bibinfo{author}{\bibfnamefont{M.}~\bibnamefont{Fraas}}, \bibnamefont{and}
  \bibinfo{author}{\bibfnamefont{G.}~\bibnamefont{Graf}}, \bibinfo{journal}{J.
  Stat. Phys.} \textbf{\bibinfo{volume}{148}}, \bibinfo{pages}{800}
  (\bibinfo{year}{2012}), ISSN \bibinfo{issn}{0022-4715}.

\bibitem[{\citenamefont{Haupt et~al.}(2013)\citenamefont{Haupt, Leijnse, Calvo,
  Classen, Splettstoesser, and Wegewijs}}]{Haupt13}
\bibinfo{author}{\bibfnamefont{F.}~\bibnamefont{Haupt}},
  \bibinfo{author}{\bibfnamefont{M.}~\bibnamefont{Leijnse}},
  \bibinfo{author}{\bibfnamefont{H.~L.} \bibnamefont{Calvo}},
  \bibinfo{author}{\bibfnamefont{L.}~\bibnamefont{Classen}},
  \bibinfo{author}{\bibfnamefont{J.}~\bibnamefont{Splettstoesser}},
  \bibnamefont{and} \bibinfo{author}{\bibfnamefont{M.~R.}
  \bibnamefont{Wegewijs}}, \bibinfo{journal}{Phys. Stat. Solidi B}
  \textbf{\bibinfo{volume}{250}}, \bibinfo{pages}{2315} (\bibinfo{year}{2013}).

\bibitem[{\citenamefont{Riwar et~al.}(2013)\citenamefont{Riwar, Splettstoesser,
  and K\"onig}}]{Riwar13}
\bibinfo{author}{\bibfnamefont{R.-P.} \bibnamefont{Riwar}},
  \bibinfo{author}{\bibfnamefont{J.}~\bibnamefont{Splettstoesser}},
  \bibnamefont{and} \bibinfo{author}{\bibfnamefont{J.}~\bibnamefont{K\"onig}},
  \bibinfo{journal}{Phys. Rev. B} \textbf{\bibinfo{volume}{87}},
  \bibinfo{pages}{195407} (\bibinfo{year}{2013}).

\bibitem[{\citenamefont{Winkler et~al.}(2013)\citenamefont{Winkler, Governale,
  and K\"onig}}]{Winkler13}
\bibinfo{author}{\bibfnamefont{N.}~\bibnamefont{Winkler}},
  \bibinfo{author}{\bibfnamefont{M.}~\bibnamefont{Governale}},
  \bibnamefont{and} \bibinfo{author}{\bibfnamefont{J.}~\bibnamefont{K\"onig}},
  \bibinfo{journal}{Phys. Rev. B} \textbf{\bibinfo{volume}{87}},
  \bibinfo{pages}{155428} (\bibinfo{year}{2013}).

\bibitem[{\citenamefont{Rojek et~al.}(2014)\citenamefont{Rojek, Governale, and
  K{\"o}nig}}]{Rojek14}
\bibinfo{author}{\bibfnamefont{S.}~\bibnamefont{Rojek}},
  \bibinfo{author}{\bibfnamefont{M.}~\bibnamefont{Governale}},
  \bibnamefont{and}
  \bibinfo{author}{\bibfnamefont{J.}~\bibnamefont{K{\"o}nig}},
  \bibinfo{journal}{Phys. Stat. Solidi B} \textbf{\bibinfo{volume}{251}},
  \bibinfo{pages}{1912} (\bibinfo{year}{2014}).

\bibitem[{\citenamefont{Landsberg}(1992)}]{Landsberg92}
\bibinfo{author}{\bibfnamefont{A.~S.} \bibnamefont{Landsberg}},
  \bibinfo{journal}{Phys. Rev. Lett.} \textbf{\bibinfo{volume}{69}},
  \bibinfo{pages}{865} (\bibinfo{year}{1992}).

\bibitem[{\citenamefont{Landsberg}(1993)}]{Landsberg93}
\bibinfo{author}{\bibfnamefont{A.~S.} \bibnamefont{Landsberg}},
  \bibinfo{journal}{Modern Physics Letters B} \textbf{\bibinfo{volume}{07}},
  \bibinfo{pages}{71} (\bibinfo{year}{1993}).

\bibitem[{\citenamefont{Andersson}(2003)}]{Andersson03thesis}
\bibinfo{author}{\bibfnamefont{S.~B.} \bibnamefont{Andersson}}, Ph.D. thesis,
  \bibinfo{school}{University of Maryland}, \bibinfo{address}{College Park}
  (\bibinfo{year}{2003}).

\bibitem[{\citenamefont{Andersson}(2005)}]{Andersson05}
\bibinfo{author}{\bibfnamefont{S.~B.} \bibnamefont{Andersson}}, in
  \emph{\bibinfo{booktitle}{IFAC Symposium on Nonlinear Control Systems}}
  (\bibinfo{year}{2005}).

\bibitem[{\citenamefont{Sinitsyn}(2009)}]{Sinitsyn09}
\bibinfo{author}{\bibfnamefont{N.~A.} \bibnamefont{Sinitsyn}},
  \bibinfo{journal}{J. Phys. A} \textbf{\bibinfo{volume}{42}},
  \bibinfo{pages}{193001} (\bibinfo{year}{2009}).

\bibitem[{\citenamefont{Sinitsyn}(2007)}]{Sinitsyn07PRB}
\bibinfo{author}{\bibfnamefont{N.~A.} \bibnamefont{Sinitsyn}},
  \bibinfo{journal}{Phys. Rev. B} \textbf{\bibinfo{volume}{76}},
  \bibinfo{pages}{153314} (\bibinfo{year}{2007}).

\bibitem[{\citenamefont{Ren et~al.}(2010)\citenamefont{Ren, H\"anggi, and
  Li}}]{Ren10}
\bibinfo{author}{\bibfnamefont{J.}~\bibnamefont{Ren}},
  \bibinfo{author}{\bibfnamefont{P.}~\bibnamefont{H\"anggi}}, \bibnamefont{and}
  \bibinfo{author}{\bibfnamefont{B.}~\bibnamefont{Li}}, \bibinfo{journal}{Phys.
  Rev. Lett.} \textbf{\bibinfo{volume}{104}}, \bibinfo{pages}{170601}
  (\bibinfo{year}{2010}).

\bibitem[{\citenamefont{Yuge et~al.}(2013)\citenamefont{Yuge, Sagawa, Sugita,
  and Hayakawa}}]{Yuge13}
\bibinfo{author}{\bibfnamefont{T.}~\bibnamefont{Yuge}},
  \bibinfo{author}{\bibfnamefont{T.}~\bibnamefont{Sagawa}},
  \bibinfo{author}{\bibfnamefont{A.}~\bibnamefont{Sugita}}, \bibnamefont{and}
  \bibinfo{author}{\bibfnamefont{H.}~\bibnamefont{Hayakawa}},
  \bibinfo{journal}{J. Stat. Phys.} \textbf{\bibinfo{volume}{153}},
  \bibinfo{pages}{412} (\bibinfo{year}{2013}).

\bibitem[{\citenamefont{Yoshii and Hayakawa}(2013)}]{Yoshii13}
\bibinfo{author}{\bibfnamefont{R.}~\bibnamefont{Yoshii}} \bibnamefont{and}
  \bibinfo{author}{\bibfnamefont{H.}~\bibnamefont{Hayakawa}}
  (\bibinfo{year}{2013}), \bibinfo{note}{arXiv:1312.3772}.

\bibitem[{\citenamefont{Esposito et~al.}(2009)\citenamefont{Esposito, Harbola,
  and Mukamel}}]{Esposito09rev}
\bibinfo{author}{\bibfnamefont{M.}~\bibnamefont{Esposito}},
  \bibinfo{author}{\bibfnamefont{U.}~\bibnamefont{Harbola}}, \bibnamefont{and}
  \bibinfo{author}{\bibfnamefont{S.}~\bibnamefont{Mukamel}},
  \bibinfo{journal}{Rev. Mod. Phys.} \textbf{\bibinfo{volume}{81}},
  \bibinfo{pages}{1665} (\bibinfo{year}{2009}).

\bibitem[{\citenamefont{Yuge et~al.}(2012)\citenamefont{Yuge, Sagawa, Sugita,
  and Hayakawa}}]{Yuge12}
\bibinfo{author}{\bibfnamefont{T.}~\bibnamefont{Yuge}},
  \bibinfo{author}{\bibfnamefont{T.}~\bibnamefont{Sagawa}},
  \bibinfo{author}{\bibfnamefont{A.}~\bibnamefont{Sugita}}, \bibnamefont{and}
  \bibinfo{author}{\bibfnamefont{H.}~\bibnamefont{Hayakawa}},
  \bibinfo{journal}{Phys. Rev. B} \textbf{\bibinfo{volume}{86}},
  \bibinfo{pages}{235308} (\bibinfo{year}{2012}).

\bibitem[{\citenamefont{Liu et~al.}(2013)\citenamefont{Liu, Agarwalla, Wang,
  and Li}}]{Liu13}
\bibinfo{author}{\bibfnamefont{S.}~\bibnamefont{Liu}},
  \bibinfo{author}{\bibfnamefont{B.~K.} \bibnamefont{Agarwalla}},
  \bibinfo{author}{\bibfnamefont{J.-S.} \bibnamefont{Wang}}, \bibnamefont{and}
  \bibinfo{author}{\bibfnamefont{B.}~\bibnamefont{Li}}, \bibinfo{journal}{Phys.
  Rev. E} \textbf{\bibinfo{volume}{87}}, \bibinfo{pages}{022122}
  (\bibinfo{year}{2013}).

\bibitem[{\citenamefont{Berry}(1984)}]{Berry84}
\bibinfo{author}{\bibfnamefont{M.~V.} \bibnamefont{Berry}},
  \bibinfo{journal}{Proc. Roy. Soc.} \textbf{\bibinfo{volume}{392}},
  \bibinfo{pages}{45} (\bibinfo{year}{1984}).

\bibitem[{\citenamefont{Simon}(1983)}]{Simon83}
\bibinfo{author}{\bibfnamefont{B.}~\bibnamefont{Simon}},
  \bibinfo{journal}{Phys. Rev. Lett.} \textbf{\bibinfo{volume}{51}},
  \bibinfo{pages}{2167} (\bibinfo{year}{1983}).

\bibitem[{\citenamefont{Cheng et~al.}(2015)\citenamefont{Cheng, Tomczyk, Lu,
  Veazey, Huang, Irvin, Ryu, Lee, Eom, Hellberg et~al.}}]{Cheng15}
\bibinfo{author}{\bibfnamefont{G.}~\bibnamefont{Cheng}},
  \bibinfo{author}{\bibfnamefont{M.}~\bibnamefont{Tomczyk}},
  \bibinfo{author}{\bibfnamefont{S.}~\bibnamefont{Lu}},
  \bibinfo{author}{\bibfnamefont{J.~P.} \bibnamefont{Veazey}},
  \bibinfo{author}{\bibfnamefont{M.}~\bibnamefont{Huang}},
  \bibinfo{author}{\bibfnamefont{P.}~\bibnamefont{Irvin}},
  \bibinfo{author}{\bibfnamefont{S.}~\bibnamefont{Ryu}},
  \bibinfo{author}{\bibfnamefont{H.}~\bibnamefont{Lee}},
  \bibinfo{author}{\bibfnamefont{C.-B.} \bibnamefont{Eom}},
  \bibinfo{author}{\bibfnamefont{C.~S.} \bibnamefont{Hellberg}},
  \bibnamefont{et~al.}, \bibinfo{journal}{Nature}
  \textbf{\bibinfo{volume}{521}}, \bibinfo{pages}{196} (\bibinfo{year}{2015}),
  \bibinfo{note}{letter}.

\bibitem[{\citenamefont{Cheng et~al.}(2016)\citenamefont{Cheng, Tomczyk, Tacla,
  Lee, Lu, Veazey, Huang, Irvin, Ryu, Eom et~al.}}]{Cheng16}
\bibinfo{author}{\bibfnamefont{G.}~\bibnamefont{Cheng}},
  \bibinfo{author}{\bibfnamefont{M.}~\bibnamefont{Tomczyk}},
  \bibinfo{author}{\bibfnamefont{A.~B.} \bibnamefont{Tacla}},
  \bibinfo{author}{\bibfnamefont{H.}~\bibnamefont{Lee}},
  \bibinfo{author}{\bibfnamefont{S.}~\bibnamefont{Lu}},
  \bibinfo{author}{\bibfnamefont{J.~P.} \bibnamefont{Veazey}},
  \bibinfo{author}{\bibfnamefont{M.}~\bibnamefont{Huang}},
  \bibinfo{author}{\bibfnamefont{P.}~\bibnamefont{Irvin}},
  \bibinfo{author}{\bibfnamefont{S.}~\bibnamefont{Ryu}},
  \bibinfo{author}{\bibfnamefont{C.-B.} \bibnamefont{Eom}},
  \bibnamefont{et~al.}, \bibinfo{journal}{Phys. Rev. X}
  \textbf{\bibinfo{volume}{6}}, \bibinfo{pages}{041042} (\bibinfo{year}{2016}).

\bibitem[{\citenamefont{Hahn}(1933)}]{BochnerLecture}
\bibinfo{author}{\bibfnamefont{H.}~\bibnamefont{Hahn}},
  \bibinfo{journal}{Monatshefte f{\"u}r Mathematik und Physik}
  \textbf{\bibinfo{volume}{40}}, \bibinfo{pages}{A27} (\bibinfo{year}{1933}).

\bibitem[{\citenamefont{Levitov et~al.}(1996)\citenamefont{Levitov, Lee, and
  Lesovik}}]{Levitov96}
\bibinfo{author}{\bibfnamefont{L.~S.} \bibnamefont{Levitov}},
  \bibinfo{author}{\bibfnamefont{H.}~\bibnamefont{Lee}}, \bibnamefont{and}
  \bibinfo{author}{\bibfnamefont{G.~B.} \bibnamefont{Lesovik}},
  \bibinfo{journal}{Journal of Mathematical Physics}
  \textbf{\bibinfo{volume}{37}} (\bibinfo{year}{1996}).

\bibitem[{\citenamefont{Schaller et~al.}(2009)\citenamefont{Schaller,
  Kie\ss{}lich, and Brandes}}]{Schaller09}
\bibinfo{author}{\bibfnamefont{G.}~\bibnamefont{Schaller}},
  \bibinfo{author}{\bibfnamefont{G.}~\bibnamefont{Kie\ss{}lich}},
  \bibnamefont{and} \bibinfo{author}{\bibfnamefont{T.}~\bibnamefont{Brandes}},
  \bibinfo{journal}{Phys. Rev. B} \textbf{\bibinfo{volume}{80}},
  \bibinfo{pages}{245107} (\bibinfo{year}{2009}).

\bibitem[{\citenamefont{Nieto}(1968)}]{Nieto68}
\bibinfo{author}{\bibfnamefont{M.~M.} \bibnamefont{Nieto}},
  \bibinfo{journal}{Phys. Rev.} \textbf{\bibinfo{volume}{167}},
  \bibinfo{pages}{416} (\bibinfo{year}{1968}).

\bibitem[{\citenamefont{Carruthers and Nieto}(1968)}]{Carruthers68}
\bibinfo{author}{\bibfnamefont{P.}~\bibnamefont{Carruthers}} \bibnamefont{and}
  \bibinfo{author}{\bibfnamefont{M.~M.} \bibnamefont{Nieto}},
  \bibinfo{journal}{Rev. Mod. Phys.} \textbf{\bibinfo{volume}{40}},
  \bibinfo{pages}{411} (\bibinfo{year}{1968}).

\bibitem[{\citenamefont{Tsui}(1993)}]{Tsui93}
\bibinfo{author}{\bibfnamefont{Y.-K.} \bibnamefont{Tsui}},
  \bibinfo{journal}{Phys. Rev. B} \textbf{\bibinfo{volume}{47}},
  \bibinfo{pages}{12296} (\bibinfo{year}{1993}).

\bibitem[{\citenamefont{{H. Grabert}}(1991)}]{Grabert91}
\bibinfo{author}{\bibnamefont{{H. Grabert}}}, \bibinfo{journal}{Z. Phys. B}
  \textbf{\bibinfo{volume}{85}}, \bibinfo{pages}{319} (\bibinfo{year}{1991}).

\bibitem[{\citenamefont{Ingold and Nazarov}(1992)}]{IngoldNazarov92}
\bibinfo{author}{\bibfnamefont{G.}~\bibnamefont{Ingold}} \bibnamefont{and}
  \bibinfo{author}{\bibfnamefont{Y.~V.} \bibnamefont{Nazarov}}, in
  \emph{\bibinfo{booktitle}{Single charge tunneling}}, edited by
  \bibinfo{editor}{\bibfnamefont{H.}~\bibnamefont{Grabert}} \bibnamefont{and}
  \bibinfo{editor}{\bibfnamefont{M.~H.} \bibnamefont{Devoret}}
  (\bibinfo{publisher}{Plenum Press}, \bibinfo{address}{New York},
  \bibinfo{year}{1992}), vol. \bibinfo{volume}{B 294} of
  \emph{\bibinfo{series}{NATO ASI Series}}, chap. \bibinfo{chapter}{Charge
  tunneling rates in ultrasmall junctions}.

\bibitem[{\citenamefont{{K. Flensberg} et~al.}(1992)\citenamefont{{K.
  Flensberg}, {S. M. Girvin}, {M. Jonson}, {D. R. Penn}, and {M. D.
  Stiles}}}]{Flensberg92}
\bibinfo{author}{\bibnamefont{{K. Flensberg}}},
  \bibinfo{author}{\bibnamefont{{S. M. Girvin}}},
  \bibinfo{author}{\bibnamefont{{M. Jonson}}},
  \bibinfo{author}{\bibnamefont{{D. R. Penn}}}, \bibnamefont{and}
  \bibinfo{author}{\bibnamefont{{M. D. Stiles}}}, \bibinfo{journal}{Phys.
  Script.} \textbf{\bibinfo{volume}{T42}}, \bibinfo{pages}{189}
  (\bibinfo{year}{1992}).

\bibitem[{\citenamefont{Levitov and Lesovik}(1993)}]{Levitov93}
\bibinfo{author}{\bibfnamefont{L.~S.} \bibnamefont{Levitov}} \bibnamefont{and}
  \bibinfo{author}{\bibfnamefont{G.~B.} \bibnamefont{Lesovik}},
  \bibinfo{journal}{JETP Lett.} \textbf{\bibinfo{volume}{58}},
  \bibinfo{pages}{230} (\bibinfo{year}{1993}).

\bibitem[{\citenamefont{Schulenborg et~al.}(2016)\citenamefont{Schulenborg,
  Saptsov, Haupt, Splettstoesser, and Wegewijs}}]{Schulenborg16a}
\bibinfo{author}{\bibfnamefont{J.}~\bibnamefont{Schulenborg}},
  \bibinfo{author}{\bibfnamefont{R.~B.} \bibnamefont{Saptsov}},
  \bibinfo{author}{\bibfnamefont{F.}~\bibnamefont{Haupt}},
  \bibinfo{author}{\bibfnamefont{J.}~\bibnamefont{Splettstoesser}},
  \bibnamefont{and} \bibinfo{author}{\bibfnamefont{M.~R.}
  \bibnamefont{Wegewijs}}, \bibinfo{journal}{Phys. Rev. B}
  \textbf{\bibinfo{volume}{93}}, \bibinfo{pages}{081411}
  (\bibinfo{year}{2016}).

\bibitem[{\citenamefont{Moyal}(1949)}]{Moyal49}
\bibinfo{author}{\bibfnamefont{J.~E.} \bibnamefont{Moyal}},
  \bibinfo{journal}{Math. Proc. Cam. Phil. Soc.} \textbf{\bibinfo{volume}{45}},
  \bibinfo{pages}{99} (\bibinfo{year}{1949}).

\bibitem[{\citenamefont{Wigner}(1932)}]{Wigner32}
\bibinfo{author}{\bibfnamefont{E.}~\bibnamefont{Wigner}},
  \bibinfo{journal}{Phys. Rev.} \textbf{\bibinfo{volume}{40}},
  \bibinfo{pages}{749} (\bibinfo{year}{1932}).

\bibitem[{\citenamefont{Bondar et~al.}(2013)\citenamefont{Bondar, Cabrera,
  Zhdanov, and Rabitz}}]{Bondar13}
\bibinfo{author}{\bibfnamefont{D.~I.} \bibnamefont{Bondar}},
  \bibinfo{author}{\bibfnamefont{R.}~\bibnamefont{Cabrera}},
  \bibinfo{author}{\bibfnamefont{D.~V.} \bibnamefont{Zhdanov}},
  \bibnamefont{and} \bibinfo{author}{\bibfnamefont{H.~A.}
  \bibnamefont{Rabitz}}, \bibinfo{journal}{Phys. Rev. A}
  \textbf{\bibinfo{volume}{88}}, \bibinfo{pages}{052108}
  (\bibinfo{year}{2013}).

\bibitem[{\citenamefont{Rammer and Smith}(1986)}]{Rammer}
\bibinfo{author}{\bibfnamefont{J.}~\bibnamefont{Rammer}} \bibnamefont{and}
  \bibinfo{author}{\bibfnamefont{H.}~\bibnamefont{Smith}},
  \bibinfo{journal}{Rep. Math. Phys.} \textbf{\bibinfo{volume}{58}},
  \bibinfo{pages}{323} (\bibinfo{year}{1986}).

\bibitem[{\citenamefont{Pluecker}(2017)}]{Pluecker_phd}
\bibinfo{author}{\bibfnamefont{T.}~\bibnamefont{Pluecker}}, Ph.D. thesis,
  \bibinfo{school}{RWTH Aachen University} (\bibinfo{year}{2017}).

\bibitem[{\citenamefont{Fock}(1928)}]{Fock1928}
\bibinfo{author}{\bibfnamefont{V.}~\bibnamefont{Fock}},
  \bibinfo{journal}{Zeitschrift f{\"u}r Physik} \textbf{\bibinfo{volume}{49}},
  \bibinfo{pages}{323} (\bibinfo{year}{1928}).

\bibitem[{\citenamefont{Ivanov and Abanov}(2010)}]{Ivanov10}
\bibinfo{author}{\bibfnamefont{D.~A.} \bibnamefont{Ivanov}} \bibnamefont{and}
  \bibinfo{author}{\bibfnamefont{A.~G.} \bibnamefont{Abanov}},
  \bibinfo{journal}{Eur. Phys. Lett.} \textbf{\bibinfo{volume}{92}},
  \bibinfo{pages}{37008} (\bibinfo{year}{2010}).

\bibitem[{\citenamefont{Ren and Sinitsyn}(2013)}]{Ren13}
\bibinfo{author}{\bibfnamefont{J.}~\bibnamefont{Ren}} \bibnamefont{and}
  \bibinfo{author}{\bibfnamefont{N.~A.} \bibnamefont{Sinitsyn}},
  \bibinfo{journal}{Phys. Rev. E} \textbf{\bibinfo{volume}{87}},
  \bibinfo{pages}{050101} (\bibinfo{year}{2013}).

\bibitem[{\citenamefont{Berry and Robbins}(1993)}]{Berry93}
\bibinfo{author}{\bibfnamefont{M.~V.} \bibnamefont{Berry}} \bibnamefont{and}
  \bibinfo{author}{\bibfnamefont{J.~M.} \bibnamefont{Robbins}},
  \bibinfo{journal}{Proc. Roy. Soc.} \textbf{\bibinfo{volume}{442}},
  \bibinfo{pages}{659} (\bibinfo{year}{1993}).

\bibitem[{\citenamefont{L\'opez et~al.}(2012)\citenamefont{L\'opez, Lim, and
  S\'anchez}}]{Lopez12a}
\bibinfo{author}{\bibfnamefont{R.}~\bibnamefont{L\'opez}},
  \bibinfo{author}{\bibfnamefont{J.~S.} \bibnamefont{Lim}}, \bibnamefont{and}
  \bibinfo{author}{\bibfnamefont{D.}~\bibnamefont{S\'anchez}},
  \bibinfo{journal}{Phys. Rev. Lett.} \textbf{\bibinfo{volume}{108}},
  \bibinfo{pages}{246603} (\bibinfo{year}{2012}).

\bibitem[{\citenamefont{S{\'a}nchez and B{\"u}ttiker}(2012)}]{Sanchez12a}
\bibinfo{author}{\bibfnamefont{R.}~\bibnamefont{S{\'a}nchez}} \bibnamefont{and}
  \bibinfo{author}{\bibfnamefont{M.}~\bibnamefont{B{\"u}ttiker}},
  \bibinfo{journal}{Eur. Phys. Lett.} \textbf{\bibinfo{volume}{100}},
  \bibinfo{pages}{47008} (\bibinfo{year}{2012}).

\bibitem[{\citenamefont{S{\'a}nchez et~al.}(2013)\citenamefont{S{\'a}nchez,
  Sothmann, Jordan, and B{\"u}ttiker}}]{Sanchez13a}
\bibinfo{author}{\bibfnamefont{R.}~\bibnamefont{S{\'a}nchez}},
  \bibinfo{author}{\bibfnamefont{B.}~\bibnamefont{Sothmann}},
  \bibinfo{author}{\bibfnamefont{A.~N.} \bibnamefont{Jordan}},
  \bibnamefont{and}
  \bibinfo{author}{\bibfnamefont{M.}~\bibnamefont{B{\"u}ttiker}},
  \bibinfo{journal}{New. J. Phys.} \textbf{\bibinfo{volume}{15}},
  \bibinfo{pages}{125001} (\bibinfo{year}{2013}).

\bibitem[{\citenamefont{Belzig and Nazarov}(2001)}]{Belzig01}
\bibinfo{author}{\bibfnamefont{W.}~\bibnamefont{Belzig}} \bibnamefont{and}
  \bibinfo{author}{\bibfnamefont{Y.~V.} \bibnamefont{Nazarov}},
  \bibinfo{journal}{Phys. Rev. Lett.} \textbf{\bibinfo{volume}{87}},
  \bibinfo{pages}{197006} (\bibinfo{year}{2001}).

\bibitem[{\citenamefont{Nazarov and Kindermann}(2003)}]{Nazarov03b}
\bibinfo{author}{\bibfnamefont{Y.~V.} \bibnamefont{Nazarov}} \bibnamefont{and}
  \bibinfo{author}{\bibfnamefont{M.}~\bibnamefont{Kindermann}},
  \bibinfo{journal}{Eur. Phys. J. B} \textbf{\bibinfo{volume}{35}},
  \bibinfo{pages}{413} (\bibinfo{year}{2003}).

\bibitem[{\citenamefont{Bagrets and Nazarov}(2003)}]{Bagrets03}
\bibinfo{author}{\bibfnamefont{D.~A.} \bibnamefont{Bagrets}} \bibnamefont{and}
  \bibinfo{author}{\bibfnamefont{Y.~V.} \bibnamefont{Nazarov}},
  \bibinfo{journal}{Phys. Rev. B} \textbf{\bibinfo{volume}{67}},
  \bibinfo{pages}{085316} (\bibinfo{year}{2003}).

\bibitem[{\citenamefont{Holevo}(2011)}]{Holevobook11}
\bibinfo{author}{\bibfnamefont{A.~S.} \bibnamefont{Holevo}},
  \emph{\bibinfo{title}{{Probabilistic and Statistical Aspects of Quantum
  Theory; 2nd ed.}}}, Publications of the Scuola Normale Superiore Monographs
  (\bibinfo{publisher}{Springer}, \bibinfo{address}{Dordrecht},
  \bibinfo{year}{2011}).

\bibitem[{\citenamefont{Chruscinski and
  Jamiolkowski}(2004)}]{Chruscinskibook04}
\bibinfo{author}{\bibfnamefont{D.}~\bibnamefont{Chruscinski}} \bibnamefont{and}
  \bibinfo{author}{\bibfnamefont{A.}~\bibnamefont{Jamiolkowski}},
  \emph{\bibinfo{title}{Geometric Phases in Classical and Quantum Mechanics}},
  Progress in Mathematical Physics (\bibinfo{publisher}{Birkh{\"a}user Boston},
  \bibinfo{year}{2004}).

\bibitem[{\citenamefont{Lukacs}(1970)}]{lukacsBook}
\bibinfo{author}{\bibfnamefont{E.}~\bibnamefont{Lukacs}},
  \emph{\bibinfo{title}{Characteristic Functions}} (\bibinfo{publisher}{Hafner
  Publishing Company}, \bibinfo{year}{1970}).

\end{thebibliography}
\end{document}